\pdfoutput=1
\documentclass[pra,aps,twocolumn,nofootinbib,pdftex]{revtex4}

\usepackage{graphicx}
\usepackage{wasysym}
\usepackage{color}

\def\inbar{\,\vrule height1.5ex width.4pt depth0pt}
\def\IR{\relax{\rm I\kern-.18em R}}
\def\IC{\relax\hbox{$\inbar\kern-.3em{\rm C}$}}

\newcommand{\bra}[1]{\left\langle #1\right|}
\newcommand{\ket}[1]{\left| #1\right\rangle}
\newcommand{\braket}[2]{\left\langle #1|#2\right\rangle}
\newcommand{\ketbra}[2]{\left| #1\right\rangle\!\left\langle#2\right|}
\newcommand{\Tr}[0]{{\rm Tr}}
\newcommand{\hv}[1]{\hat{\vec{#1}}}
\newcommand{\pr}[0]{{\rm pr}}
\newcommand{\iea}[0]{{\it et al.~}}
\newcommand{\ieac}[0]{{\it et al., }}
\newcommand{\eqref}[1]{(\ref{#1})}
\newcommand{\eeqref}[1]{Eq.~(\ref{#1})}
\newcommand{\id}{\mathbf{I}}

\definecolor{MyGreen}{rgb}{0.0,0.5,0.1}

\begin{document}
\title{Continuous-variable optical quantum state tomography}

\author{A. I. Lvovsky}

\affiliation{Department of Physics and Astronomy\\ University of Calgary,\\ Calgary,
Alberta T2N 1N4, Canada\\ URL: www.iqis.org}

\author{M. G. Raymer}

\affiliation{Department of Physics and Oregon Center
for Optics\\ University of Oregon\\ Eugene, Oregon 97403, USA
\\email: raymer@uoregon.edu}

\begin{abstract}

This review covers latest developments in continuous-variable quantum-state tomography of optical fields and photons, placing a special accent on its practical aspects and applications in quantum information technology. Optical homodyne tomography is reviewed as a method of reconstructing the state of light in a given optical mode. A range of relevant practical topics are discussed, such as state-reconstruction algorithms (with emphasis on the maximum-likelihood technique), the technology of time-domain homodyne detection, mode matching issues, and engineering of complex quantum states of light. The paper also surveys quantum-state tomography for the transverse spatial state (spatial mode) of the field in the special case of fields containing precisely one photon.

\end{abstract}
\date{\today}

\maketitle
\tableofcontents

\newpage

\section{INTRODUCTION}
\label{sec:intro}

\subsection{The concept of quantum tomography}
A quantum state is what one knows about a
physical system. The known information is codified in a state vector
$\left| \psi  \right\rangle $, or in a density operator $\hat\rho $, in a way that enables the observer to make the best possible
statistical predictions about any future interactions (including
measurements) involving the system. Such a definition has a
comfortable interpretation within information theory, and so appears
natural in the context of research in quantum information (QI).

Imagine that an experimentalist, Alice, uses a well-characterized
procedure to prepare an individual particle in a particular physical state. Since Alice
possesses the information about the procedure, she can make
definite predictions about the particle's behavior under various
conditions, and is thus fully aware of the particle's state.

Now suppose Alice sends the prepared particle to another party, Bob, who is not aware of the preparation procedure, but wishes to determine the state of the particle.
By making observations on the particle, Bob can obtain information about the physical state prepared by Alice by observing how it interacts with other well-characterized
systems, such as a measurement apparatus\footnote{We can interpret the quantum state as a belief, or confidence
level, that a person has in his or her knowledge and ability to
predict future outcomes concerning the physical system (Fuchs, 2002).
No measurements can, generally speaking, provide full information on Alice's preparation procedure.}.
The amount and nature of this information depends strongly on whether the particle is macroscopic or microscopic.
In the macroscopic, classical case, Bob can observe the individual particle's
trajectory without disturbing it, and determine its state.


In quantum mechanics, on the contrary, it is impossible to learn the quantum
state of any individual physical system. Each observation, no matter how subtle, will disturb
its state just enough to prevent further observations from yielding
enough information for a state determination. This is the basis of
quantum key distribution for cryptography (Bennett and Brassard, 1984).

If Alice provides Bob with an ensemble of identically prepared systems,
then he can measure the same variable for each
system, and build up a histogram of outcomes, from which a
probability density can be estimated. According to the Born rule of
quantum theory, this measured probability density will equal the
square-modulus of the state-vector coefficients, represented in the
state-space basis corresponding to the measuring apparatus. This by
itself will not yet yield the full state information, since the phase of
the complex state-vector coefficients will be lost. 

As an example, measure the position $x$
 of each of 100,000 identically prepared electrons, which can move
only in one dimension. This yields an estimate of the position
probability density, or the square-modulus $\left| {\psi (x)}
\right|^2 $
 of the Schr\"odinger wave function. If the wave function has the form
$\left| {\psi (x)} \right|\exp [i\phi (x)]$, where $\phi (x)$
 is a spatially dependent phase, then we will need more information
than simply $\left| {\psi (x)} \right|^2 $
 in order to know the wave function. If we are able to measure the
momentum $p$
 of a second group of identically prepared electrons, then we can
estimate the probability density $\left| {\widetilde\psi (p)}
\right|^2$, where
\begin{equation}\label{ipx}
\widetilde\psi (p) = \int {\psi (x)\exp ( - i{\kern 1pt} x{\kern 1pt}
p/\hbar )dx}
\end{equation}
is the Fourier transform of the spatial wave function. If we know \emph{a
priori} that the ensemble can be described by a pure state, then we
can determine, by numerical methods, the complex wave function $\psi
(x)$, up to certain symmetry transformations (such as a complex
conjugation) just from these two types of measurement. This is a classic example of phase retrieval (Gerchberg and Saxon, 1972).

In the typical case, however, we do not know ahead of time if the system's
state
is pure or mixed. Then we must make \emph{many} sets of measurements on many
sub-ensembles, every time modifying the apparatus so that sets of
projection statistics associated with a different basis can be acquired. One
can then combine these results to reconstruct the density matrix of
the state. The data do not yield the state directly, but rather
indirectly through data analysis (i.e., a logical inference process).
This is the basis of \emph{quantum-state tomography} (QST). A set of observables whose measurements provide tomographically complete information about a quantum system is called a \emph{quorum} (Fano, 1957).

Niels Bohr (1958) seems to have had an intuitive idea of QST, when he said,
``A completeness of description like that aimed at in classical
physics is provided by the possibility of taking every conceivable
arrangement into account". A more rigorous concept was developed in theoretical
proposals (Newton and Young, 1968; Band and Park, 1970,
1971, 1979; Bertrand and Bertrand, 1987; Vogel and Risken, 1989), followed by
 the first
experiments determining the quantum state of a light field (Smithey \ieac 1993a,b).
Nowadays, quantum tomography has been applied to a variety of quantum systems and has become a
standard tool in QI research (Paris and \v{R}eh\'{a}\v{c}ek, 2004).

To continue the example of an electron moving in one dimension, a
quorum of variables can be constructed by measuring different groups
of electrons' positions $x' $
 after a variable length of time has passed. For example, in free
space this is $x'  = x + p{\kern 1pt} t/m$, where $m$
 is the electron mass. The wave function at time $t$
 is
\begin{equation}\label{elecprop}
\psi (x' ,t) = \int {G(x' ,x;t)\,\psi (x)dx},
\end{equation}
where
\begin{equation}\label{massprop}
G(x' ,x;t)=\sqrt{\frac{m}{2\pi i \hbar t}}\exp\left(\frac{im(x-x')^2}{2\hbar t}\right)
\end{equation}
 is the quantum propagator appropriate to the wave equation for the
particle. The experimentally estimated probability densities $
{\rm pr} (x' ,t) = \left| {\psi (x' ,t)} \right|^2 $, for all $t$
(positive and negative), provide sufficient information to invert
Eq.(\ref{elecprop}) and determine the complex state function $\psi (x)$
(assuming the functions ${\rm pr} (x' ,t)$
 are measured with very high signal-to-noise ratio). Note that
Eq.(\ref{elecprop}) can be interpreted as a generalization of Eq.(\ref{ipx}). As such,
it corresponds to a change of basis.

If the state is not known beforehand to be pure (that is, the
physical system's state is entangled with some other system), then it
is described by a density matrix, $\rho (x'_1 ,x'_2;t)$, in which case
the probability densities correspond to the average,
\begin{eqnarray}\label{dmprobdens}
\rho (x'_1,x'_2;t)&=&\int \int \,dx_1\,dx_2
\\ \nonumber&\times&G^*(x'_1 ,x_1;t)G(x'_2 ,x_2;t)\rho (x_1,x_2;0).
 \end{eqnarray}
Through inversion of Eq.~(\ref{dmprobdens}), the set of the measured probability
functions ${\rm pr} (x' ;t)$
 determines the density matrix $\rho (x_1,x_2;0)$.

 This procedure works in principle for a Schr\"odinger equation with
an arbitrary, known potential-energy function. Such a
method was proposed (Raymer \ieac 1994;
Janicke and Wilkens, 1995; Leonhardt and Raymer, 1996; Raymer, 1997b) and implemented
(Kurtsiefer \ieac 1997) for the transverse spatial state of an ensemble of
helium atoms and the classical light beam (McAlister \ieac 1995).


\subsection{Quantum tomography of light}

The current interest in QST is motivated by recent developments in QI processing, which requires {\it inter alia} a technique for detailed characterization of quantum states involved (Paris and \v{R}eh\'{a}\v{c}ek, 2004). Additionally, significant progress has been made in measurement technologies, which now allow experimenters to measure a set of observables sufficiently diverse to allow reliable state
reconstruction from the data.

Among many physical systems in which QI processing can be implemented, light is of particular significance because it is mobile and thus irreplaceable as an information carrier in quantum communication networks. In this article, we review the basis of and methods for QST of
optical fields.

Even specialized to light, quantum tomography is too vast a field to be fully covered in a single review paper. Here we choose to concentrate on optical QST which involves measuring continuous degrees
of freedom: field amplitude and/or spatial distribution. We study two tomographic problems which at first appear to have little in common. The first deals with the case in which the mode of the field is known (or chosen) \emph{a priori}, and the state of this mode is to be determined. The second deals with the case that the full field is known to contain exactly one photon, but the manner in which this photon is distributed among spatial and spectral modes is to be determined. A careful analysis shows that these problems are strongly related in their mathematical methods for reconstructing a quantum state.

\subsubsection{Optical homodyne tomography}\label{OTHintrosec}
The first problem is the characterization of the state of the optical field in a certain spatiotemporal mode. The Hamiltonian of an electromagnetic mode is equivalent to that of the harmonic oscillator.
Quantum states of light can thus be reconstructed similarly to motional states of massive particles discussed in the previous section. This is done by measuring quantum noise statistics of the field amplitudes at different optical phases (Leonhardt, 1997). The procedure of this reconstruction is known as \emph{optical homodyne tomography} (OHT).

It is interesting, in the context of this article, that homodyne tomography was the first
experimental demonstration of optical QST.
Using balanced homodyne detection (BHD), Smithey {\it et al.} (1993a)
 measured a set of probability densities for the quadrature
amplitudes of a squeezed state of light. These histograms were
inverted using the inverse Radon transform, familiar from medical
tomographic imaging, to yield a reconstructed Wigner distribution and
density matrix for a squeezed state of light. This 1993 paper
introduced the term ``tomography'' into quantum optics.

OHT is the subject of the next four sections of this article. In Sections II and III, we discuss the concept of homodyne tomography and the methods of reconstructing the state's Wigner function and density matrix from a set of experimental data. Special attention is paid to the likelihood-maximization technique, which is now most commonly used. Section IV is devoted to technical issues arising in experimental OHT. In Section V, we discuss applications of OHT in experiments of engineering and characterizing specific quantum states of light, such as photons, qubits, and ``Schr\"odinger cat" states.

In the context of applications, it is instructive to compare OHT with another technique: determining the quantum state of a system of dual-rail optical qubits\footnote{In the dual-rail qubit, the logical value is assigned to a single photon being in
one of two orthogonal modes $A$ or $B$:
$\ket{\tilde 0}=\ket{1_A,0_B},\ \
\ket{\tilde 1}=\ket{0_A,1_B}$, where the right-hand
side is written in the photon number (Fock) basis for each mode.} by measuring relative photon number statistics in each mode and in their various linear superpositions. Due to its relative simplicity, this approach has been popular in a wide variety of experiments (see Altepeter \ieac 2004 for a review).

A textbook example of the above is the work of James \iea (2001). In this experiment the polarization state of a pair of entangled photons $A$ and $B$ generated in type-II parametric down-conversion was analyzed by measuring photon coincidence count statistics in sixteen polarization projections. Tomographic analysis has revealed the photons to be almost perfectly in the state
\begin{equation} \label{bellpc}
\ket{\Psi}=\frac 1 {\sqrt 2}\left(\ket{H_AV_B}+\ket{V_AH_B}\right),
\end{equation}
where H and V indicate horizontal and vertical polarization. It is tempting to say that such a state has high entanglement.

This analysis does not reveal, however, that the photon pair is generated not ``on demand'', but with some probability $\varepsilon^2$, which is usually low. A more complete representation of the state of the optical modes analyzed could be
\begin{eqnarray} \label{bellfull}
\ket{\Psi} &=& \left| 0_{AH}0_{AV}0_{BH}0_{BV} \right\rangle   \\ \nonumber
&+&\varepsilon \left( {\left| 1_{AH}0_{AV}1_{BH}0_{BV} \right\rangle   + \left| 0_{AH}1_{AV}0_{BH}1_{BV} \right\rangle } \right)\\ \nonumber  &+&O(\varepsilon^2),
\end{eqnarray}
where, for example, $\ket 1_{AV}$ indicates a one-photon state present in the vertical polarization mode of channel $A$.
The bipartite entanglement is of the order $\left| {\varepsilon ^2 \log (\varepsilon )} \right|$, which is much less than 1.

Eqs.~\eqref{bellpc} and \eqref{bellfull} reveal a significant limitation of the photon counting approach. This method works well if it is {\it a priori} known that the modes
involved are in one of the qubit basis states or their linear combination. In practice, however, this is not always the case: photons can be lost, or multiple photons can be present where we expect only one. Such events compromise performance of quantum logical gates, but usually go unrecognized by the photon counting approach; they are simply eliminated from the analysis\footnote{An important exception is the work by Chou \iea (2005), where the photon-counting approach is used, but the vacuum contribution is accounted for when evaluating the entanglement of a dual-rail qubit.}. As a result, characterized is not the true quantum
state of the carrier modes, but its \emph{projection} onto the qubit
subspace of the optical Hilbert space. This may lead to false estimation of gate performance benchmarks (van Enk \ieac 2007).

OHT, on the contrary, permits \emph{complete} characterization of the field state in a particular spatiotemporal mode, taking into account the entire Hilbert space of quantum optical states. It thus provides more reliable information about the performance and scalability of an optical QI processor. However, it is also more technically involved as it requires matched local oscillators, sophisticated detection electronics, larger POVMs and measurement data sets. It is thus less suitable for characterizing multimode states: to date, the largest qubit systems measured using photon counting contained six qubits (Lu \ieac 2007) while with OHT, only single dual-rail qubits were reconstructed (Babichev \ieac 2004b).


\subsubsection{Optical mode tomography}

An altogether different use of QST arises when a
light field is known to contain a definite number of photons,
but their distribution over spatial and/or spectral modes is unknown. If the set
of modes is \emph{discrete}, (e.g. in the case of polarization qubits),
characterization can be done using the photon counting method discussed above (Altepeter {\it et al.,} 2004). But if the distribution of light particles over electromagnetic modes is described by a \emph{continuous} degree of freedom,
methods of continuous-variable QST become irreplaceable.

The problem of reconstructing the modal distribution of a field state is largely analogous to determining the spatial
wave function $\psi (\vec r )$
 of a massive particle, as described by Eq.(\ref{elecprop}). It turns
out that the procedure outlined via Eqs.(\ref{ipx})--(\ref{dmprobdens}) also applies to
QST for an ensemble of single photons, and is in fact quite similar to the method of OHT.  This is consistent with the adoption
of a sensible definition of a photon's spatial wave function (Sipe, 1995; Bialynicki-Birula, 1996), in which the Schr\"odinger equation
is replaced by the Maxwell equations (since a photon cannot be
strictly localized in space, some subtleties must be taken into account). In Section \ref{spatialtomo} of this review we analyze various techniques and recent experimental progress in reconstructing spatial optical modes of single photons as well as entangled pairs.

\section{The principles of homodyne tomography} \label{part2}
\subsection{Balanced homodyne detection} \label{BHDsec}

The technique of balanced homodyne detection (BHD) and homodyne tomography has been extensively described
in the literature, for example, in the textbook of
Leonhardt (1997) and in recent reviews by Raymer and
Beck (2004), Zhang (2004) and Zavatta \iea (2005). Here, we present only a brief introduction with concentration on theoretical aspects of mode matching between the local oscillator and the signal field.

\begin{figure}
\includegraphics[keepaspectratio,width=0.45\textwidth]{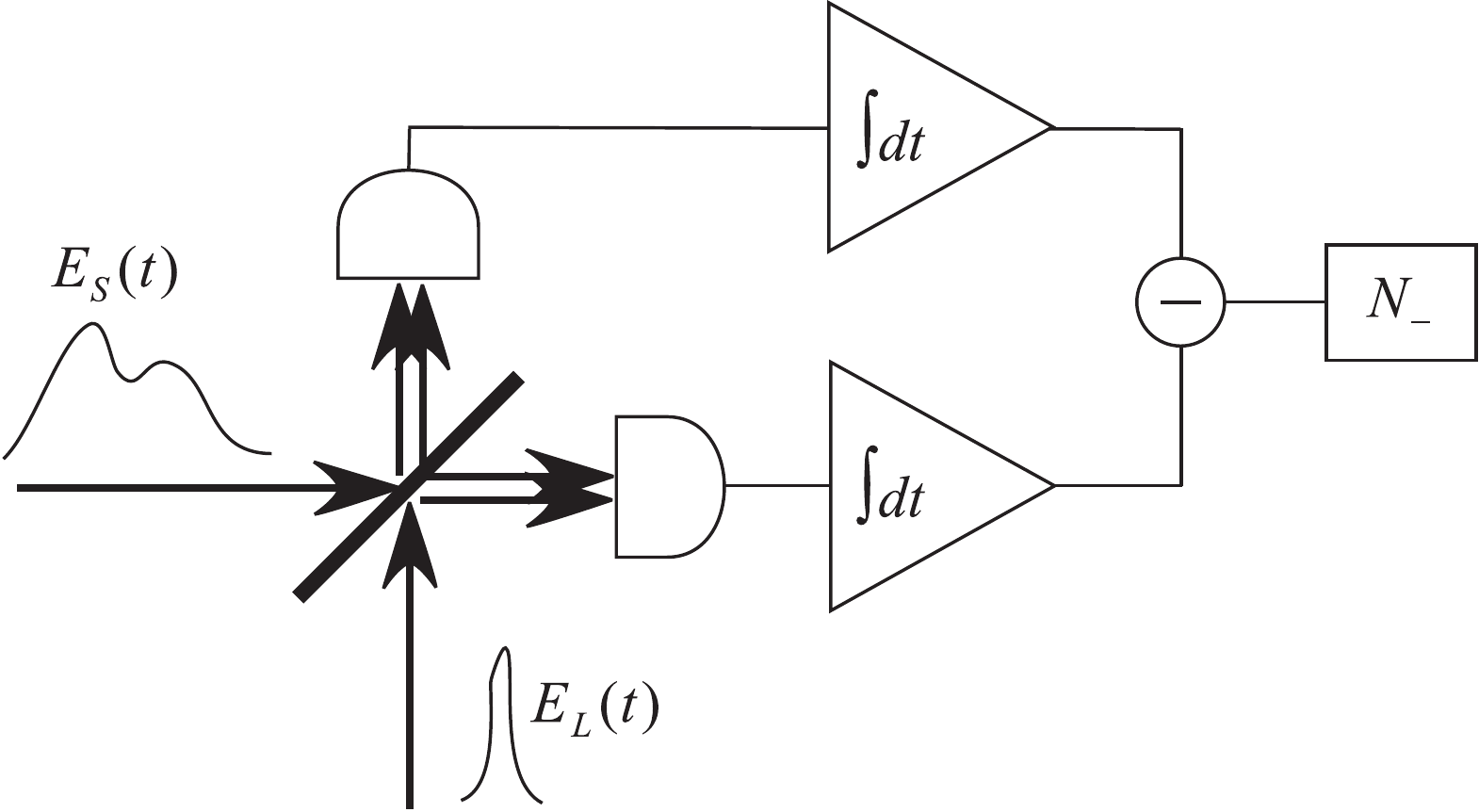}
\vspace*{8pt}
\caption{\label{figbhd}Balanced homodyne detection}
\end{figure}

Figure \ref{figbhd} illustrates balanced homodyne detection, which is a means to measure the amplitude of any phase component of a light mode. In BHD, the weak signal field $\hv{E}_S (t)$  (which may be multimode) and a strong coherent local oscillator (LO) field $\hv{E}_L (t)$
 are overlapped at a 50\% reflecting beam splitter, and the two interfered fields are detected, temporally integrated, and subtracted.

The signal electric field operator is written as a sum of positive- and
negative-frequency parts, which are conjugates of one another,
${\hv E} _S  = {\hv E} _S^{( + )}  + {\hv E} _S^{( - )} $. The positive-frequency part can be decomposed into plane waves according to Dirac's quantization scheme
\begin{equation}\label{Esrt}
{\hv E} _S^{( + )} (\vec r ,t) = i\,\sum\limits_j^{} \sqrt {\frac{\hbar \omega _j }{2\varepsilon _0 V }}\hat b_j \vec\epsilon _j\exp( i\vec k_j\vec r- i\omega _j t),
\end{equation}
 where $\omega_j$, $\vec k_j$, and $\vec\epsilon_j$ are, respectively, the mode frequency, wave vector, and the unit polarization vector; the creation and annihilation operators obey the commutator
$[\hat b_j ,\hat b_{j'}^\dag  ] = \delta _{jj'}$ and are defined in some large volume V (which may be taken to infinity later). It is convenient to consider the signal field in the paraxial approximation with $z$ being the propagation axis. In this case $\omega _j\approx ck_{jz}$ and the polarization $\vec\epsilon_j$ is along either $x$ or $y$.

The LO field is treated classically, and at each photodiode face ($z=0$) is assumed to be a strong coherent pulse propagating along the $z$ axis,
\begin{eqnarray}\label{ELO}
&&{\vec E }_L^{( + )}  (\vec r,t)
\\ \nonumber&&\hspace{0.5 cm}=i\sqrt {\frac{\hbar \omega _L }{2\,\varepsilon _0 V }} \alpha _L \vec\epsilon_L v _L (x,y)\,g_L (t)\exp(ik_Lz-i\omega_Lt),
\end{eqnarray}
where the coherent-state amplitude is $\alpha _L  = \left| {\alpha _L } \right|e^{i\theta } $, and
$v_L(x,y)g_L (t)$ is the normalized spatiotemporal mode.

The local oscillator and the signal fields meet at a beam splitter, where they undergo the transformation
\begin{equation}
(\vec E_L,\vec E_S) \to \left(\frac{\vec E_L+\vec E_S}{\sqrt 2},\frac{\vec E_L-\vec E_S}{\sqrt 2}\right).
\end{equation}
The difference of the numbers of photoelectrons recorded in the two beam splitter outputs is then, assuming a perfect detection efficiency (see Raymer {\it et al.}, 1995; Raymer and Beck, 2004 for details and more general considerations),
\begin{eqnarray}\label{Nminus}
\hat N_- &=&\int\limits_{\rm Det}\int\limits_{\Delta t}\frac{\varepsilon_0 V}{c\hbar\omega}(2\vec E_L\hv E_s) dt\,dx\,dy\\ \nonumber
&=&  \left| {\alpha _L } \right|\left( {\hat a\,e^{ - i\theta }  + \hat a^\dag  \,e^{i\theta } \,} \right),
\end{eqnarray}
where the integration is done over the detector sensitive area and the measurement time $\Delta t$. Assuming that the above fully accommodate the local oscillator pulse, all integration limits in Eq.~(\ref{Nminus}) can be assumed infinite. The photon creation operator $\hat a^\dag$ associated with the detected spatiotemporal mode is given by
\begin{equation}\label{adag}
\hat a^\dag   = \sum\limits_j {C_j } \hat b_j^\dag,
\end{equation}
where the $C_j$'s  equal the Fourier coefficients for the LO pulse,
\begin{eqnarray}\label{Cj}
C_j&=&\vec\epsilon\,^*_L\vec\epsilon_j \int \int v^*_L(x,y)g^*_L (t)\hspace*{3cm} \\
\nonumber&\times&\exp\left( ik_{jx}x+ik_{jy}y- ic(k_{jz}-k_L) t\right)dt\,dx\,dy.
\end{eqnarray}

When using a pulsed LO field $E_L (t)$, the concept of a light mode needs to be generalized beyond the common
conception as a monochromatic wave. As first discussed by Titulaer and Glauber (1966),
a polychromatic light wave packet can be considered a mode with a well defined spatial-temporal shape,
whose quantum state is described in the usual way using photon creation and annihilation operators.
For example, a one-photon wave-packet state is created by $\ket{1_{\hat a}}=\hat a^\dag  \ket{\rm vac}$ (more on this in Sec.~\ref{photonwfsec}).
The meaning of Eqs.(\ref{adag}), (\ref{Cj}) is that the
\emph{BHD detects the state of the electromagnetic field in the
spatial-temporal mode defined by the LO pulse} (Smithey \ieac 1993a; Raymer \ieac 1995;
Raymer and Beck, 2004).
This allows temporal and spatial selectivity, or gating, of the signal field (not the signal intensity).
This gating technique (linear-optical sampling) has application in ultrafast signal characterization (Dorrer \ieac 2003; Raymer and Beck, 2004).


As usual, the mode's annihilation operator can be expressed as a sum of
Hermitian operators $\hat a = e^{i\,\theta } ( {\hat Q_\theta   + i\hat P_\theta  } )/\sqrt{2}$,
called quadrature amplitudes, with\footnote{Some authors use the convention $[\hat  Q_\theta  ,\hat  P_\theta  ] = i/2$. All quadrature-dependent plots used in this article have been (re-)scaled to comply with the uniform convention $[\hat  Q_\theta  ,\hat  P_\theta  ] = i$.} $[\hat  Q_\theta  ,\hat  P_\theta  ] = i$.
For zero phase, $\hat Q_\theta  ,\,\hat P_\theta$ are denoted $\hat Q,\,\hat P$, respectively (so $\hat Q_\theta=\hat Q \cos\theta+\hat P\sin\theta$),
and are analogous to position and momentum variables for a massive harmonic oscillator.
For the LO phase equal to $\theta$, BHD measures the quadrature amplitude
\begin{equation}\label{Qtheta}
\hat  N_ -  /(|\alpha _L |\sqrt{2} ) = (\hat  a\,e^{ - i\,\theta }  + \hat  a^\dag  \,e^{i\,\theta } )/\sqrt{2} = \hat  Q_\theta  .
\end{equation}
According to quantum mechanics, the probability density for observing the
quadrature equal to $Q_\theta$ for the field in the signal mode given by the density operator $\hat\rho$ is
\begin{equation}\label{prsimple}
{\rm pr}(Q_\theta,\theta )=\bra{Q_\theta}\hat\rho\ket{Q_\theta},
\end{equation}
where $\ket{Q_\theta}$ is the quadrature eigenstate. These probability densities, also known as \emph{marginal distributions}, are histograms of the field amplitude noise samples measured with the homodyne detector. The optical phase plays the role of time in Eqs.~(\ref{ipx})--(\ref{dmprobdens}).
When it is varied over one complete cycle, quadrature amplitudes $Q_\theta$
form a quorum for QST (Vogel and Risken, 1989).

In a
practical experiment, the photodiodes in the homodyne detector are
not 100\% efficient, i.e. they do not transform every incident
photon into a photoelectron. This leads to a distortion of the
quadrature noise behavior which needs to be compensated for in the
reconstructed state. We present, without derivation, a generalization of the above
expression for detectors with a non-unitary quantum efficiency $\eta$ (Raymer and Beck, 2004;
Raymer \ieac 1995):
\begin{equation}\label{prcomplex}
{\rm pr}(Q_\theta,\theta ) = \langle :\frac{{\exp[ - (Q_\theta/\eta   - \hat  Q_\theta  )^2 /2\sigma ^2 ]}}{{ \sqrt{2\pi \sigma ^2 }  }}:\rangle
\end{equation}
where $2\sigma ^2  = 1/\eta$ and the brackets indicate a quantum expectation value. The double dots indicate normal operator ordering (annihilation operators to the right of creation operators).

\subsection{Wigner function}
\begin{figure}
\includegraphics[keepaspectratio,width=0.45\textwidth]{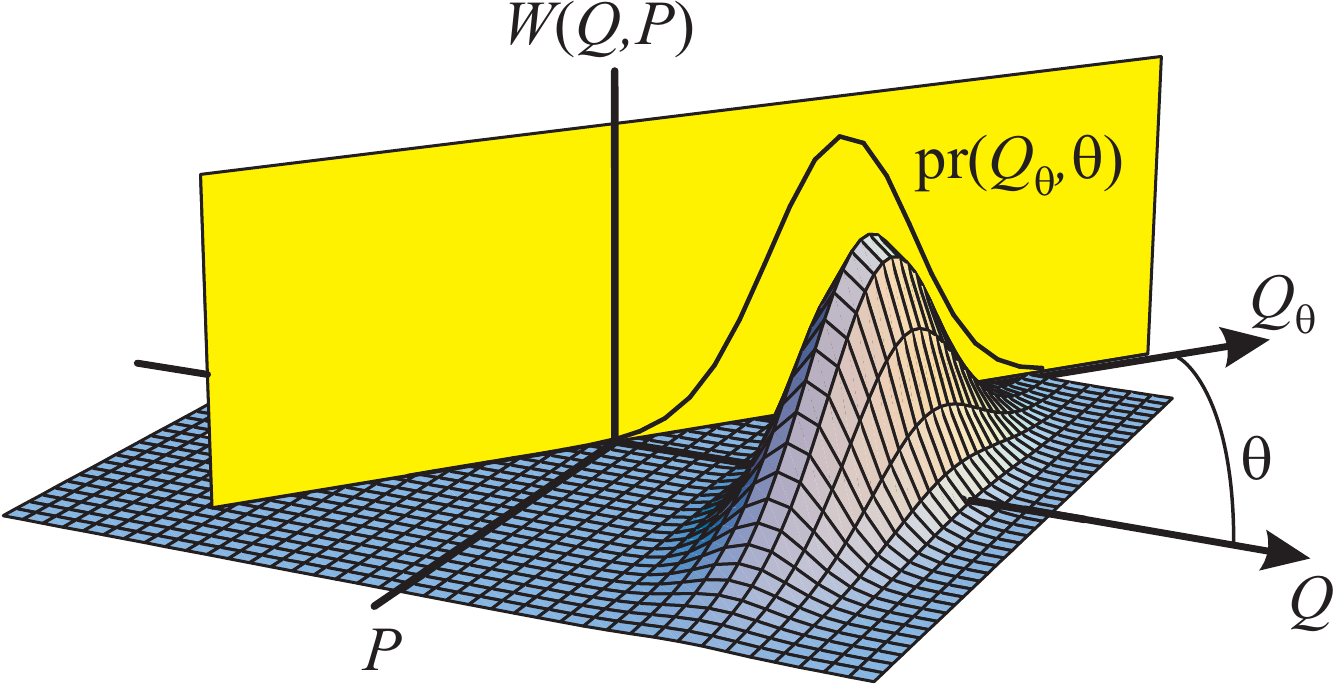}
\vspace*{28pt}
\caption{\label{wigfig}The Wigner function. The experimentally measured field quadrature probability density ${\rm pr} (Q_\theta  ,\theta )$ is the integral projection of the Wigner function $W(Q,P)$ onto a vertical plane defined by the phase of the local oscillator.}
\end{figure}

Because the optical state reconstructed using tomography is generally non-pure, its canonical representation is in the form of a density matrix, either in the quadrature basis or in the photon-number (Fock) basis. In the case of homodyne tomography it is convenient to represent the reconstructed state in the form of the phase-space quasiprobability density, the Wigner function (Wigner, 1932).
\begin{equation}\label{wigdef}
W_{\hat\rho}(Q,P) = \frac{1}{{2\pi }}\int_{ - \infty }^\infty  {\langle Q + \frac{1}{2}Q'|\hat  \rho |Q - \frac{1}{2}Q'\rangle \,e^{ - i\,P\,Q\,'} } dQ'.
\end{equation}
This object uniquely defines the state and, at the same time, is directly related to the quadrature histograms (\ref{prsimple}, \ref{prcomplex}) measured experimentally (Raymer \ieac 1995;
Raymer and Beck, 2004) via the integral
\begin{eqnarray}\label{prwig}
{\rm pr} (Q_\theta  ,\theta ) \hspace{7cm} &&\\ \nonumber
= \int\limits_{-\infty}^{+\infty}\int\limits_{-\infty}^{+\infty} \delta (Q_\theta- Q \cos\theta  - P  \sin\theta)\,W_{\rm Det} (Q,P)  dQ\,dP \hspace{3mm} && \\ \nonumber
= \int\limits_{ - \infty }^\infty  W_{\rm Det} (Q_\theta  \cos\theta  - P_\theta  \sin\theta ,Q_\theta  \sin\theta  + P_\theta  \cos\theta ) dP_\theta.\hfill &&
\end{eqnarray}
In other words, the histogram ${\rm pr}(Q_\theta,\theta )$ is the integral projection of the Wigner function onto a vertical plane oriented at angle $\theta$ to the $Q$ axis (Fig.~\ref{wigfig}). The ``detected" Wigner function $W_{\rm Det}$ corresponds to the ideal Wigner function (\ref{wigdef}) for a loss-free detector,
and for a detector with quantum efficiency $\eta$ it is obtained from the latter via a convolution (Leonhardt and Paul, 1993; Kuhn \ieac 1994; Raymer \ieac 1995; Leonhardt, 1997)
\begin{eqnarray}
\label{wigdet}
&&W_{\rm Det} (Q,P)=  \frac{1}{\pi (1-\eta)} \int\limits_{-\infty}^{+\infty}\int\limits_{-\infty}^{+\infty} W(Q',P')\\
&&\hspace{0.5cm}\times
 {\exp \left[ - \frac{(Q - Q'\sqrt{\eta})^2  + (P - P'\sqrt{\eta})^2 }{1-\eta} \right]\;}  dQ'dP'.\nonumber
\end{eqnarray}

\section{Reconstruction algorithms}
A homodyne tomography experiment yields a set of pairs $(Q_m,\theta_m)$, which can be binned up to form marginal distributions $\pr(Q,\theta)$ for several local oscillator phases. Our next task is to develop mathematical methods that can be used to convert the experimental data into the state's density matrix and/or Wigner function. This is the subject of this section of the review.

Mathematical methods of OHT can be divided into two categories.
The so-called \emph{inverse linear transform} techniques (Sec.~\ref{detsec}) use the fact that the experimentally measured marginal distributions are integral projections of the Wigner function. Because integration is a linear operation, one can reverse it and reconstruct the Wigner function from the set of marginals in a procedure that somewhat resembles solving a system of linear equations of the form \eqref{prwig}. We discuss this and related methods in Sec.~\ref{detsec}; they are reviewed in more detail in (Paris and \v{R}eh\'{a}\v{c}ek, 2004; Leonhardt, 1997; D'Ariano, 1997; Welsch \ieac 1999; Raymer and Beck, 2004).

For reasons discussed later (Sec.~\ref{maxliksec}), inverse linear transform methods are rarely used in modern OHT. More frequently, we employ methods of \emph{statistical inference}, whose classical versions have been
developed in traditional statistics and data analysis (Paris and \v{R}eh\'{a}\v{c}ek, 2004, chapters 2,3,6, and 10). A popular method is likelihood maximization (MaxLik), which looks for the most
probable density matrix that will generate the observed data. It
is discussed in detail in Sec.~\ref{maxliksec}. Another statistical inference method, entropy maximization, is briefly reviewed in Sec.~\ref{otherstat}.
\subsection{State reconstruction via inverse linear transformation}\label{detsec}
\subsubsection{Inverse Radon transformation}
The projection integral (\ref{prwig}), known as the \emph{Radon transform} (Herman, 1980), can be inverted numerically using the back-projection algorithm,
 familiar from medical imaging (Herman, 1980; Leonhardt, 1997) to reconstruct the phase-space density $W_{\rm Det} (Q,P)$:
 \begin{eqnarray}\label{invRadon}
W_{\rm Det}(Q,P)&=&\frac{1}{2\pi^2}\int\limits_0^\pi\int\limits_{-\infty}^{+\infty}{\rm pr}(Q_\theta,\theta)\\ \nonumber &\times& K(Q\cos\theta+P\sin\theta-Q_\theta)\,dQ_\theta\,d\theta,
\end{eqnarray}
with the integration kernel
\begin{equation}\label{fbpakernel}
K(x)=\frac{1}{2}\int\limits_{-\infty}^{+\infty}|\xi|\exp(i\xi x)d\xi=-\mathcal{P}\frac{1}{x^2},
\end{equation}
where $\mathcal{P}$ denotes a principle value integration.

\begin{figure}[b]
\includegraphics[keepaspectratio,width=0.41\textwidth]{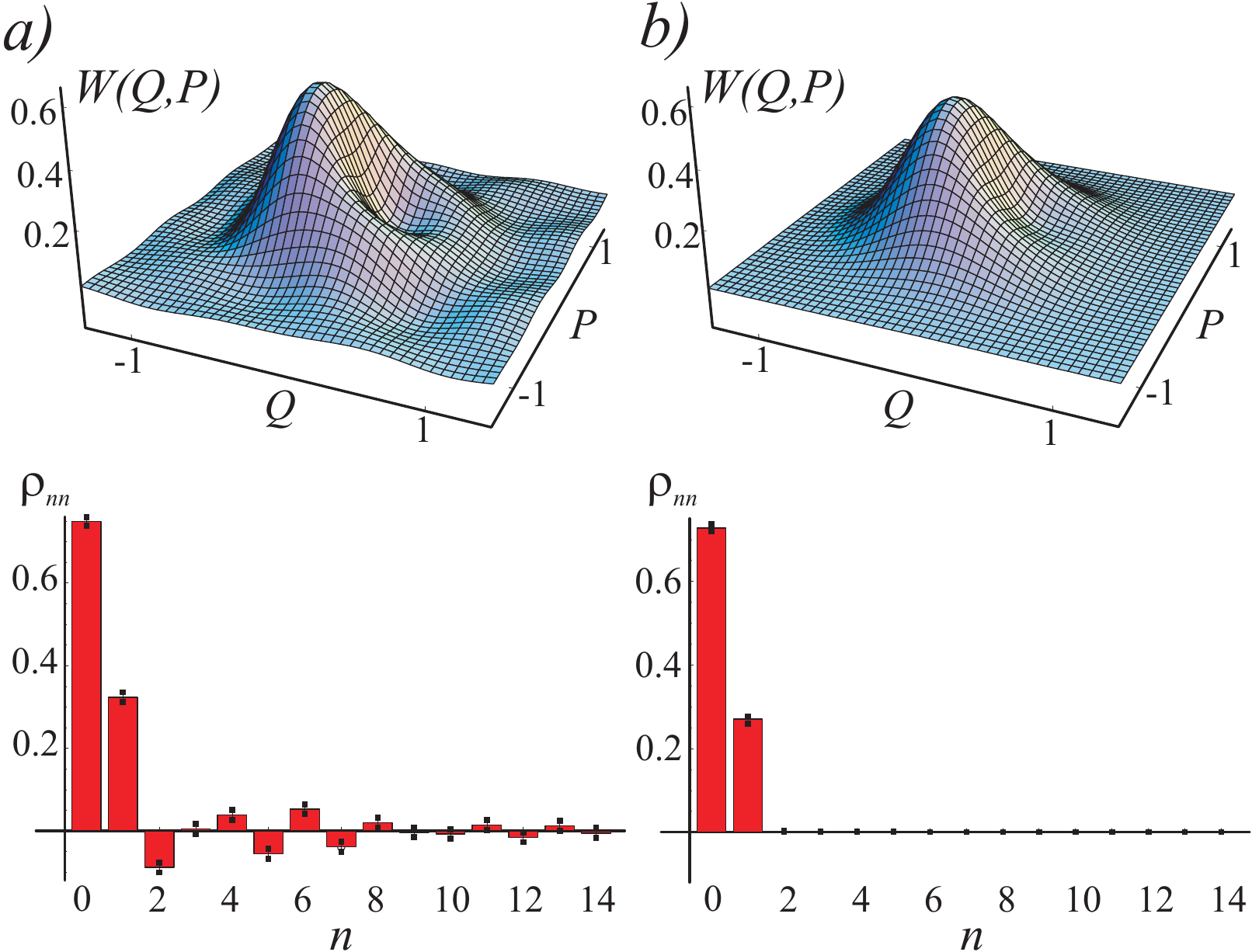}
\vspace*{8pt}
\caption{\label{CompFig} Quantum optical state estimation from a
set of 14152 experimental homodyne measurements (Lvovsky and Mlynek, 2002)
by means of the inverse Radon transformation and the pattern--function method (a) and the
likelihood maximization algorithm (b). The Wigner function and the
diagonal elements of the reconstructed density matrix are shown.
The inverse Radon transformation in (a) was performed by means of
the filtered back-projection algorithm. The statistical uncertainties in (b) were determined by
means of a Monte-Carlo simulation (see text).}
\vspace{10cm}
\end{figure}

The kernel is infinite at $x=0$, so in numerical implementations of the inverse Radon transformation it is subjected to low pass filtering: the infinite integration limits in Eq. (\ref{fbpakernel}) are replaced by $\pm k_c$, with $k_c$ chosen so as to reduce the numerical artifacts associated with the reconstruction while keeping the main features of the Wigner function [see e.g. Fig.~\ref{CompFig}(a)]. This method is known as the \emph{filtered back-projection algorithm}.

This strategy was used in the first QST experiments (Smithey \ieac 1993a; Dunn \ieac 1995). In later implementations of this algorithm (Lvovsky and Babichev, 2002), the intermediate step of binning the data
and calculating individual marginal distributions associated with
each phase was bypassed: the summation of Eq.~(\ref{invRadon})
was applied directly to acquired pairs $(\theta_m,Q_m)$:
\begin{equation}
W_{\rm Det}(Q,P)\cong\frac{1}{2\pi^2N}\sum\limits_{m=1}^NK(Q\cos\theta_m+P\sin\theta_m-Q_m),
\end{equation}
with phases $\theta_m$ uniformly spread over the $2\pi$ interval.

A nonclassical state of light\footnote{See, for example, Lvovsky and Shapiro (2002), as well as Zavatta \iea (2007) for a review of definitions and measurable criteria of a nonclassical nature of a state of light.}, after undergoing an optical loss, becomes nonpure. Therefore, typically, the reconstructed state is not pure, that  is, $\Tr[\hat  \rho _{\rm Det} ^2 ] \ne 1$. A special case is that of a coherent state, which is not nonclassical and remains pure under losses. For such a state, one can reconstruct the Schroedinger wave function or a state vector, as demonstrated in Smithey {\it et al.} (1993b).

A more general method for reconstructing the Wigner function of a state that has undergone optical losses is proposed in Butucea \iea ~(2005), where Eqs.~(\ref{invRadon}) and (\ref{fbpakernel}) are modified so as to incorporate the effect of nonunitary efficiency. This paper, as well as Gu\c t\u a and Artiles (2006), also perform a minimax analysis of the error in the evaluation of the Wigner function via the inverse Radon transformation.

Given the experimentally reconstructed Wigner function, we can inverse Fourier transform \eeqref{wigdef} to compute the density operator in the quadrature basis, and, subsequently, in any other basis.
 This scheme was applied to reconstruct photon-number statistics
 $\left\langle n \right|\hat \rho \left| n \right\rangle$,
 as well as quantum-phase statistics for squeezed and for
 coherent light (Beck et al., 1993; Smithey et al., 1993b;
Smithey et al., 1993c). This calculation can however be significantly simplified, as discussed below.

\subsubsection{Pattern functions}

If the goal is to reconstruct the density operator of the ensemble, we can exploit the overlap formula
\begin{equation}\label{wigoverlap}
\Tr(\hat\rho\hat A)=2\pi\int\limits_{-\infty}^{+\infty}\int\limits_{-\infty}^{+\infty}W_{\hat\rho}(Q,P)W_{\hat A}(Q,P)\,dQ\,dP
\end{equation}
valid for any operator $\hat A$ and the associated density matrix $W_{\hat A}(Q,P)$ as defined by Eq. (\ref{wigdef}) with $\hat\rho$ replaced by $\hat A$. For example, given $\hat A_{mn}=\ketbra{m}{n}$ with $\ket{m}$ and $\ket{n}$ being the Fock states, we write $\rho_{mn}=\Tr(\ketbra{m}{n}\hat\rho)$ and use Eq.~(\ref{wigoverlap}) to determine, one by one, the elements of the density matrix in the Fock basis.

The intermediate step of
reconstructing the Wigner function can however be sidestepped using an improved deterministic scheme
introduced by D'Ariano {\it et al.} (1994), and refined several times to the present optimal form (Leonhardt \ieac 1996, 1997; D'Ariano {\it et al.,} 2004).
We combine Eqs.~(\ref{invRadon}) and (\ref{wigoverlap}) to write
\begin{eqnarray}\label{patterngen}
\Tr(\hat\rho\hat A)&=&\int\limits_0^\pi\int\limits_{-\infty}^{+\infty}{\rm pr}(Q_\theta,\theta)F_{\hat A}(Q_\theta,\theta)\,dQ_\theta\,d\theta \\ \nonumber
&=&\langle F_{\hat A}(Q_\theta,\theta)\rangle_{Q_\theta,\theta},
\end{eqnarray}
where averaging is meant in the statistical sense over all acquired values of $(Q_\theta,\theta)$, and
\begin{eqnarray}\label{FAgen}
F_{\hat A}(Q_\theta,\theta)&&\\ \nonumber &&\hspace{-1.5cm}=\frac{1}{\pi}\iint_{-\infty}^{+\infty}K(Q\cos\theta+P\sin\theta-Q_\theta)W_{\hat A}(Q,P)\,dQ\,dP
\end{eqnarray}
is the \emph{sampling function}.
Given a specific operator $\hat A$, the function $F_{\hat A}$ does not depend on the experimental histogram ${\rm pr}(Q_\theta,\theta)$, but only on the operator itself. It thus need be calculated only once, prior to the experiment, and substituted into Eq.~(\ref{patterngen}) once the data become available.

Specializing to the Fock basis, $F_{mn}(Q,\theta)=(1/\pi) e^{i(m-n)\theta}M_{mn}(Q)$, with $M_{mn}(Q)$ being so-called \emph{pattern functions} (D'Ariano \ieac 1994; Paul \ieac 1995; Leonhardt and Raymer, 1996):
\begin{equation}\label{patcalc}
M_{mn}(Q)=-\mathcal{P}\int\limits_{-\infty}^{+\infty}\frac{\psi_m(x)\psi_n(x)}{(Q-x)^2}dx,
\end{equation}
where
\begin{equation}\label{psix}
\psi_n(x)=\braket{n}{x} =
\left(\frac{1}{\pi}\right)^{1/4}\frac{H_n(x)}{\sqrt{2^n
n!}} \ \exp\left(-\frac{x^2}{2}\right),
\end{equation}
are the Fock state wave functions --- that is, wavefunctions of energy eigenstates of a harmonic oscillator. $H_n$ denote the Hermite polynomials.
Figure \ref{CompFig}(a) shows an example of calculating the density matrix using the pattern function method.

Efficient numerical algorithms for computing the pattern functions were given by Leonhardt {\it et al.} (1996) and Leonhardt (1997). In our experience, the most practical algorithm involves the irregular wave functions $\varphi_n(x)$, which are alternative, non-normalizable solutions of the time-independent Schr\"odinger equation for the harmonic oscillator. These functions obey a recursion
\begin{equation}
\varphi_{n+1}(x)=\frac{1}{\sqrt{2n+2}}[x\varphi_n(x)-\varphi_n'(x)]
\end{equation}
with
\begin{equation}
\varphi_0(x)=\pi^{3/4}\exp\left(-\frac{x^2}{2}\right){\rm erfi}(x)
\end{equation}
and are all readily expressed through the error function ${\rm erfi}(x)$. Once the desired number of the irregular wave functions have been calculated, the pattern functions are obtained using
\begin{equation}
M_{mn}(x)=\left\{\begin{array}{l}\partial[\psi_m(x)\varphi_n(x)]/\partial x\ \ {\rm for}\ n\ge m \\ \partial[\psi_n(x)\varphi_m(x)]/\partial x\ \ {\rm for}\ n<m.\end{array}\right.
\end{equation}

The pattern function method can be extended to \emph{direct sampling}, or quantum estimation (Paul \ieac 1995; Munroe \ieac 1995), of the expectation value of \emph{any} operator directly without first reconstructing the state. In many cases, this requires fewer probability functions to be measured, since less complete information is being asked for. Indeed, since Wigner functions are linear with respect to their generating operators, we conclude from Eq.~(\ref{FAgen}) that for any operator $\hat A=\sum A_{mn}\ket{m}\bra{n}$,
\begin{eqnarray}\label{FAFmn}
F_{\hat A}(Q_\theta,\theta)&=&\sum\limits_{m,n}F_{mn}(Q,\theta)\hat A_{mn}\\ \nonumber
&=&\frac{1}{\pi}\sum\limits_{m,n}^{} {\left\langle n \right|} \hat A\left| m \right\rangle \,M_{mn} (q)\exp[i(m - n)\theta ].
\end{eqnarray}
The expectation value $\langle\hat A\rangle=\Tr(\hat\rho\hat A)$ can then be calculated according to Eq. (\ref{patterngen}).

For example, if we desire the photon-number probability ${\rm pr} (j)$,
we choose $\hat A_j = \left| j \right\rangle \left\langle j \right|$.
Then $F_{\hat A_j}(Q,\theta ) = (1/\pi) M_{jj} (Q)$, which is independent of phase $\theta $,
and Eq.(\ref{patterngen}) becomes (Munroe \ieac 1995)
\begin{equation}\label{prjpat}
{\rm pr} (j) = Tr\left( {\hat \rho \left| j \right\rangle \left\langle j \right|} \right) = \int_{ - \infty }^\infty  {dQ} \,M_{jj} (Q)\langle{{\rm pr}(Q,\theta )}\rangle_{\theta}
\end{equation}
This is a convenient result, since only a single probability function need be measured, while sweeping or randomizing the phase. Demonstrations of this technique can be found in Munroe {\it et al.} (1995), Schiller {\it et al.} (1996), as well as Raymer and Beck (2004).

Although the techniques of OHT have been generalized to fields involving more than one optical mode (spatial, polarization, or temporal) (Raymer \ieac 1996; Opatrn\'{y} \ieac 1996; D'Ariano \ieac 2000), their practical application is challenging. This is a particular case where direct sampling is handy. We can apply it if our task is to determine the expectation values of certain observables, but full reconstruction of the multimode state is not necessary. An example is the acquisition of correlated photon number statistics of two-mode fields (McAlister and Raymer, 1997b;
Vasilyev \ieac 2000; Blansett \ieac 2001; Voss \ieac 2002; Blansett \ieac 2005).



\subsection{Maximum-likelihood reconstruction}
\label{maxliksec}
\subsubsection{Why maximum likelihood?}
Quantum state reconstruction can never be perfect, due to statistical and systematic uncertainties in the estimation of the measured statistical distributions. In both discrete- and continuous-variable domains, inverse linear transformation methods work well only when these uncertainties are negligible, i.e. in the limit of a very large number of data and very precise measurements. Otherwise the errors in the ``right-hand sides'' of the system of linear equations we are trying to solve can lead to inaccurate, even seemingly
unphysical, features in the reconstructed state. For example,
negative values may be found on the diagonal of the reconstructed
density matrix and its trace is not guaranteed to equal one [Fig.~\ref{CompFig}(a)].

In the case of continuous-variable tomography, there is an additional complication: a harmonic oscillator is a quantum system of infinite dimension, and no finite amount of measurement data will constitute a quorum. In order to achieve reconstruction, one needs to make certain assumptions that limit the number of free parameters defining the state in question. For example, the filtered
back-projection imposes low pass filtering onto the Fourier image
of the Wigner function, i.e. assumes the ensemble to possess a
certain amount of ``classicality" (Vogel, 2000). Such smoothing reduces the accuracy of the reconstruction (Herman, 1980; Leonhardt, 1997) and introduces characteristic ripples (Breitenbach \ieac 1997) on the reconstructed
phase-space density [Fig.~\ref{CompFig}(a)].


Although errors cannot be eliminated completely, we would like a reconstruction method that guarantees a physically plausible ensemble and minimizes artifacts. This requirement is satisfied by the \emph{Maximum
Likelihood} (MaxLik) approach, which aims to find, among the variety of all possible
density matrices, the one which maximizes the probability of
obtaining the given experimental data set \emph{and} is physically plausible. Because this method is relatively new, but is rapidly gaining popularity, we here present its relatively detailed description. A yet more comprehensive review on quantum MaxLik (limited to the discrete domain) is given in Hradil \iea ~(2004).

\subsubsection{Classical algorithm}
We begin with a brief discussion of the classical expectation-maximization method. Consider a certain system characterized a set of parameters $\vec{r}$ (such that $r_i>0$ and $\sum_i r_i=1$), which we need to determine. We are allowed to subject the system to a measurement with a random outcome. The probability of each possible result (indexed by $j$) is related to $\vec{r}$ linearly:
\begin{equation} \label{linpos}
\pr_{\vec{r}}(j)=\sum_i r_i h_{ij},
\end{equation}
where all $h_{ij}$ are known positive numbers. The measurement is repeated $N$ times, of which each outcome occurs $f_j$ times. The goal is to infer the parameter set $\vec r$ from the set of measurement results $\vec f$.

The ideal inference is the one that satisfies the system of linear equations
\begin{equation} \label{linposf}
\frac{f_j}{N}=\sum_i r_i h_{ij}.
\end{equation}
However, a solution to this system exists only if the number of parameters is larger than the number of equations. Otherwise, we have to settle for less: find the distribution $\vec r$ which would maximize the probability (likelihood)
\begin{equation} \label{Lgen}
\mathcal{L}({\vec r})=\prod_j\left[\pr_{\vec{r}}(j)\right]^{f_j},
\end{equation}
of the observed measurement result. This approach has a very large variety of applications ranging form image de-blurring to investment portfolio optimization.

The maximum-likelihood parameter set is determined by the so-called expectation-maximization (EM) algorithm which consists of sequential iterations (Dempster \ieac 1977; Vardi and Lee, 1993):
\begin{equation}\label{classiter}
r_i^{(n+1)}=r_i^{(n)}\sum_j\frac{h_{ij}r^{(n)}_j}{\pr_{\vec{r}^{(n)}}(j)},
\end{equation}
initialized with some positive vector $\bf{r}$. Each single iteration step is known to increase the likelihood. Furthermore, because the likelihood is a convex function (i.e. for any two distributions ${\vec r}_1$ and ${\vec r}_2$ holds $\mathcal{L}(\frac{{\vec r}_1+{\vec r}_2}{2})\ge \frac{\mathcal{L}({\vec r}_1)+\mathcal{L}({\vec r}_2)}{2}$), the iterations will approach the global likelihood maximum.

\subsubsection{The discrete quantum case} \label{discML}
A quantum tomographic procedure can be associated with a positive operator-valued measure (POVM), with each possible measurement result described by a positive operator $\hat\Pi_j$, which occurs with a probability
\begin{equation}
\pr_{\hat\rho}(j)=\Tr[\hat\Pi_j\hat{\rho}].
\end{equation}
Here, again, we are dealing with a linear inversion problem, because the probabilities are proportional to the density matrix elements. However, the latter are not necessarily positive (not even real) and their sum is not equal to one, so the EM algorithm in its original form has only limited application to the quantum case.

In order to reconstruct a quantum state, we introduce the non-negative operator
\begin{equation} \label{Rgen}
\hat R(\hat{\rho})=\frac 1 N \sum_j\frac{f_j}{\pr_{\hat\rho}(j)}\hat\Pi_j.
\end{equation}
As shown by Hradil (1997), the state that maximizes the likelihood \eqref{Lgen} obeys the extremal equation \begin{equation}
\label{Rrho}\hat R(\hat{\rho}_0)\hat\rho_0=\hat\rho_0\hat
R(\hat{\rho}_0)=\hat\rho_0,
\end{equation}
as well as \begin{equation}  \label{RrhoR} \hat
R(\hat{\rho}_0)\hat\rho_0\hat R(\hat{\rho}_0)=\hat\rho_0.
\end{equation} One can intuitively understand these equations as follows: when $\hat\rho$ is the maximum-likelihood state, we have $f_j/N\approx\pr_j$, so the operator $\hat R$ becomes $\sum_j \hat\Pi_j$, which is normally unity.

The analogy to the classical scheme would suggest an iterative procedure $\hat\rho^{(k+1)}=
\hat R(\hat\rho^{(k)}) \hat\rho^{(k)}$ based on \eeqref{Rrho}. However, unfortunately, such iteration does not preserve positivity of the density matrix [unless it is guaranteed to be diagonal in some basis, in which case the iteration reduces to \eeqref{classiter} --- such as in Banaszek (1998a, b)]. A possible
solution is to apply the expectation-maximization iteration to the
diagonalized density matrix followed by re-diagonalization (\v{R}eh\'{a}\v{c}ek {\it et al.}, 2001; Artilles \ieac 2005).

A more common approach to constructing the iterative algorithm relies on \eeqref{RrhoR} (Hradil \ieac 2004). We
choose some initial denstity matrix as, e.g.,
$\hat\rho^{(0)}=\mathcal{N}[\hat 1]$, and apply repetitive
iterations
\begin{equation}\label{iterhomo}
\hat\rho^{(k+1)}=\mathcal{N}\left[\hat
R(\hat\rho^{(k)})\hat\rho^{(k)}\hat R(\hat\rho^{(k)})\right],
\end{equation}
where $\mathcal{N}$ denotes normalization to a unitary trace. Hereafter we refer this scheme as the ``$R\rho R$ algorithm''.

This iteration ensures positivity of the density matrix and has shown fast convergence in a variety of experiments. However, there is no guarantee for monotonic increase of the likelihood in every
iteration; on the contrary, there exists a (somewhat pathological) counterexample (\v{R}eh\'{a}\v{c}ek {\it et al.}, 2007).  There remains a risk
that the algorithm could fail for a particular experiment.

The remedy against this risk is proposed in Hradil \iea ~(2004) and further elaborated in \v{R}eh\'{a}\v{c}ek {\it et al.} (2007). These publications present a ``diluted'' linear iteration
\begin{equation}\label{modified}
\hat\rho^{(k+1)}=\mathcal{N}\left[\frac{\hat\id+\epsilon \hat
R}{1+\epsilon}\hat\rho^{(k)}\frac{\hat\id+\epsilon \hat R}{1+\epsilon}\right],
\end{equation}
which depends on a single
parameter $\epsilon$ that determines the ``length'' of the step in the
parameter space associated with one iteration. For $\epsilon\to\infty$, the iteration becomes $R\rho R$.
On the other hand, in the limit of $\epsilon\to 0$, there is a proof that the
likelihood will monotonically increase and the iterations will converge to the maximum-likelihood state.
We thus obtain a backup algorithm for the case the likelihood fails to increase in the $R\rho R$ iteration. In practice, however, this situation is not likely.

In some tomography schemes,
one or more possible measurement results may not be accessible and, consequently,
$\hat G\equiv \sum_j \hat\Pi_j$ is not equal to the unity operator.
Then the extremal map (\ref{RrhoR}) should be replaced by
\begin{equation}\label{biased}
\hat G^{-1}\hat R(\hat \rho_0) \hat \rho_0
\hat R(\rho_0)\hat G^{-1}=\hat \rho_0
\end{equation} to avoid
biased results (\v{R}eh\'{a}\v{c}ek {\it et al.}, 2001; Hradil \ieac 2006; Mogilevtsev \ieac 2007). This issue may become significant in homodyne tomography reconstruction, which we discuss next.

\subsubsection{Iterative scheme for homodyne tomography}
The applications of MaxLik to homodyne tomography have been pioneered by Banaszek (1998a, b), who reconstructed the
photon-number distribution (the diagonal density matrix elements
which correspond to a phase-randomized optical ensemble) from a
Monte-Carlo simulated data set by means of the classical EM algorithm. This idea was then extended to reconstructing the Wigner function point-by-point (Banaszek, 1999) by applying
phase-dependent shifts to the experimental data.
In a subsequent
publication, Banaszek {\it et al.} (1999) discussed direct MaxLik
estimation of the density matrix, but presented no specific algorithm. More recently, the $R\rho R$ iterative algorithm was adapted to OHT (Lvovsky, 2004) and has since been widely used in experiments on homodyne reconstruction. We describe this adaptation below.

For a given local oscillator phase $\theta$, the probability to detect a particular
quadrature value $Q_\theta$ is proportional to
\begin{equation}\label{pr}
{\rm pr}_{\hat\rho}(Q_\theta,\theta)\propto{\rm Tr} [\hat\Pi(Q_\theta,\theta)\hat\rho],
\end{equation} where $\hat\Pi(Q_\theta,\theta)$ is the projector
onto this quadrature eigenstate, expressed in the Fock
basis as
\begin{equation}\label{projFock}
\bra{m}\hat\Pi(Q_\theta,\theta)\ket{n}=\braket{m}{Q_\theta,\theta}\braket{Q_\theta,\theta}{n},
\end{equation} where the wavefunction
$\braket{m}{Q_\theta,\theta} = e^{im\theta}\psi_m(Q_\theta)$
is given by Eq. (\ref{psix}).

Because a homodyne measurement generates a number from a continuous range, one
cannot apply the iterative scheme (\ref{iterhomo}) directly to the
experimental data. One way to deal with this difficulty is to
discretize the data by binning it up according to $\theta$ and $Q_\theta$
and counting the number of events $f_{Q_\theta,\theta}$ belonging to each
bin. In this way, a number of histograms, which represent the
marginal distributions of the desired Wigner function,
can be constructed. They can then be used to implement the iterative
reconstruction procedure.

However, discretization of continuous experimental data will
inevitably lead to a loss of precision\footnote{In fact, recent research shows the precision loss due to binning to be insignificant. On the other hand, binning greatly reduces the number of data and thus expedites the iterative reconstruction algorithm (D. Mogilevtsev, 2007, private communication)}. To lower this loss, one
needs to reduce the size of each bin and increase the number
of bins. In the limiting case of infinitely small bins,
$f_{Q_\theta,\theta}$ takes on the values of either 0 or 1, so the
likelihood of a data set $\{(Q_i,\theta_i)\}$ is given by
\begin{equation} \label{Lhomo}
\mathcal{L} = \prod_i {\rm pr}_{\hat\rho}(Q_i,\theta_i),
\end{equation}
and the iteration operator (\ref{Rgen}) becomes
\begin{equation}\label{Rhomo}
\hat R(\hat\rho) = \sum_i\frac{\hat\Pi(Q_i,\theta_i)}{{\rm
pr}_{\hat\rho}(Q_i,\theta_i)},
\end{equation}
where $i=1\dots N$ enumerates individual measurements. The
iterative scheme (\ref{iterhomo}) can now be applied to find the
density matrix which maximizes the likelihood (\ref{Lhomo}).

In practice, the iteration algorithm is executed with the density
matrix in the photon number representation. In order to limit the number of unknown parameters, we truncate the Hilbert space by excluding
Fock terms above a certain threshold. This is equivalent to assuming that the signal field intensity is limited. In many experimental
situations, application of this assumption is better justified than the low-pass filtering used in the filtered back-projection algorithm.

Fig.~\ref{CompFig} compares the inverse linear transform and MaxLik reconstruction methods in application to the experimental data from Lvovsky and Mlynek (2002). The data set consists of 14152 quadrature
samples of an ensemble approximating a coherent superposition of
the single-photon and vacuum states. We see that the MaxLik method eliminates unphysical features and artefacts that are present in the inverse Radon reconstruction.


\subsubsection{Error handling}\label{maxlikerror}
A homodyne detector of non-unitary efficiency
$\eta$ can be modeled by a perfect detector preceded by an absorber. In transmission through this absorber, photons can be lost, and the optical state undergoes a so-called generalized Bernoulli
transformation (Leonhardt, 1997). If $\eta$ is known, the Bernoulli transformation can be incorporated into the matrices of the POVM elements $\hat\Pi(Q_\theta,\theta)$ (Banaszek \ieac 1999; Lvovsky, 2004). These operators can then be used to construct the matrix $\hat R$, so the iterative algorithm will automatically yield the density matrix corrected for detector inefficiencies.

Theoretically, it is also possible to correct for the detector
inefficiencies by applying the inverted Bernoulli transformation
\emph{after} an efficiency-uncorrected density matrix has been
reconstructed (Kiss \ieac 1995). However, this may give rise to
unphysically large density matrix elements associated with high
photon numbers. Similar concerns about possible numerical instability
arise when the detector inefficiency is being accounted for in the pattern-function
reconstruction (Kiss \ieac 1995). With the inefficiency correction incorporated, as
described above, into the MaxLik reconstruction procedure, this issue does
not arise (Banaszek, 1998b).

Another source of error in OHT MaxLik estimation can be the incomplete character of the homodyne measurements: the sum of the projection operators $\hat G=\sum_i\hat\Pi(Q_i,\theta_i)$ is not equal identity operator (even in the truncated Fock space). Mogilevtsev \iea(2007) found that the deviation can be quite significant. This issue can be resolved by employing the iteration based on the biased extremal equation \eqref{biased} instead of \eeqref{RrhoR}, such as in the experimental work by Fernholz \iea (2008).

Finally, we discuss statistical uncertainties of the reconstructed
density matrix. In generic MaxLik algorithms, they are typically
estimated as an inverse of the Fisher information matrix (Rao \ieac 1945;
Cram\'{e}r, 1946). This method can be generalized to the quantum case (Hradil \ieac 2004; Usami \ieac 2003). In application to OHT, calculating the Fisher information appears quite complicated due to a very large number of independent measurements involved.

A sensible alternative is offered by a clumsy, yet simple and
robust technique of {\it simulating} the quadrature data that
would be associated with the estimated density matrix $\hat\rho_{ML}$ if
it were the true state. One generates a large number of random
sets of homodyne data according to Eq. (\ref{pr}), then applies
the MaxLik reconstruction scheme to each set and obtains a series
of density matrices $\hat\rho'_k$, each of which approximates the
original matrix $\hat\rho_{ML}$. The average difference $\langle|
\hat\rho_{ML} - \hat\rho'_k|\rangle_k$ evaluates the statistical
uncertainty associated with the reconstructed density matrix.

\subsection{Maximum-entropy reconstruction}
\label{otherstat}
The maximum-entropy (MaxEnt) method is applied in the situation opposite to that of the MaxLik approach: when the number of equations in system \eqref{linposf} (i.e. the number of available data) is \emph{smaller} than the number of unknown parameters (Bu\v zek, 2004). In this case, the solution is not unique, and MaxEnt looks for the \emph{least
biased} solution, i.e. the one that maximizes the von Neumann entropy $S=-\Tr(\hat\rho\log\hat\rho)$.

In OHT, one usually collects a very large ($10^4$--$10^6$) number of data points, hence the situation where the MaxEnt method is applicable is uncommon. Nevertheless, Bu\v zek and Drobn\'y (2000) have elaborated application of this method to homodyne tomography and performed reconstruction of various simulated data sets. They found the results to be significantly better than those obtained by inverse linear transform, particularly in situations of incomplete tomographic data (marginal distributions available for a small number of phases or measured on short intervals). 
\section{Technical aspects}






\subsection{Time-domain homodyne detection}\label{tdhd}
When homodyne detection was first introduced to quantum optical measurements in mid-1980s, it was used for evaluating field quadrature noise rather than full state tomography. Such measurements are convenient to perform in the \emph{frequency domain}, observing a certain spectral component (usually around 1-10 MHz where the technical noise is minimized) of the photocurrent difference signal using an electronic spectral analyzer. Frequency-domain detection was used, for example, to observe quadrature squeezing (Slusher \ieac 1985;
Wu \ieac 1986).

Quantum-information applications require measurement of optical modes that are localized in time. Homodyning has to be performed in the \emph{time domain}: difference photocurrent is observed in real time and integrated over the desired temporal mode to obtain a single value of a field quadrature. Repeated measurements produce a quantum probability distribution associated with this quadrature.

In this section, we discuss the design of time-domain balanced detectors that operate with pulsed local oscillators. The first such detector was implemented by Smithey {\it et al.} (1992,1993a) in their original quantum tomography experiments. Among subsequent schemes we note that of Hansen {\it et al.} (2001) which features a higher bandwidth and a signal-to-noise ratio as well as that of Zavatta {\it et al.} (2002), exhibiting a further significant bandwidth increase at a cost of a somewhat poorer noise characteristics.

\begin{figure}[htb]
\includegraphics[keepaspectratio,width=0.45\textwidth]{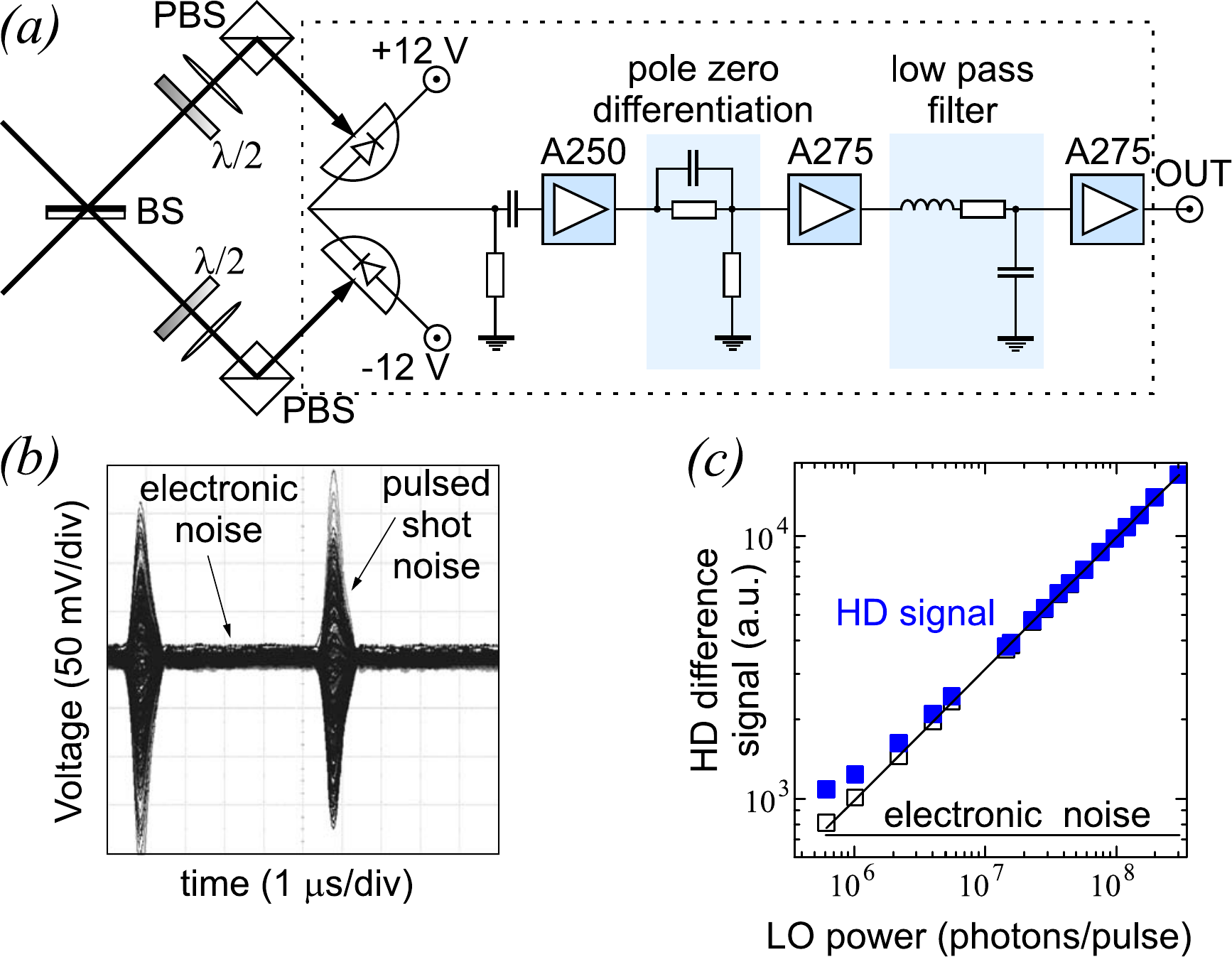}
\vspace*{8pt}
\caption{(a) Electro-optical scheme of the homodyne detector. (b) A superposition of multiple oscilloscope traces of the detector output. Each pulse produces a time-resolved quantum noise sample. (c) RMS peak amplitude of the noise pulses as a function of the LO power showing the expected square root power dependence up to the LO intensities of $3\times10^8$ photons per local oscillator pulse. Filled squares show the measured noise variances, open squares have the electronic noise background corresponding to 730 electrons/pulse subtracted. From Lvovsky \iea (2001). \label{bhdscheme}}
\end{figure}

Figure \ref{bhdscheme}(a) shows the main elements of the circuit of Hansen {\it et al.} (2001), which are typical for today's pulsed, time-domain HDs. A pair of high-efficiency photodiodes are wired in series to subtract their output
currents, and this difference signal is amplified by a charge-sensitive transimpedance amplifier, followed by a pulse-forming network. The optics in front of the photodiodes permits variable attenuation of the input to each photodiode (alternatively, a setting with two polarizing beam splitters and a half-wave plate between them can be used for combining the local oscillator and the signal). Thorough balancing of the photodiodes' photocurrents is essential for the proper operation of the BHD.

With each LO pulse, the detector produces a burst of amplified subtraction photocurrent [Fig.~\ref{bhdscheme}(b)]. Because the response time of the detector is much slower than the width of the laser pulse, the generated signal is proportional to the time integral of the photocurrent over the pulse duration. This is a single sample of the field quadrature noise in the spatiotemporal optical mode of the local oscillator pulse.

To prove that the pulsed noise generated by the homodyne detector with a vacuum signal input is indeed the shot noise, one needs to verify that the output rms noise scales as the square root of the LO power\footnote{This follows from Eq.~(\ref{Qtheta}), because $|\alpha _L |=\sqrt{N_{LO}}$ and $Q_\theta$ varies on the scale of 1.} [Fig.~\ref{bhdscheme}(c)]. This is a signature distinguishing the shot noise from the classical noise (proportional to the local oscillator intensity) and the electronic noise (which is constant) (Bachor and Ralph, 2004).


Design of time-domain BHD is more technically challenging than its frequency-domain counterpart. First, the electronics must ensure time separation of responses to individual laser pulses. The shot-noise difference charge must be low-noise amplified within a bandwidth exceeding the local oscillator pulse repetition rate. Second, precise subtraction of photocurrent is necessary in order to eliminate the classical noise of the local oscillator. There is a competition between this requirement, which is easier satisfied at lower LO energies, and that of a sufficiently strong subtraction signal $N_-$, which increases with the LO power. The compromise is achieved on the scale of $N_-\sim 10^3$--$10^6$ photoelectons. Finally, the measured quadrature values must not be influenced by low-frequency noises. The detector must thus provide ultra-low noise, high subtraction and flat amplification profile in the \emph{entire frequency range} from (almost) DC to at least the LO pulse repetition rate.

A typical dilemma faced by a BHD designer is a trade-off between the signal-to-noise (more precisely, shot-to-electronic noise) ratio and the bandwidth (Raymer and Beck, 2004). An amplifier with a higher bandwidth usually exhibits poorer noise characteristics (Nicholson, 1974; Radeka, 1988). An additional bandwidth limitation arises from the intrinsic capacitance of photodiodes, which may cause instability in the amplification circuit. Technologically, this capacitance is determined by the thickness of the photodiode PIN junction; reducing this thickness compromises the quantum efficiency. A homodyne detector with a time resolution capable to accommodate a typical repetition rate of a mode-locked, pulsed Ti:Sapphire laser (around 80 MHz) has been demonstrated by Zavatta \iea (2002, 2005b).

Suppression of the homodyne detector electronic noise is important for quantum state reconstruction. As shown by Appel \iea (2007), presence of the noise leads to an equivalent optical loss of $1/S$, where $S$ is the detector's signal-to-noise ratio.

We briefly note that time-domain homodyne detection finds its applications not only in quantum tomography, but also in other fields of quantum and classical technology. One example is shot-noise-limited absorption measurements at subnanowatt power levels achievable thanks to the very low technical noise (Hood \ieac 2000). Another is ultrafast, ultrasensetive linear optical sampling for characterizing fiber optical systems (Dorrer \ieac 2003). Time-domain homodyning is also an essential element of continuous-variable quantum cryptography (Silberhorn \ieac 2002; Grosshans and Grangier, 2002;
Funk, 2004; Lodewyck \ieac 2007).

\subsection{Matching the mode of the local oscillator} \label{MMsec}
\subsubsection{The advanced wave}
In homodyne detection, the spatiotemporal optical mode to be measured is determined by that of the local oscillator. In this way, OHT provides indirect information on the modal structure of the signal field. This is useful for evaluating quantum optical information processing systems, which require that interacting optical qubits be prepared in identical, pure optical modes. On the other hand, \emph{achieving} the mode matching between the local oscillator and the signal, or even preparing the signal state in a well-defined, pure spatiotemporal mode, can be challenging. In this section, we discuss the mode matching techniques, specializing to an important particular case of the signal state being a heralded single photon. 



\begin{figure}    
\includegraphics[keepaspectratio,width=0.45\textwidth]{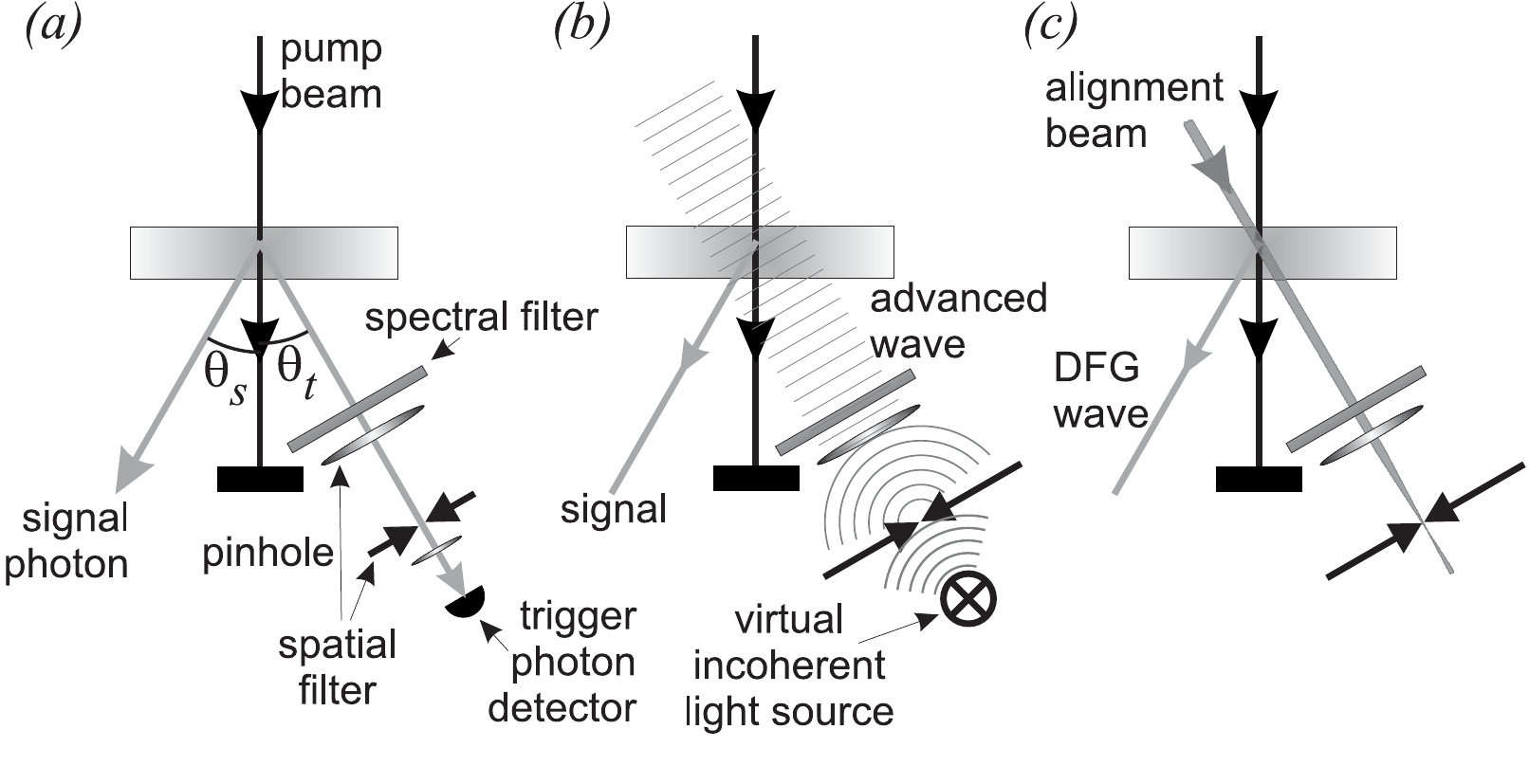}
\vspace*{8pt}
\caption{Parametric down-conversion and the advanced wave model. (a) \label{AbbPDC1}Preparation of single photons by
conditional measurements on a biphoton state. (b) The Klyshko advanced wave model. The trigger detector is replaced with an incoherent light source, which generates an incoherent advanced wave propagating backwards in space and time. Nonlinear interaction of this wave
with the pump produces a difference-frequency pulse that mimics that of the conditionally prepared photon.
(c) In an experiment, a laser beam, aligned for maximum transmission
through all the filters, can model the advanced wave. From Aichele \iea (2002). \label{modematching}}
\end{figure}

In order to prepare a heralded photon, a parametric down-conversion (PDC) setup is pumped relatively weekly so it generates, on average, much less than a single photon pair per laser pulse (or the inverse PDC bandwidth).  The
two generated photons are separated into two emission channels
according to their propagation direction, wavelength and/or
polarization. Detection of a photon in one of the emission channels
(labeled \emph{trigger} or \emph{idler}) causes the state of the photon pair to collapse,
projecting the quantum state in the remaining (\emph{signal}) channel into
a single-photon state [Fig.~\ref{modematching}(a)]. Proposed and tested experimentally in 1986 by
Hong and Mandel (1986) as well as Grangier {\it et al.} (1986), this technique has become a workhorse for many quantum optics experiments.

The biphoton is a complex entangled state with many parameters (spectrum, direction, polarization, etc.) of the two photons highly correlated:
\begin{equation} \label{pdcout}
\ket{\Psi_{st}}=\int\Psi(\omega_s,\omega_t,\vec k_s,\vec k_t)\ket{1_{\omega_s,\vec k_s}}\ket{1_{\omega_t,\vec k_t}}d\omega_s d\omega_t d\vec k_s d\vec k_t,
\end{equation} where $\omega$ and $\vec k$ denote the frequencies and wavevectors of the signal and trigger photons. If the trigger photon is measured with any uncertainty in one of these parameters, the signal photon will be prepared in a non-pure state
\begin{equation}\label{condphoton}
\rho_s={\rm Tr}_t T(\omega_t,\vec k_t)\ketbra{\Psi_{si}}{\Psi_{si}},
\end{equation} where $T(\omega_t,\vec k_t)$ is the transmission function of the filters in the trigger channel defining the measurement uncertainty. 

Formation of the heralded mode is nicely illustrated by the heuristic concept of \emph{advanced waves} proposed by Klyshko (1998a, b, c) and further advanced by Aichele {\it et al.} (2002). According to this concept,
the trigger photon detector is replaced with a fictitious light source, which,
at the moment of detection, produces a classical incoherent electromagnetic wave traveling backwards in space and time [see Fig.~\ref{modematching}(b)].
When propagating through the trigger channel filters, the advanced wave
acquires some degree of spatiotemporal coherence, quantified
by the filters' width. It then enters the down-conversion
crystal and undergoes nonlinear interaction with the pump pulse,
generating a difference-frequency pulse. This pulse turns out to be
\emph{completely identical}, in its modal characteristics, to the optical mode of the conditionally prepared single photon (Aichele \ieac 2002), and is thus helpful for visualizing many of its properties.

Suppose, for example, that the pump is pulsed (femto- or picosecond). Because the timing jitter of the photon counter event is typically on the order of a nanosecond, the exact moment of when the photon pair has been emitted is uncertain, so the advanced wave can be assumed continuous in time. The nonlinear interaction between the advanced wave and the pump is however restricted by the spatiotemporal window determined by the coherent pump
pulse. If the latter is much narrower than the coherence time and coherence width of the advanced
wave, the difference-frequency pulse will be almost transform limited, both in the spatial and temporal dimensions.

We conclude that narrow spatial and spectral filtering of the trigger photon can be used to obtain the signal photon in a pure spatiotemporal mode. To our knowledge, for the first time this matter has been investigated theoretically by Zukowski {\it et al.} (1995), independently by Rarity (1995), and later confirmed in a more detailed study by Ou (1997). Specifically in the context of OHT, theoretical treatment was given by Grosshans and Grangier (2001) as well as Aichele {\it et al.} (2002).

Mode purity of the signal photon does not by itself guarantee its matching to the local oscillator.
The advanced wave model suggests the following experimental procedure for achieving this matching. Although the
advanced wave propagates backwards in space and time and is thus a
purely imaginary object, it can be modeled by a forward-going \emph{alignment
beam} inserted into the trigger channel so that it overlaps
spatially and temporally with the pump beam inside the crystal and
passes through the optical filters [Fig.~\ref{modematching}(c)].
Nonlinear interaction of such an alignment beam with the pump wave
will produce difference frequency generation into a spatiotemporal
mode similar (albeit no longer completely identical) to that of
the conditionally prepared single photon. If one observes and optimizes the interference pattern between this wave and the local oscillator, one can be sure that, after blocking the alignment beam, the mode of the signal photon will be matched to that of the local oscillator (Aichele \ieac 2002).

Controlling the \emph{spatial} mode of the signal photon is simplified if a single-mode optical fiber is used as an optical filter instead of a pinhole arrangement. Such a filter automatically selects a pure spatial mode in the trigger channel, which transforms to a spatially pure signal photon. It is also advantageous in terms of the pair production rate [Ourjoumtsev \iea (2006a), supporting material] Unfortunately, there is no similar arrangement possible for the spectral (temporal) mode matching.

\subsubsection{Decorrelating photons}
Reducing the spectral line width of the trigger filter will improve the mode purity of the heralded photons, but also reduce their production rate. This compromise would be avoided if we could arrange the PDC setup in such a way that the trigger and signal photons in the output of the down-converter are uncorrelated: the function $\Psi$ in Eq.~(\ref{pdcout}) could be written as
\begin{equation}\label{uncorrpdc}
\Psi(\omega_s,\omega_t)=\psi_s(\omega_s)\times\psi_t(\omega_t).
\end{equation}
In this case, detection of \emph{any} photon in the trigger channel signifies that the signal photon has been emitted into a pure spatiotemporal mode defined by the function $\psi_s$.

The first detailed theoretical inquiry into preparation of uncorrelated down-conversion spectra was made in Grice \iea (2001), based on general theoretical analysis of Keller and Rubin (1997). This theory was further elaborated in U'Ren \iea (2005, 2007). The configuration of the correlation function $\Psi$ depends primarily on the energy-conservation condition
\begin{equation}\label{energyconservation}
\omega_s+\omega_t=\omega_p
\end{equation}
and the phase-matching condition
\begin{equation}\label{phasematch}
\vec k_s+\vec k_t\cong\vec k_p.
\end{equation}
For any generated pair of photons with parameters ($\omega_s,\vec k_s,\omega_t,\vec k_t$) there must exist a pump photon ($\omega_p,\vec k_p$) for which the above equations are satisfied.

Suppose the PDC occurs in an almost collinear configuration and the crystal is aligned so that Eqs.~\eqref{energyconservation} and \eqref{phasematch} simultaneously hold for the central pump frequency $\omega_{p0}$ and some signal and idler frequencies $\omega_{s0}$ and $\omega_{t0}$, respectively. The frequency and the wavevector are connected through dispersion relations: $d\omega=v_{gr}d|\vec k|$, where $v_{gr}$ is the wave's group velocity.
Neglecting dispersion orders higher than one (analysis beyond this approximation is made in U'Ren \ieac 2005), we cast \eeqref{phasematch} into the form
\begin{equation}\label{pmspectral}
\frac{\omega_s-\omega_{s0}}{v_{gr,s}}+\frac{\omega_t-\omega_{t0}}{v_{gr,t}}=\frac{\omega_p-\omega_{p0}}{v_{gr,p}}.
\end{equation}
Equations \eqref{energyconservation} and \eqref{pmspectral}, plotted in the  $(\omega_{s},\omega_{t})$ plane with $\omega_p=\omega_{p0}$, form straight lines crossing at $(\omega_{s0},\omega_{t0})$.

\begin{figure}
\includegraphics[keepaspectratio,width=0.45\textwidth]{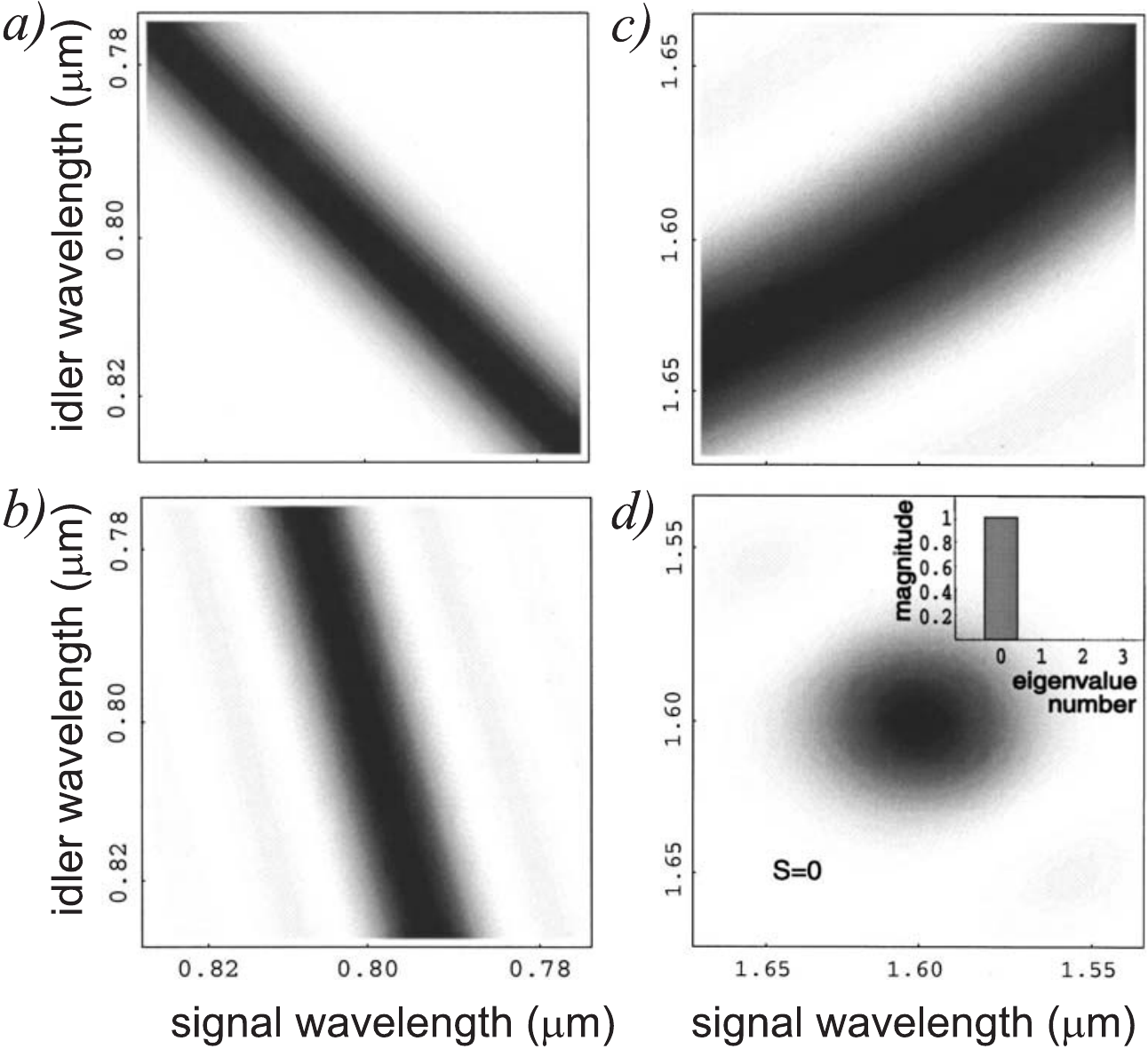}
\vspace*{8pt}
\caption{\label{gricefig} Shape of the biphoton correlation spectrum determined by (a): the energy conservation condition (\ref{energyconservation}), (b): generic phase matching (\ref{pmspectral}) in the case of collinear type II PDC in a BBO crystal pumped at 400 nm, (c): engineered phase-matching (collinear type II PDC in a BBO crystal pumped at 800 nm), (d): cumulative effect of conditions (a) and (c). The insert in (d) shows the Schmidt decomposition of the biphoton spectrum $\Psi(\omega_s,\omega_i)=\sum_m\sqrt{\lambda_m}u_m(\omega_s)v_m(\omega_i)$; with a proper combination of geometrical paremeters of the experiment the Schmidt decomposition contains only one term, i.e. the spectrum (d) is uncorrelated. Reproduced with permission from Grice \ieac (2001).}
\end{figure}

Considered more accurately, the lines defined by these equations are not infinitely narrow. This is because the pump is pulsed, so it contains photons not only at $\omega_{p0}$, but in a finite frequency range $|\omega_p-\omega_{p0}|\lesssim \pi/\tau_p$ determined by the inverse pump pulse width $\tau_p$. The phase-matching condition also has a tolerance: $|\vec k_s+\vec k_t-\vec k_p| \lesssim \pi/L_c$, where $L_c$ is the crystal length that limits the region of nonlinear optical interaction. The down-conversion spectrum is thus determined by the overlap of two band-shaped areas in the $(\omega_{s},\omega_{t})$  plane [Fig. \ref{gricefig}(a,b)].

As can be seen, the spectrum does not automatically uphold \eeqref{uncorrpdc}. However, by choosing the crystal length and other parameters of PDC, one can \emph{engineer} the tilt angle and the width of the spectral region in which phase matching is satisfied, such that the overlap area can be expressed in the product form (\ref{uncorrpdc}), implying uncorrelated photon spectra. The condition that has to be fulfilled takes the form (Grice \ieac 2001)
\begin{equation}\label{conduncorr}
\frac 1 {\sigma^2}=-0.048L^2\left(\frac1{v_{gr,p}}-\frac1{v_{gr,s}}\right)\left(\frac1{v_{gr,p}}-\frac1{v_{gr,t}}\right),
\end{equation}
where $L$ is the crystal length and $\sigma$ is the pump spectrum width.

There exist several theoretical proposals on shaping the phase matching region. Grice \iea (2001) calculate that \eeqref {conduncorr} satisfies in a BBO crystal for degenerate collinear type-II down-conversion if the pump wavelength is set to 800 nm [Fig.~\ref{gricefig}(c,d)]. They calculate a number of alternative down-conversion configurations with decorrelated spectra. U'Ren \iea (2003) propose to implement PDC in a slightly non-collinear configuration and impose additional restrictions onto the signal-idler spectrum by collecting only the photons emitted at certain angles. In theoretical works by Walton \iea (2003, 2004), down conversion occurs in a nonlinear waveguide, pumped almost orthogonally to the guided direction. This also allows restricting the transverse components of the signal and idler photon momenta, but without compromising the pair production rate and with the possibility to choose the central wavelength of each photon. Torres \iea (2005) proposes to engineer the down-conversion spectrum by employing a chirped, tilted pump and utilizing the Poynting vector walk-off effect. Raymer \iea (2005) put forward the idea of placing the down-conversion crystal into a microcavity, whose linewidth is much narrower than that allowed by the energy conservation and phase matching conditions. This leads a spectrally uncorrelated biphoton in a spatial mode defined by the cavity.  U'Ren \iea (2006, 2007) theoretically show that the group delays can be controlled by means of a periodic assembly (superlattice) of nonlinear crystals and birefringent spacers, and perform a proof-of-principle experiment to this effect, albeit without actually achieving an uncorrelated spectrum.

The only experimental demonstration of a virtually uncorrelated down-conversion spectrum to date is offered by Mosley \iea (2007). They use a relatively long potassium-dihydrogen-phosphate (KDP) crystal and a pump wavelength of 415 nm. Under these conditions, the pump will propagate with the same group velocity as the idler photon. Then the region allowed by the phase-matching condition [Fig.~\ref{gricefig}(b)] becomes vertical and narrow, so the overlap region exhibits almost no correlation. This is confirmed by observing high-visibility Hong-Ou-Mandel interference of heralded signal photons from separate crystals.

\subsubsection{The continuous-wave case}
\begin{figure}
\includegraphics[keepaspectratio,width=0.45\textwidth]{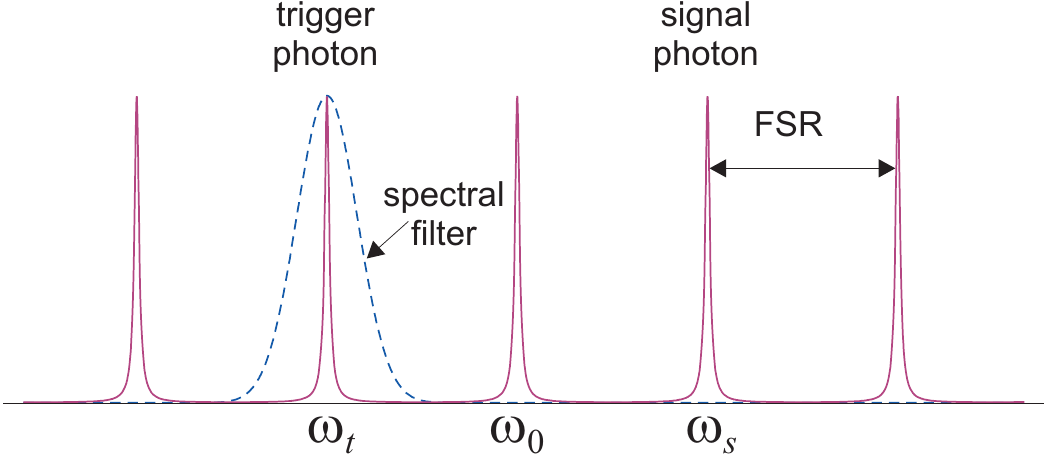}
\vspace*{8pt}
\caption{\label{opafilter} Spectrum of the Fabry-Perot parametric cavity. If one of the spectral modes ($\omega_t$) is selected in the trigger channel, a click of the trigger detector heralds production of the signal photon in a pure cavity mode at $\omega_s=2\omega_0-\omega_t$.}
\end{figure}

A completely different approach to producing heralded photons must be taken if the pump is monochromatic (continuous) and down-conversion occurs in an optical cavity, such as in Neergaard-Nielsen \iea (2007). \emph{Spatial} mode matching is simplified in this configuration because both photons are prepared in the spatial mode of the cavity. The biphoton \emph{spectrum} is determined by the cavity transmission spectrum: it consists of narrow (a few MHz) equidistant modes separated by the cavity free spectral range (FSR)\footnote{The free spectral range of a cavity equals the inverse roundtrip time of a photon inside the cavity.}. If the down-converter is pumped at a frequency $2\omega_0$ (where $\omega_0$ coincides with one of the cavity resonances), down-converted photons will be generated at frequencies $\omega_0\pm n\times{\rm FSR}$ (Fig.~\ref{opafilter}).

Under these circumstances, the time uncertainty of the trigger detector event is insignificant, so the advanced wave can be considered to be a short pulse. This eliminates the need for narrow filtering of the trigger photon; it is sufficient to apply a spectral filter that would transmit one of the cavity modes (e.g. $\omega_t=\omega_0-{\rm FSR}$). The advanced wave pulse will be filtered by the cavity. Its interaction with the pump will produce a difference-frequency pulse at $\omega_s=\omega_0+{\rm FSR}$ which, in turn, gets filtered by the cavity. Because the cavity spectrum is approximately Lorentzian, the temporal shape of the conditionally prepared mode is given by
\begin{equation} \label{polziksmode}
f(t,t_c)=e^{\gamma|t-t_c|},
\end{equation}
where $t_c$ is the moment of the trigger event and $\gamma$ is the HWHM cavity linewidth [Fig.~\ref{moelmersfig}(a)]\footnote{\eeqref{polziksmode} has to be modified if the weak pumping limit is not applicable (M{\o}Lmer, 2006; Sasaki and Suzuki, 2006).}.

Homodyne detection of the field in this mode requires a local oscillator pulse with the temporal shape [$g_L(t)$ in \eeqref{ELO}] identical to $f(t,t_c)$. Such a pulse can be ``tailored'' from a continuous laser field by means of acousto- or electro-optical amplitude modulation. An alternative, more practical, procedure, consists on using a continuous local oscillator ($g_L(t)=1$) and continuously acquiring the difference photocurrent as a function of time. The acquired photocurrent is then post-processed by multiplying it by $f(t,t_c)$ and subsequently integrating over time. As evidenced by \eeqref{Nminus}, the integrated difference charge is the same as that obtained with a pulsed LO. This idea was first utilized in Neersgaard-Nielsen \iea(2006) and subsequently in Wakui \iea(2006) and Neersgaard-Nielsen \iea(2007).

\subsubsection{Strong pumping mode}
\begin{figure}
\includegraphics[keepaspectratio,width=0.45\textwidth]{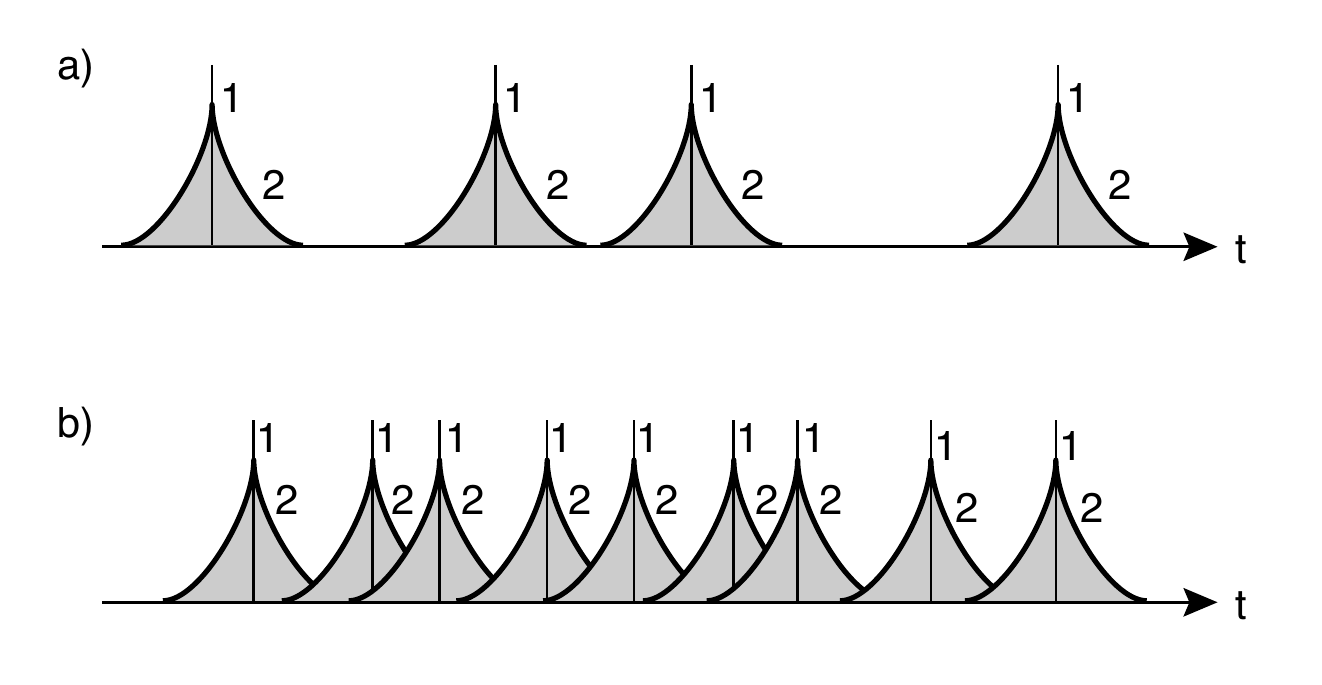}
\vspace*{8pt}
\caption{\label{moelmersfig} Temporal modes of the trigger (1) and signal (2) photons. Assuming that the time resolution of the trigger detector is very high, the signal mode is defined by \eeqref{polziksmode}. Cases (a) and (b) correspond to the weak and strong pumping regimes, respectively.
Reproduced with permission from M{\o}lmer (2006)}
\end{figure}

We now briefly discuss the regime of strong-pump PDC, such that the number of pairs generated within the time period corresponding to the inverse down-conversion bandwidth is not negligible. This case is complicated and largely uninvestigated for the following reason: if there exists any correlation in the joint time- or spectral distributions of the signal and trigger photons, narrow filtering of the trigger channel does not guarantee purity of the signal state. For example, in the monochromatic-pump case, a trigger event at time $t_c$ heralds the presence of a photon in the mode $f(t,t_c)$ but does not ensure that no incoherent contributions are present from photons in ``nearby" modes $f(t,t'_c)$ [Fig.~\ref{moelmersfig}(b)]. In order to prepare a high-purity state, one needs to have a high-efficiency trigger detector, which will not only trigger a pair production event, but ensure there are no more events nearby. If the state to be prepared is more complex than a Fock state, such as a the Schr\"odinger kitten state, the situation is even more complicated.

The strong pumping case in the context of preparing single and multiple heralded photons was investigated by M{\o}lmer (2006) as well as Nielsen and M{\o}lmer (2007a,b). Sasaki and Suzuki (2006) report a comprehensive theoretical study employing mode expansion in the basis of prolate spherical functions and obtain analytical expressions for a few limiting cases.

\emph{Experimental} handling of the mode mismatch in the strong pumping case is, on the other hand, relatively straightforward (Ourjoumtsev \ieac 2006a, 2006b, 2007). One introduces an empiric probability $\xi$ that the state heralded by a click in the trigger detector belongs to the mode analyzed by the homodyne detector. With probability $1-\xi$, the heralded state belongs to an orthogonal mode, which is equivalent to a dark count event. The actual value of $\xi$ can be found by matching to the experimental data statistics.

\section{Applications in quantum technology}
\label{homodisc}

Implementation of light for the purposes of QI
technology relies on our ability to \emph{synthesize, manipulate,}
and \emph{characterize} various quantum states of the
electromagnetic field. OHT is used in optical quantum information as a way to solve the last of the above tasks. In this section, we discuss applications of OHT to ``discrete-variable'' quantum-optical information and quantum-optical technology in general. We review the new states of light that have been created in the last few years, methods of their preparation, and their tomographic reconstruction.

It is convenient to restrict this review by the temporal boundaries of the present century. The only nonclassical state of light investigated by OHT prior to 2001 was the squeezed state (Smithey \ieac 1993; Breitenbach \ieac 1997). The last few years, on the contrary, have shown a technology boom, resulting in a plethora of new quantum optical states (Table I), some of which are significant not only to QI technology, but to the very roots of quantum physics.


\begin{table}[ht]
\begin{tabular}{|l|l|}
\hline\hline
 Reference& State\\
\hline\hline
Lvovsky \iea (2001) & Single-photon Fock state \\
Zavatta \iea (2004b)& $\ket 1$\\
Neergaard-Nielsen (2007) & \\
\hline
Lvovsky and Babichev (2002) & Displaced single photon\\
\hline
Lvovsky and Mlynek (2002) & Single-rail qubit \\
Babichev \iea (2003) & $\alpha\ket 0+\beta\ket 1$\\
Babichev \iea (2004a) & \\
Zavatta \iea (2006a)& \\
\hline
Babichev \iea (2004b) & Dual-rail qubit \\
Zavatta \iea (2006a,b)& $\alpha\ket{0,1}+\beta\ket{1,0}$\\
\hline
Zavatta \iea (2004a) & Photon-added coherent state \\
Zavatta \iea (2005a) & $\hat a^\dag\ket\alpha$ \\
\hline
Zavatta \iea (2007) & Photon-added thermal state \\
Parigi \iea (2007) & \\
\hline
Ourjoumtsev \iea (2006b) &  Two-photon Fock state $\ket 2$\\
\hline
Wenger \iea (2004b) & Photon-subtracted squeezed \\
Neergaard-Nielsen \iea (2006) & state (``Schr\"odinger kitten") \\
Ourjoumtsev \iea (2006a) & $\ket 1 + \alpha \ket 3$, where $|\alpha|\ll 1$ \\
Wakui \iea (2007) & \\
\hline
Ourjoumtsev \iea (2007) & Squeezed ``Schr\"odinger cat"\\
\hline\hline
\end{tabular}
\caption{Quantum states recently characterized by OHT}
\end{table}

\subsection{Fock state tomography} \label{Focksec}
The first non-Gaussian state to be studied by OHT (Lvovsky \emph{et al.}, 2001) is the single photon. This is not surprising, given the role this state plays in basic and applied quantum optics. Another important motivation for this experiment was to demonstrate reconstruction of an optical state whose Wigner function takes on negative values.

\begin{figure}
\includegraphics[keepaspectratio,width=0.45\textwidth]{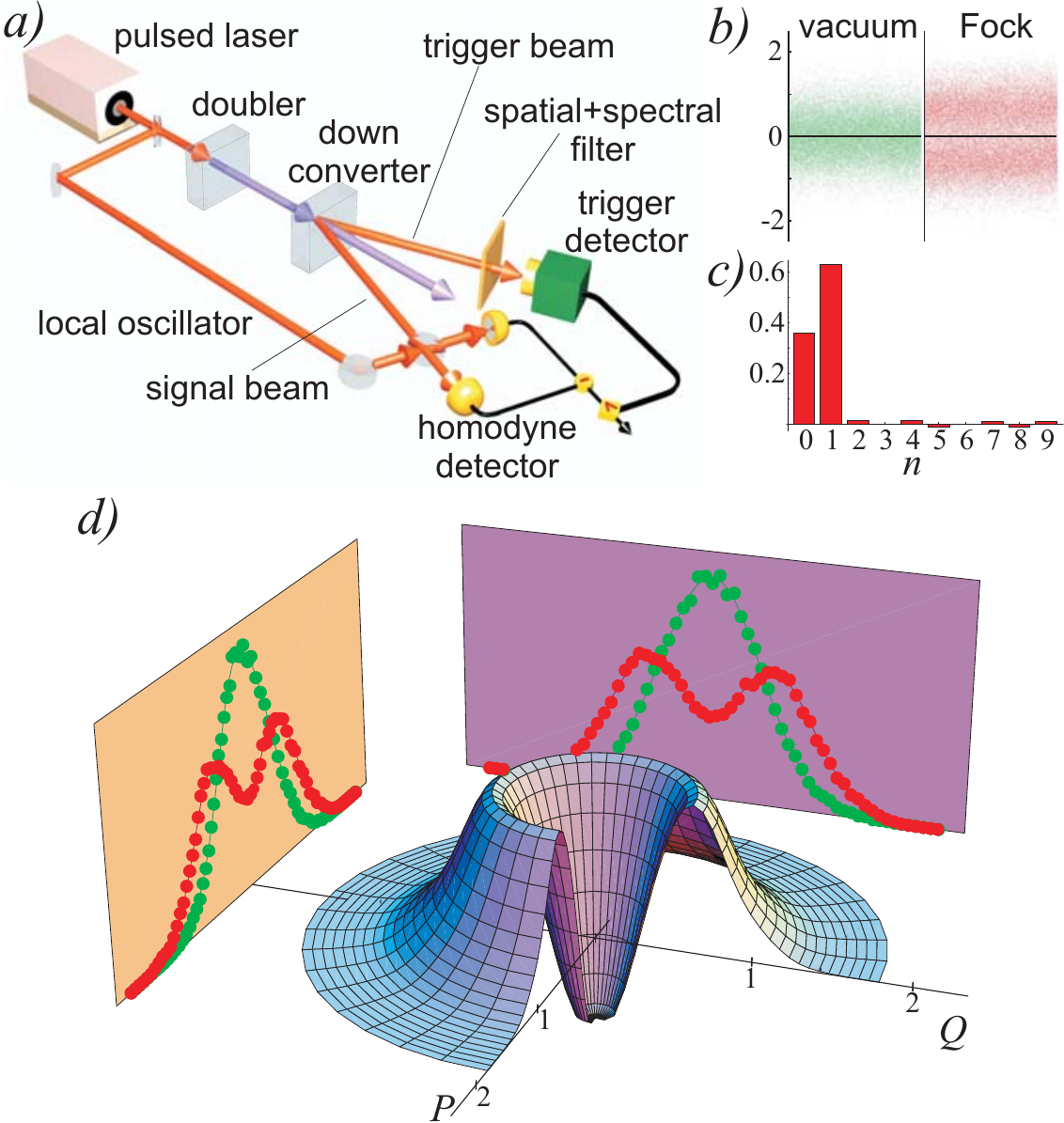}
\vspace*{8pt}
\caption{\label{focktomo} The experiment on quantum tomography of the single-photon
Fock state.
(a) Simplified scheme;
(b) 45000 raw quadrature noise samples for the vacuum state and the Fock state;
(c) the density matrix (diagonal elements) reconstructed using the quantum state sampling method;
(d) the reconstructed Wigner function is negative near the origin point because the measurement efficiency reaches 62\%. Side projections show phase-randomized marginal distributions for the measured vacuum and Fock states. From Lvovsky \iea (2001); Lvovsky and Babichev (2002).}
\end{figure}

The schematic and results of the experiment are shown in Fig.~\ref{focktomo}. The experiment employed a picosecond Ti:sapphire laser at a 790-nm wavelength. Pulsed single photons were prepared by conditional measurements on a biphoton state generated via parametric down-conversion (in the weak pumping regime). Narrow spatiotemporal filtering of the trigger photon was used as outlined in Sec.~\ref{MMsec}. The field state in the signal channel was characterized by
means of optical homodyne tomography.

Remarkably, all imperfections of the experiment (losses in transmission of the signal photon, quantum
efficiency of the HD, trigger dark counts, mode matching of the signal photon and the local oscillator, and spatiotemporal coherence of the signal photon) had a similar effect on the reconstructed state: admixture of the vacuum $\ket 0$ to the ideal Fock state $|1\rangle$:
\begin{equation}\label{rhomeas}\rho_{\rm
meas} = \eta|1\rangle\langle 1| + (1-\eta) |0\rangle\langle 0|.
\end{equation}
The greater the efficiency $\eta$,
the deeper the ``well" in the Wigner function; classically impossible
negative values are obtained when $\eta>0.5$. The original 2001 experiment showed $\eta=0.55\pm0.01$; later, this value was improved to $0.62$.

An interesting feature of the optical single-photon state reconstruction is
that the technique of homodyne tomography can be fully understood
in the framework of classical physics. This
measurement could have been conducted (and interpreted) by someone
who does not believe in quantum mechanics. Yet the result of negative quasi-probabilities
would appear absurd, incompatible with classical
physics --- thus providing a very strong evidence of
``quantumness" of our world.


A version of the single-photon Fock state tomography experiment, featuring a fast homodyne detector, allowing measurements at a full repetition rate of the pulsed laser (82 MHz), was reported by Zavatta \iea (2004b).


Tomography of the two-photon Fock state $\ket 2$ was reported by Ourjoumtsev \iea (2006b). The experimental arrangement is similar to the of Fig.~\ref{focktomo}(a), but the trigger channel is split and directed into two single-photon detectors, whose simultaneous click triggers a homodyne measurement. In order to obtain a sufficient rate of such events, the parametric gain has to be non-negligible, which significantly complicates the analysis of the experiment. Ourjoumtsev and co-workers found that the experimentally observed state can be fit by a theory taking into account five experimental parameters:
\begin{itemize}
\item gain of the down-converter;
\item excess gain of a fictitious phase-independent amplifier placed behind the (ideal) down-converter;
\item BHD efficiency (including optical losses);
\item BHD electronic noise (whose effect is identical to optical loss, see Sec.~\ref{tdhd});
\item probability $\xi$ that the single-photon detection events corresponds to a heralded mode that matches the local oscillator (see Sec.~\ref{MMsec}).
\end{itemize}
By optimizing these parameters, negative values of the experimental Wigner function were obtained.

\subsection{The optical qubit}
\subsubsection{The dual-rail qubit}\label{dualrail}
\begin{figure}
    \begin{center}
\includegraphics[keepaspectratio,width=0.45\textwidth]{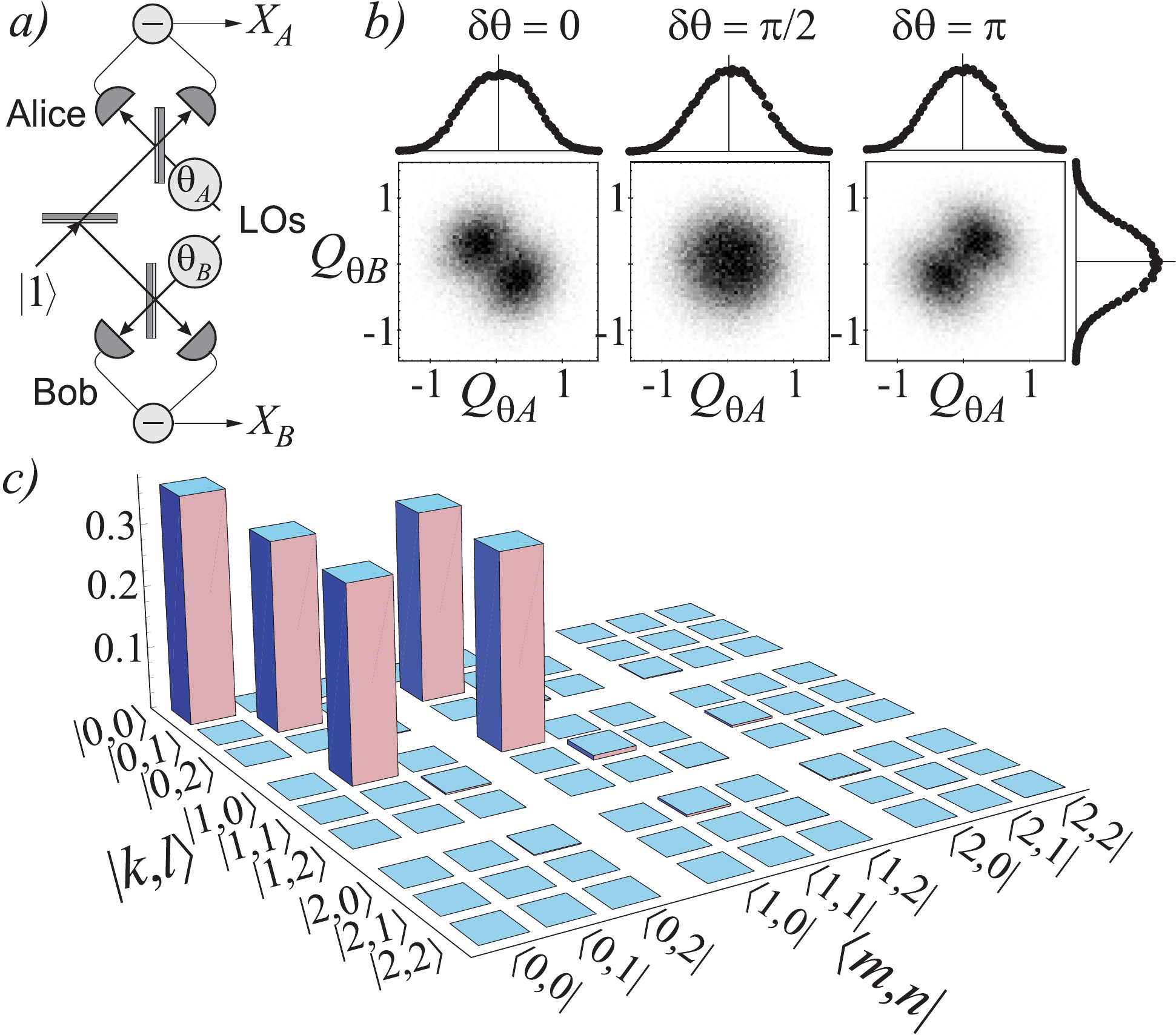}
\vspace*{8pt}
        \caption{\label{qubittomopic} The experiment on homodyne tomography of the dual-rail qubit. (a) Scheme of the experimental setup. (b)
        Histograms of the experimental quadrature statistics ${\rm
        pr}_{\delta\theta}(Q_{\theta A},Q_{\theta B})$ for a symmetric beam splitter.
        Phase-dependent quadrature correlations are a consequence
        of the entangled nature of the state $\ket{\Psi_{\rm qubit}}$. Also
        shown are individual histograms of the data measured by
        Alice and Bob, which are phase-independent. (c) Density matrix (absolute values) of the measured ensemble in
        the photon number representation. From Babichev \iea (2004b)}
        \label{corr}
    \end{center}
\end{figure}

As discussed in Sec.~\ref{OTHintrosec}, one application of OHT, where it can be of advantage compared to other state characterization methods, is measurement of systems of dual-rail optical qubits. Tomography of one dual-rail qubit was performed by Babichev {\it et al.} (2004b)\footnote{A deterministic OHT scheme for two-mode state
reconstruction in the Fock basis was first proposed and tested
numerically in Raymer {\it et al.} (1996). A detailed theoretical analysis of different aspects of
such an experiment was also made by Jacobs and Knight (1996) as well as Grice and Walmsley (1996).}.
A dual-rail qubit, described by the state
\begin{equation}\label{psimin}
  \ket{\Psi_{\rm dual-rail}}=\tau\ket{1_A,0_B}-\rho\ket{0_A,1_B},
\end{equation} is generated when a single photon $\ket{1}$
incident upon a beam splitter with transmission $\tau^2$ and reflectivity $\rho^2$, entangles itself with the vacuum
state $\ket{0}$ present in the other beam splitter input.
 To perform tomography measurements, BHDs (associated with fictitious
observers Alice and Bob) were placed into each beam splitter output channel
[Fig.~\ref{qubittomopic}(a)]. With every incoming photon, both detectors made
measurements of field quadratures $Q_{\theta A}$ and $Q_{\theta A}$ with the local
oscillators' phases set to $\theta_A$ and $\theta_B$,
respectively. Fig.~\ref{qubittomopic}(b) shows histograms of these
measurements, which are the marginal
distributions of the four-dimensional Wigner function of the
dual-rail state. They have been used to determine the state via the maximum-likelihood technique (Sec.~\ref{maxliksec}), resulting in the density matrix shown in Fig.~\ref{qubittomopic}(c). As expected, the reconstruction reveals \emph{all} the terms in the density matrix, including those (e.g. double-vacuum $\ketbra{0,0}{0,0}$) usually missed by the photon-counting method.

\emph{Time-encoded} dual-rail optical qubits were prepared and characterized by Zavatta \iea (2006a). The trigger channel of the PDC entered a fiber Michaelson interferometer whose path length difference was equal to the optical path inside the cavity of the master mode-locked Ti:Sapphire laser. The trigger detector, placed at the output of the interferometer, is then unable to distinguish between a photon generated by some $n$th pump pulse that has traveled the long path of the interferometer and a photon generated by the $(n+1)$th pulse that has taken the shorter path. The conditional state was thus prepared in a coherent superposition
\begin{equation}
\ket\Psi=\frac 1 {\sqrt 2} \left(\ket{1^{(n)},0^{(n+1)}}+e^{-i\phi}\ket{0^{(n)},1^{(n+1)}}\right),
\end{equation}
where the phase $\phi$ could be controlled by one of the interferometer mirrors. For characterizing the above state, a single homodyne detector suffices, but its signal has to be acquired at two different moments in time.

\subsubsection{Nonlocality of the single photon} \label{nonlocalitysec}
Whether the state (\ref{psimin}) can be considered entangled is a widely debated issue.
This controversy seems to be related to the wave-particle duality of light. If the photon is viewed as a \emph{state} of the electromagnetic oscillator, the notation (\ref{psimin}) is valid and denotes an entangled entity (Van Enk, 2005). If, on the other hand a photon is considered to be a particle, i.e. not a state but a \emph{carrier of a state}, e.g. of a polarization state, the dual rate qubit should be written as a superposition of two localizations of one photon, which may not be seen as entangled. Advocates of the former view proposed experiments on using the split single photon to demonstrate quantum nonlocality (Oliver and Stroud, 1989;
Tan \ieac 1991; Banaszek and Wodkiewicz, 1999;
Jacobs and Knight, 1996; Hessmo \ieac 2004); others disputed them (Greenberger \ieac 1995; Vaidman, 1995).

It is in the inherent nature of OHT to interpret the photon as a state of a field rather than a particle ``in its own right". In experiments on homodyne tomography of the delocalized photon, Babichev \iea (2004b) and Zavatta \iea (2006b) present different arguments that OHT characterization of the dual-rail qubit (\ref{psimin}) can be interpreted to violate
Bell-type inequalities, albeit with loopholes. In Babichev \iea (2004), quadrature measurements have been converted to a dichotomic format by means of a fictitious discriminator. Correlations between the discriminator outputs acquired by Alice and Bob exhibited a Bell-like interference pattern. For sufficiently high threshold values, its amplitude exceeds $1/\sqrt{2}$ and the Bell inequality is violated. Zavatta \iea (2006b) determined the Wigner function of the measured dual-rail state and showed it to violate the Bell test of Banaszek and Wodkiewicz (1999).

Further evidence of the entangled nature of the delocalized single photon is its applicability as a resource in quantum communication protocols such as quantum teleportation and remote state preparation (RSP)\footnote{Both teleportation (Bennett {\it et al.}, 1993) and RSP (Lo, 2000) are quantum communication protocols allowing disembodied transfer of quantum information
between two distant parties by means of a shared entangled
resource and a classical channel. The difference between them is that in
teleportation, the sender (Alice)
possesses one copy of the source state, while in RSP she is instead aware of its full
classical description.}.

\subsubsection{Remote state preparation using the nonlocal single photon state}\label{RSPsec}
To implement RSP, Alice performs a
measurement on her share of the entangled resource in a basis
chosen in accordance with the state she wishes to prepare.
Dependent on the result of her measurement, the entangled ensemble
collapses either onto the desired state at the receiver (Bob's)
location, or can be converted into it by a local unitary operation.


The experiment on tomography of the dual-rail qubit can be interpreted as an
implementation of the remote preparation protocol in the continuous basis (Babichev \ieac 2004a, Zavatta \ieac 2006a).
By performing a homodyne
measurement on her part of the entangled state (\ref{psimin}) and detecting a particular quadrature value $Q_{\theta A}$ at the local
oscillator phase $\theta$, Alice projects the entangled resource
(\ref{psimin}) onto a quadrature eigenstate $\bra{Q_{\theta A}}$:
\begin{eqnarray}
    \ket{\psi_B}&=&\braket{Q_{\theta A},\theta_A}{\Psi} \\ \nonumber
    &=&\tau \braket{Q_{\theta A},\theta_A}{1}_A \ket{0}_B -\rho \braket{Q_{\theta A},\theta_A}{0}_A\ket{1}_B,
\end{eqnarray}
which is just a coherent superposition of the single-photon and
vacuum states, i.e. a \emph{single-rail optical qubit}. By choosing her LO phase $\theta_A$ and postselecting a particular
value of $Q_{\theta A}$, Alice can control the coefficients in the superposition, i.e. remotely prepare any arbitrary state within the single-rail qubit subspace.

\subsubsection{Teleportation using the nonlocal single photon state}
Entanglement contained in the delocalized single photon state is between the single-photon and vacuum states. It allows rudimentary teleportation of single-rail qubits $\ket{\psi_{\rm single-rail}}=\alpha\ket 0+\beta \ket 1$ by means of a modified Bennet \iea (1993) protocol. Alice performs
a Bell-state measurement on the source state and her share of $\ket{\Psi_{\rm dual-rail}}$ by overlapping them on a beam splitter and sending both beam splitter outputs to single-photon detectors. If one of these detectors registers the vacuum state, and the other detects one photon, the input state of the Bell-state analyzer is projected onto $\ket{\Psi_{\rm dual-rail}}$ and
Bob's channel obtains a copy of the source state (Pegg \ieac 1998; \"Ozdemir \ieac 2002). If the input state contains terms outside of the single-rail qubit subspace, these terms will be removed from the teleported ensemble. This is known as the ``quantum
scissors" effect.

Although the implementation of the protocol requires highly-efficient, number-resolving photon detectors, its conceptual demonstration can be done with standard commercial units. Babichev \emph{et al.} (2003) performed this experiment by using a weak pulsed coherent state as the source. The teleported state was characterized by
means of OHT.
The teleportation fidelity approached unity for low input state amplitudes but with this parameter increasing, it quickly
fell off due to the effect of ``quantum scissors".

\subsubsection{Quantum-optical catalysis}
\begin{figure}
\includegraphics[keepaspectratio,width=0.45\textwidth]{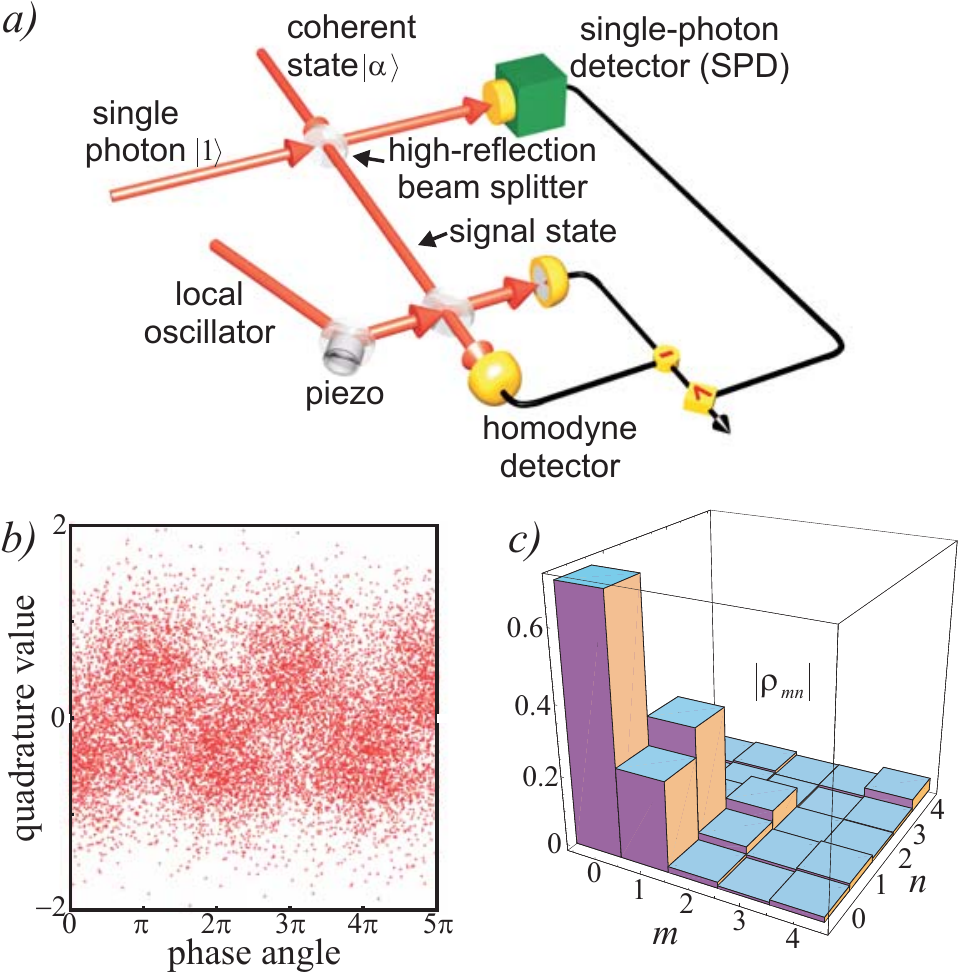}
\vspace*{8pt}
\caption{\label{catfig} The ``quantum-optical catalysis" experiment. (a) The scheme. Measurements by
the HD are conditioned on the
single-photon detector registering a photon; (b) 14153 raw
quadrature data; (c) absolute values of the density matrix
elements in the Fock representation for $\alpha\approx 0.3$. The beam splitter transmission is $t^2=0.075$. From Lvovsky and Mlynek (2002).}
\end{figure}

The two subsections above demonstrated how a single-rail optical qubit can be prepared by conditional measurements and linear-optical operations on a single photons. Another way of achieving the same goal was reported by Lvovsky and Mlynek (2002). A single-photon state $\ket 1$ and a coherent state $\ket\alpha$ were overlapped on a high-reflection beam splitter. One of the beam splitter outputs was subjected to a measurement via a single-photon
detector [Fig.~\ref{catfig}(a)]. In the event of a ``click", the other beam splitter output is projected onto a single-rail qubit $
t|0\rangle+\alpha|1\rangle$, $t^2$ being the beam splitter transmission [Figs.~\ref{CompFig} and \ref{catfig}(b,c)].

This result is somewhat counter-intuitive: a classical coherent input state is converted into a nonclassical single-rail qubit even though the input single photon emerges ``intact" at the output (hence the name {\it
quantum-optical catalysis}). This transformation is an example of optical nonlinearity induced by a conditional optical measurement, the key principle behind linear-optical quantum computation (Knill {\it et al.}, 2001; Koashi \emph{et al.}, 2001; Kok \ieac 2007).

We note that if the signal channel is analyzed without conditioning
on the single-photon detection event, it approximates another
important nonclassical state of light, the displaced Fock state (Lvovsky and Babichev, 2002, and references therein).

\subsection{``Schr\"odinger cats" and ``kittens"}
The Schr\"odinger cat (Schr\"odinger, 1935) is a famous \emph{Gedankenexperiment} in quantum physics, in which a macroscopic object is prepared in a coherent superposition of two classically distinguishable states. It brightly illustrates one of the most fundamental questions of quantum mechanics: at which degree of complexity does a quantum superposition of two states stops being a superposition and probabilistically becomes one of its terms?

In quantum optics, the Schr\"odinger cat usually means a coherent superposition $\ket\alpha\pm\ket{-\alpha}$  of coherent states of relatively large amplitude and opposite phase (Bu\v zek and Knight, 1995). In addition to the above fundamental aspect, these states are useful for many quantum information protocols such as quantum teleportation (van Enk and Hirota, 2001), quantum computation (Ralph, 2003), and error correction (Cochrane \ieac 1999). It is thus not surprising that experimental synthesis of Schr\"odinger cats has been an object of aspiration for several generations of physicists. Recent years have marked a breakthrough: invention and experimental realization of two schemes that permit preparation of optical Schr\"odinger cats of arbitrarily high amplitudes.

\subsubsection{Squeezing of single photons}
\begin{figure}
\includegraphics[keepaspectratio,width=0.45\textwidth]{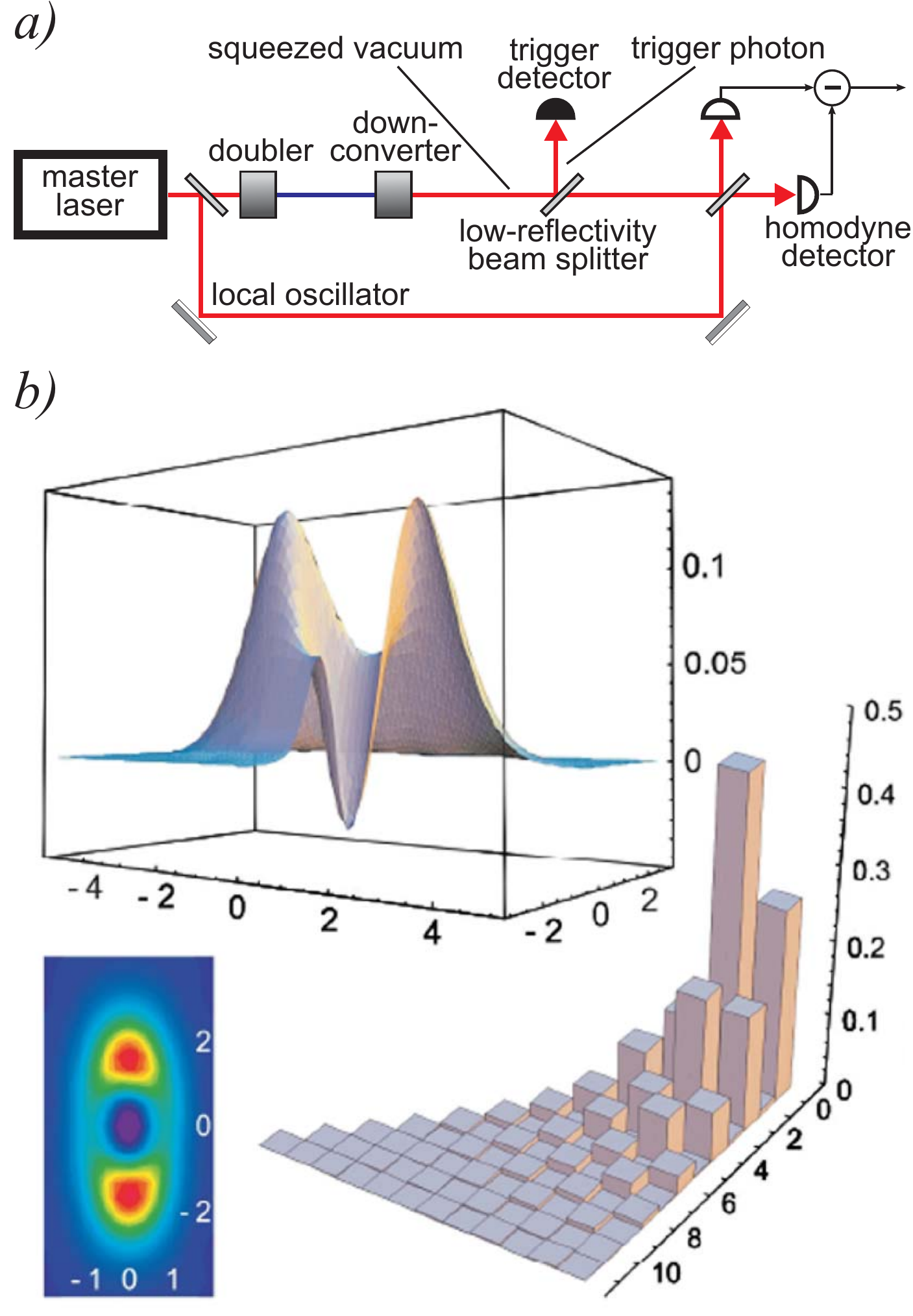}
\vspace*{8pt}
\caption{\label{kittenfig}(a) A generic scheme of conditional preparation of the squeezed single-photon state (``Schr\"odinger kitten''). (b) The Wigner function and the density matrix reconstructed from the experimental data. Part (b) is reproduced with permission from Neergaard-Nielsen \iea (2006).}
\end{figure}

The first scheme was proposed by Dakna \iea (1997). An odd Schr\"odinger cat state of low amplitude can be decomposed into the Fock basis as follows:
\begin{equation}\label{oddcat}
\ket\alpha-\ket{-\alpha}\ \propto\ \alpha\ket 1 + \frac{\alpha^3}{\sqrt 6}\ket 3+\ldots\,.
\end{equation}
For $\alpha\lesssim 1$, this state is approximated, with very high fidelity (Lund \ieac 2004), by the squeezed single-photon state. Experimentally this state can be obtained by removing one photon from the squeezed vacuum
\begin{equation}
\mid\Psi_s\rangle \approx \ket 0 + \frac 1 {\sqrt 2} \zeta \ket 2 + \sqrt{\frac 3 2} \zeta^2 \ket 4+\ldots,
\end{equation}
where $\zeta$ is the small squeezing parameter (Dakna \ieac 1997). The photon removal procedure consists of transmitting the state through a low-reflection beam splitter and sending the reflected mode to a single-photon detector [Fig.~\ref{kittenfig}(a)]. A click in this detector indicates that at least one photon has been removed from $\ket{\Psi_s}$. Because the reflectivity of the beam splitter is small, the probability of removing more than one photon is negligible. The conditional state in the transmitted channel of the beam splitter is then approximated by
\begin{equation} \label{condstate}
\mid\Psi_{\rm cond}\rangle \propto \frac 1 {\sqrt 2} \zeta \ket 1 + \sqrt 3 \zeta^2 \ket 3+\ldots\,.
\end{equation}
Setting $\alpha^2=6\zeta$ yields the ``Schr\"odinger kitten'' \eqref{oddcat}.

The first experiment implementing of this protocol was performed by Wenger {\it et al.}~(2004b) and later improved by Ourjoumtsev \iea (2006a). Parametric deamplification of 150-fs, 40-nJ pulses at 850 nm in a 100 $\mu$m non-critically phase matched KNbO$_3$ crystal was used to generate pulsed squeezed vacuum (Wenger \ieac 2004a). The heralded beam splitter output state was subjected to pulsed OHT, showing preparation of a Schr\"odinger kitten of size $|\alpha^2|=0.79$ with a 70-\% fidelity. Experimental imperfections could be modeled by the same five parameters as in Ourjoumtsev \iea (2006b) (see Sec.~\ref{Focksec}).

In the continuous-wave regime, a similar procedure has been independently demonstrated by Neergaard-Nielsen \iea (2006) [the results of this experiment are shown in Fig.~~\ref{kittenfig}(b)] and by Wakui \iea (2007).

\begin{figure}
\includegraphics[keepaspectratio,width=0.45\textwidth]{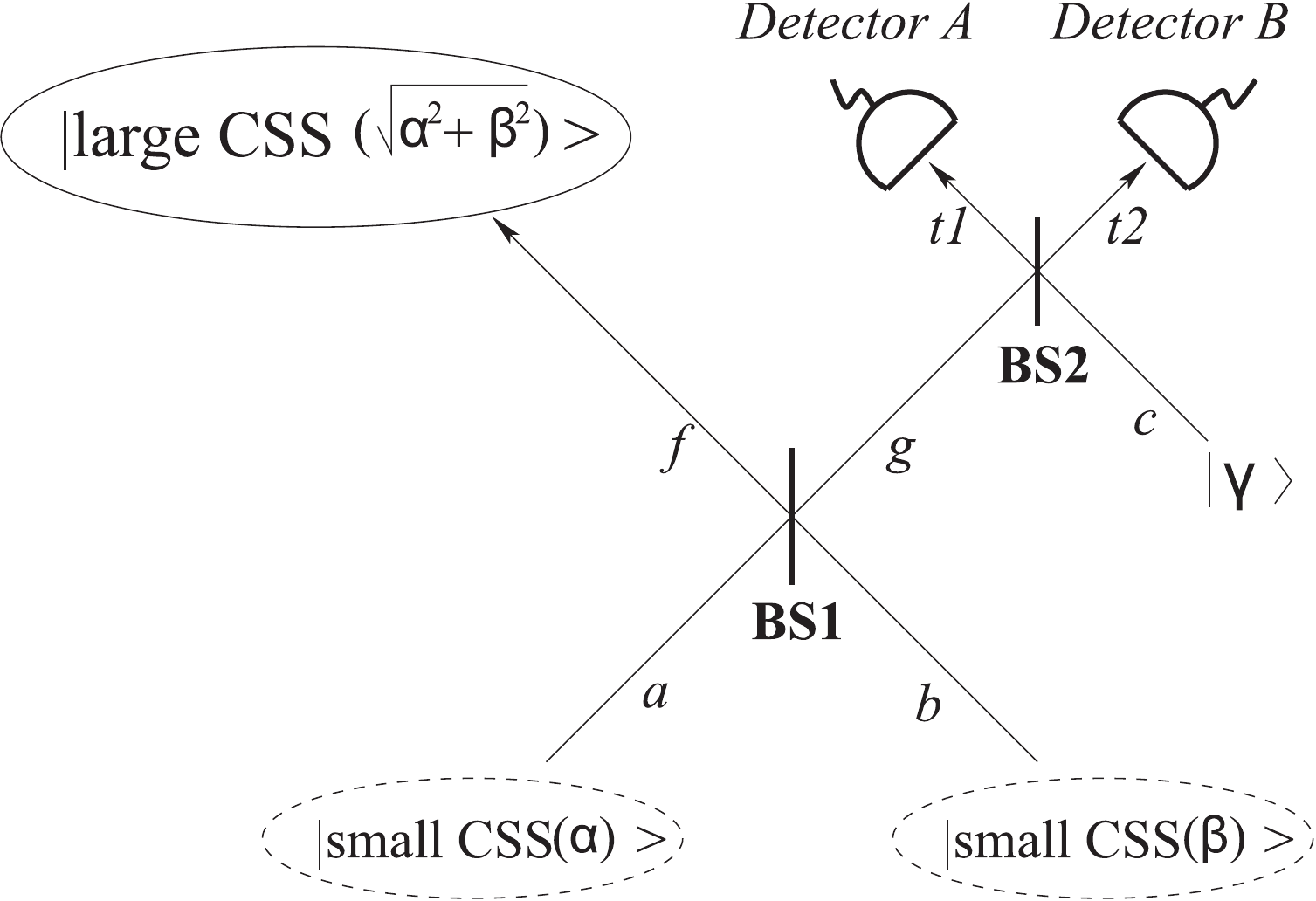}
\vspace*{8pt}
\caption{\label{breedfig} A schematic for amplification of the ``Schr\"odinger cat" state. The amplified cat emerges in channel $g$ if detectors $A$ and $B$ click in coincidence. In the text, $\alpha=\beta=\gamma/\sqrt 2$ is assumed. Reproduced with permission from Lund \iea (2004)}
\end{figure}

These experiments, which combine, for the first time, the techniques of conditional preparation of single photons and pulsed squeezing, are a significant technological breakthrough in quantum optical information technology. However, the photon-subtracted squeezed state resembles a cat state only for small $\alpha$'s. An interesting method for generating Schr\"odinger cats of larger amplitudes has been proposed by Lund {\it et al.} (2004). Two small odd cat states\footnote{Under ``even'' and ``odd'' cat states we understand superpositions $\ket\alpha+\ket{-\alpha}$ and $\ket\alpha-\ket{-\alpha}$, respectively.} $\ket{{\rm CSS}(\alpha)}$, which we assume to be of equal amplitudes ($\alpha=\beta$), overlap on a symmetric beam splitter BS1 (Fig.~\ref{breedfig}), which transforms them into an entangled superposition
\begin{eqnarray}\label{breeding}
\ket{{\rm CSS}(\alpha)}_a\ket{{\rm CSS}(\alpha)}_b&&\\ \nonumber
&&\hspace*{-3cm}\to\ket 0_f\left(\ket{\sqrt 2 \alpha}+\ket{\sqrt 2 \alpha}\right)_g-\left(\ket{\sqrt 2 \alpha}+\ket{\sqrt 2 \alpha}\right)_f\ket 0_g
\end{eqnarray}
of a cat state of amplitude $\sqrt 2 \alpha$ in one of the output channels and vacuum in the other. Now if we measure the state in channel $g$ and find it to be \emph{not} the cat state, the channel $f$ will be projected onto the cat state. Such conditional measurement is implemented by overlapping channel $g$ with a coherent state of amplitude $\gamma=\sqrt 2 \alpha$ on an additional beam splitter BS2. If channel $g$ contains a cat state, the interference will cause all optical energy to emerge at only one side of BS2. Therefore, detecting coincident photons in both outputs of BS2 indicates that channel $g$ contained vacuum, and thus channel $f$ is prepared in the Schr\"odinger cat state of amplitude $\sqrt 2 \alpha$.

By applying this linear optical protocol repeatedly, we can ``breed" Schr\"odinger cat states of arbitrarily high amplitude. A remarkable practical advantage of this technique is that it requires neither null single photon detection nor photon number discrimination. It does however require a high degree of mode matching among the interfering optical channels; otherwise the fidelity will rapidly decrease (Rohde and Lund, 2007).

\subsubsection{Generating ``Schr\"odinger cats" from Fock states}
\begin{figure}
\includegraphics[keepaspectratio,width=0.45\textwidth]{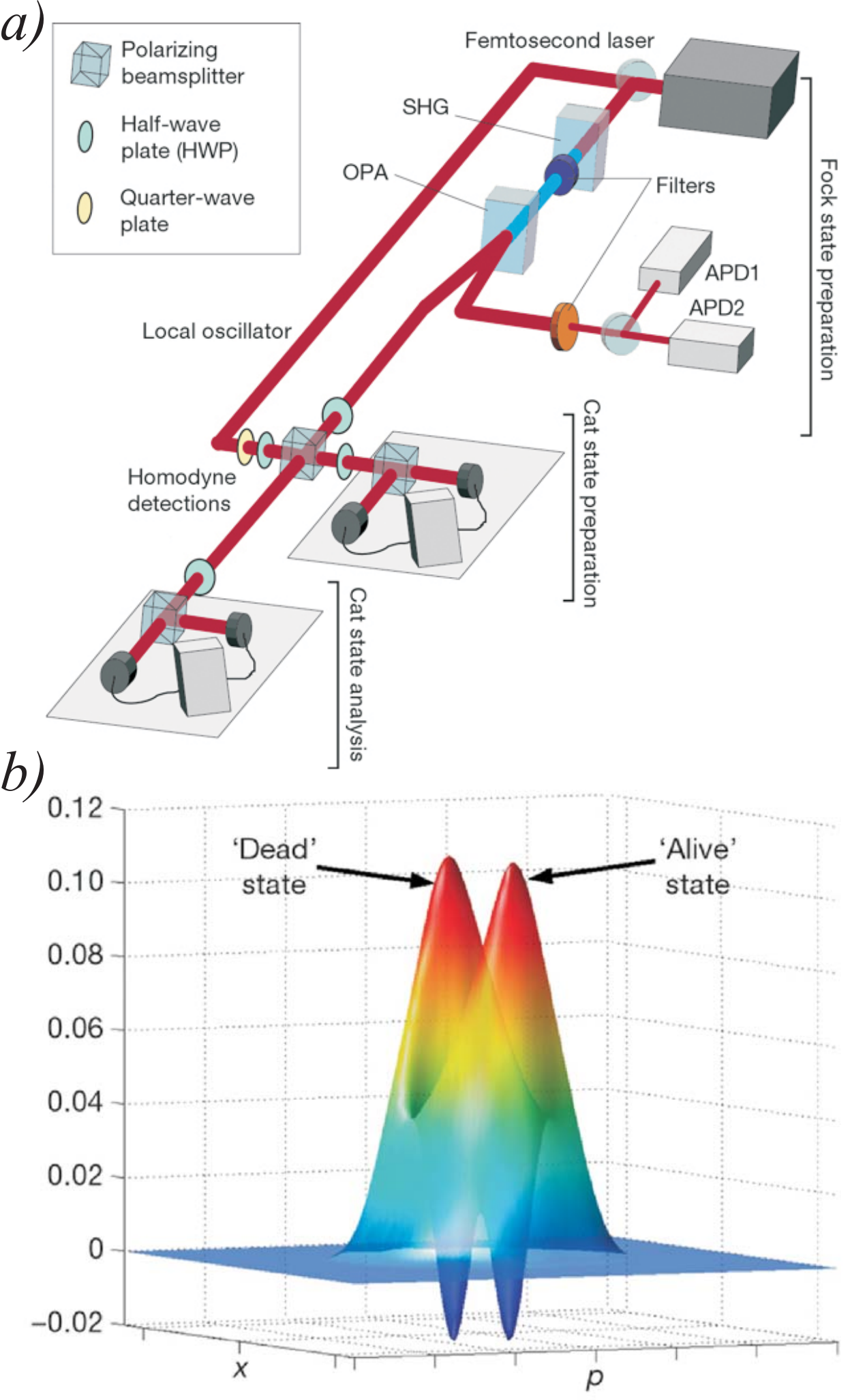}
\vspace*{8pt}
\caption{\label{bigcatfig}(a) A scheme of the experiment on conditional preparation of the ``Schr\"odinger cat'' state from photon number states. (b) The experimental Wigner function. Reproduced with permission from Ourjoumtsev \iea (2007).}
\end{figure}

An alternative way of generating optical cat states was recently proposed and implemented by Ourjoumtsev \iea (2007). This technique employs Fock states, rather than the squeezed state, as the primary resource. The procedure is remarkably simple: an $n$-photon number state is split on a symmetric beam splitter, and one of the output channels is subjected to homodyne detection. Conditioned on this measurement producing approximately zero, the other beam splitter output mode will contain an approximation of the squeezed Schr\"odinger cat of amplitude $\alpha\approx\sqrt n$ [Fig.~\ref{bigcatfig}(a)].

To gain some insight into this method, let us assume that the quadrature measured by the preparation homodyne detector is the momentum $P$. The wavefunction of the initial Fock state is given by
\begin{equation}\label{psip}
\psi_n(P)=\braket{n}{P} =H_n(P)\exp\left(-\frac{P^2}{2}\right).
\end{equation}
In writing the above equation, we used \eeqref{psix}, neglected normalization factors and remembered that the Fock state is phase-independent (i.e. its wave function is the same for all quadratures). ``Splitting" the state $\ket n$ means entangling it with the vacuum, which has the wavefunction $\psi_0(P_0)=\exp(-{P_0^2}/{2})$, via transformation $P\to(P-P_0)/\sqrt 2,\ P_0\to(P+P_0)/\sqrt 2$. Accordingly, the two-mode wavefunction of the beam splitter output is given by
\begin{eqnarray}\label{psioutdual}
\tilde\phi(P,P_0)&=&\psi_n(\frac{P-P_0}{\sqrt 2})\psi_0(\frac{P+P_0}{\sqrt 2})\\ \nonumber
&=&H_n(\frac{P-P_0}{\sqrt 2})\exp\left(-\frac{P^2+P_0^2}{2}\right),
\end{eqnarray}
Detecting the momentum quadrature value $P_0=0$ in one of the modes has an effect similar to that discussed in Sec.~\ref{RSPsec}: it ``remotely'' prepares the other mode in the state with wavefunction $\tilde\phi_{\rm cond}(P)=H_n((P-P_0)/{\sqrt 2})e^{-P^2/2}$.

This state is easier to analyze in the position quadrature representation. Making a Fourier transform of $\tilde\phi_{{\rm cond}}(P)$, we find
\begin{equation}\label{psioutsingle}
\phi_{{\rm cond}}(Q)=Q^ne^{-Q^2/2}.
\end{equation}
This function has two peaks at $Q=\pm\sqrt n$ and vanishes at $Q=0$ and $Q=\pm\infty$. The wave function of the coherent state $\ket\alpha$, on the other hand, has a single maximum at $\alpha\sqrt 2$. Thus the wavefunction $\phi_{{\rm cat}(\alpha)}(Q)$ of the even Schr\"odinger cat momentum-squeezed by a factor $S$ has two peaks located at $Q=\pm S\alpha\sqrt 2$. Good matching between $\phi_{{\rm cond}}(Q)$ and $\phi_{{\rm cat}(\alpha)}(Q)$ obtains when their peaks have the same position and the same width, which happens when $\alpha=\sqrt n,\ S=1/\sqrt 2$. Amazingly, the fidelity of this matching \emph{increases} with $n$, reaching the value of 99 \% already at $n=3$.

Experimental implementation of this protocol with $n=2$ is almost identical to tomography of the two-photon Fock state (Ourjoumtsev \ieac 2006b), except that two homodyne detectors are required. Reconstruction of the output state bears close resemblance to the squeezed Schr\"odinger cat [Fig.~\ref{bigcatfig}(b)], with differences mainly caused by technical limitations, which are similar to those in previous experiments by this group.



\subsection{Photon-added states}\label{PAsec}
\emph{Photon-added states} (Agarwal and Tara, 1991) are generated when the photon creation operator acts on an arbitrary state $\ket{\psi}$ of light: $\ket{\psi,m}=(\hat a^\dagger)^m\ket{\psi}$. These states are nonclassical due to a vanishing probability of finding $n<m$ photons (Lee, 1995). Recently, two important photon-added states were experimentally generated and characterized via OHT.
\subsubsection{Single-photon-added coherent states}
These states are of interest because in the limit of large $\alpha$, they approximate highly classical coherent states $\ket{\alpha}$ while for $\alpha\to 0$  they become highly nonclassical Fock states $\ket{m}$. Therefore, photon-added coherent states can be interpreted as a link between the particle and wave aspect of the electromagnetic field.

\begin{figure}
\includegraphics[keepaspectratio,width=0.45\textwidth]{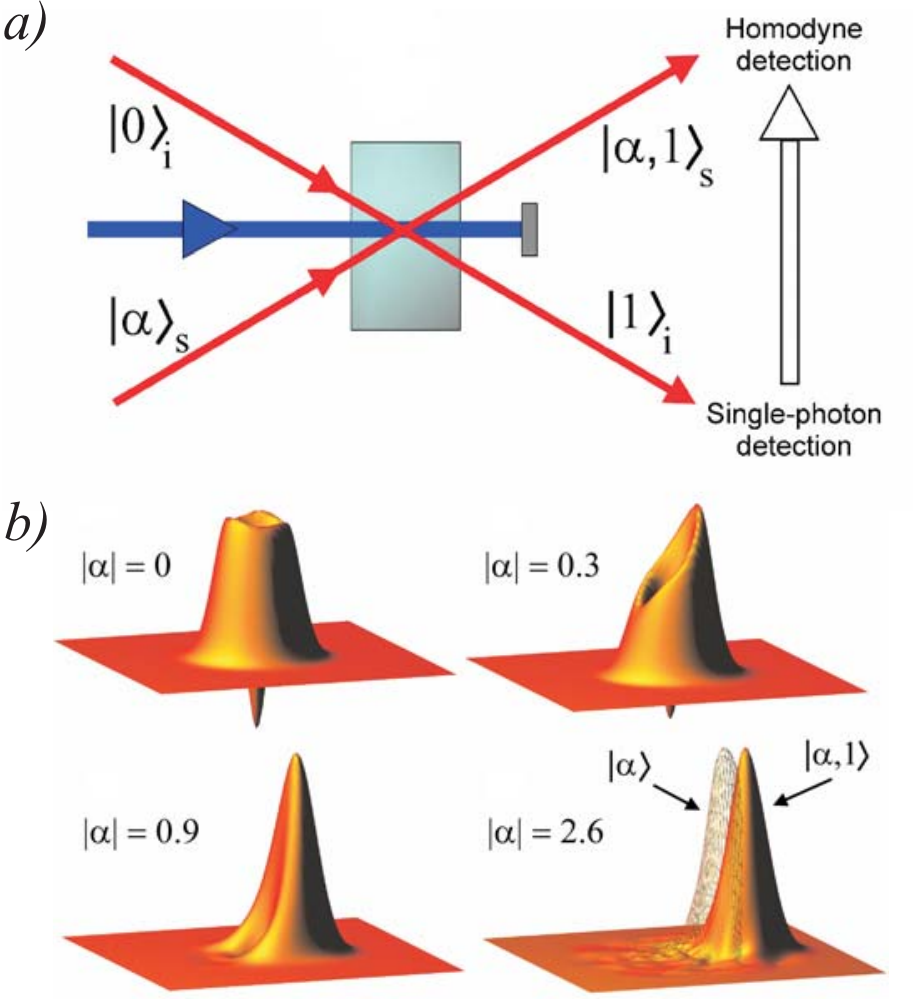}
\vspace*{8pt}
\caption{\label{bellini}(a) Conditional preparation of the single-photon-added coherent state. (b) With increasing $\alpha$, the Wigner function of the reconstructed SPACS gradually evolves from a highly nonclassical to a highly classical shape. For $|\alpha|=2.6$, also the Wigner function of the unexcited seed coherent state is shown. Reproduced with permission from Zavatta \ieac (2004a).}
\end{figure}

Experimentally, photon addition can be implemented using a procedure opposite to photon subtraction described in the previous section. Instead of passing through a beam splitter, the target state is transmitted through a signal channel of a parametric down-conversion setup [Fig.~\ref{bellini}(a)]. If a photon pair is generated in the down-converter, a photon is added to the target state. This event, which is heralded by a single photon emerging in the trigger channel, can be followed by an OHT measurement of the signal ensemble.

This scheme was implemented, for the first time, by Zavatta {\it et al.} (2004a, 2005a, 2005b). Thanks to high-bandwidth time-domain homodyne detection (Zavatta \ieac 2002), no pulse-picking was necessary so the setup could be made highly compact and phase stable. Fig.~\ref{bellini}(b) demonstrates how increasing amplitude of the input coherent state results in gradual transition from the Fock state to an approximation of a coherent state. An interesting feature observed in SPACS of moderate amplitudes is quadrature squeezing (up to 15\%) associated with certain phases.

\subsubsection{Single-photon-added thermal states}
The thermal state is a phase-symmetric ensemble with Bose-Einstein photon number statistics. By itself, it is a classical state, but shows a high degree of nonclassicality when acted upon by the photon creation operator. This was shown experimentally by Zavatta \iea (2007). The thermal state was simulated by sending a coherent laser beam through a rotating ground glass disk and collecting a fraction of the scattered light with a single-mode fiber. It was then subjected to photon addition as described above, and subsequently to homodyne tomography. The measured state is verified to be highly nonclassical according to several criteria.

In a related work, Parigi \iea (2007) apply a sequence of photon addition and subtraction operators to the thermal state. They find, contrary to classical intuition, but in full agreement with quantum physics, that the effects of these operators do not cancel each other, and, furthermore, depend on the sequence in which they are applied. This provides a direct evidence of noncommutativity of these operators, which is one of the cardinal concepts of quantum mechanics.

\section{Spatial quantum-state tomography} \label{spatialtomo}

\subsection{Spatial mode of the one-photon field}\label{photonwfsec}
As we discussed, homodyne tomography typically is used to measure the quantum state of light occupying a single selected optical mode, which is defined by the local oscillator pulse. If our goal is to characterize the field state in multiple modes, homodyne tomography becomes increasingly difficult.

There is a special multi-mode situation, on the other hand, that is amenable to full characterization: If it is known \emph{a priori} that only one photon (elementary excitation) of the field exists in a certain space-time volume, it is sensible to ask what is the temporal-spatial wave-packet mode that describes this photon. This task is close to that of finding the \emph{wave function} of the photon treated as a massless particle. This notion is known to be controversial (see Smith and Raymer, 2007, for a review). However, if one restricts attention to the photon's transverse degrees of freedom in the paraxial approximation, the subtleties that arise can be circumvented.

Assuming a constant polarization, a single-photon state of the quantized field can be represented by a superposition [{\it cf.} Eq.~(\ref{adag})]
\begin{eqnarray}
\ket{1_{\hat a}}  = \hat a^\dag  \left| {{\rm{vac}}} \right\rangle  &=& \int {d^3 kC(\vec k )} \widehat b^\dag_{\vec k} \left| {{\rm{vac}}} \right\rangle
\\ \nonumber &=& \int {d^3 kC(\vec k )} \ket{1_{\vec k}} ,
\end{eqnarray}
where
\begin{equation}
\ket{1_{\vec k}}\equiv\ket{1_{\vec k}}\otimes\prod\limits_{\vec k'\ne\vec k}\ket{{\rm vac}_{\vec k'}}
\end{equation}
is a one-photon state occupying a plane-wave mode with definite momentum $\vec p  = \hbar \vec k $. The function $C(\vec k )$ defines the spatial mode of the photon in the momentum representation.
In the position representation for free space propagation, the matrix element
\begin{eqnarray}\label{Ert}
 \vec E(\vec r ,t)&=&\left\langle {{\rm{vac}}} \right|\hat{\vec E }^{( + )} (\vec r,t)\ket{1_{\hat a}}  \\ \nonumber
  &=& i\vec\epsilon\int {d^3 k\sqrt {\frac{{\hbar \omega_{\vec k}}}{{(2\pi)^3\varepsilon _0 }}} } \,C(\vec k ) \,\exp( i\vec k \vec r- i\omega_{\vec k}t\,),
 \end{eqnarray}
defines the spatial distribution of the photon's field. The goal of spatial QST is to reconstruct the state by determining the function $\vec E(\vec r ,t)$.

Suppose a photon is created with a narrowly defined frequency $\omega _0$ and is propagating along the $z$ axis with the wavenumber $k_0=\omega_0/c$. In the paraxial approximation, $k_z\gg k_x,k_y$ and
\begin{equation}
k_z=\sqrt{k_0^2-k_x^2-k_y^2}\approx k_0-\frac{k_x^2}{2k_0}-\frac{k_x^2}{2k_0},
\end{equation}
so we can rewrite Eq.~(\ref{Ert}) as
\begin{equation}
\vec E (\vec r,t) =i\vec\epsilon \sqrt {\frac{{\hbar \omega (k)}}{{2\,\varepsilon _0 }}}\exp[ - i(\omega _0 \,t - k_0 z)\,]\;u (\vec r ).
\end{equation}
in which we define the spatial mode [with $\vec x \equiv (x,y), \vec k_x \equiv (k_x,k_y)$]
\begin{equation}\label{photonwf}
u(\vec r)= u(\vec x,z) \cong \int  C(\vec k_x )
\exp\left[i\vec k_x \vec x-i\frac{\vec k_x^2}{2k_0}z\right]\,{d^2k_x }.
\end{equation}

We compare the above expression with the Schr\"odinger evolution of a free particle of mass $m$ in two dimensions, initially in a superposition $\ket{\psi(0)}=\int {d\vec k_x } \,C(\vec k_x )\ket{k_x}$:
\begin{equation}\label{masswf}
\psi(\vec x,t)\cong\int C(\vec k_x )\exp\left[i\vec k_x\vec x-i\hbar\frac{\vec k_x^2}{2m}t\right]\,{d^2k_x },
\end{equation}
where $m$ is the mass of the particle. Expressions (\ref{photonwf}) and (\ref{masswf}) become equivalent if one sets $z=ct$ and $m=\hbar k_0/c$.

We can utilize this equivalence by applying the program set forth in Eqs.~(\ref{ipx})--(\ref{dmprobdens}) in the Introduction in order to determine the transverse wave function of a photon. We measure the beam intensity profile $I(\vec x,z)=\langle|E(\vec x,z)|^2\rangle$ in different planes along the beam propagation direction. The transverse degrees of freedom of the wave evolve during propagation, allowing inversion of measured intensity (probability) distributions using the propagator (\ref{massprop}), which now takes the form
\begin{equation}\label{photonprop}
G(\vec x' ,\vec x;t)\propto\exp\left(\frac{ik_0|\vec x-\vec x'|^2}{2z}\right),
\end{equation}
to determine the transverse wavefunction $u(\vec x,z)$ of the photon.

If the transverse state of the photon is not pure, it is defined by the density matrix
\begin{equation}\label{fielddm} \rho (\vec x_1 ,\vec x_2 )=\left\langle {E(\vec x_1 )E^*(\vec x_2 )} \right\rangle,\end{equation}
with the angle brackets implying an ensemble average over all statistical realizations of the photon wave function. Here we notice that the above definition is completely analogous to that of the classical field correlation function determining the degree of its spatiotemporal coherence\footnote{In Sec.~\ref{MMsec}, we discussed the identity between the mode of the conditionally prepared photon and the classical difference-frequency signal generated by the advanced wave. This identity is expressed in terms of definition (\ref{fielddm}) valid for both single photons and classical fields.}. Therefore, the tomography procedure we have developed for single photons is also applicable to classical fields, making them a useful ``testing ground" for single-photon QST procedures.

One can also introduce the transverse, two-dimensional spatial Wigner distribution at a particular plane in the fashion analogous to Eq.(\ref{wigdef}):
\begin{equation}W(\vec x,\vec k_x ) = \frac{1}{4\pi^2 }\int {\rho(\vec x + \frac{1}{2}\vec\xi,\vec x - \frac{1}{2}\vec\xi)}  \,e^{ - i\,\vec k_x\cdot\vec\xi} d\xi^2
\end{equation}
where $\vec k_x $ is the transverse-spatial wave-vector component. The transverse Wigner function is reconstructed from a set of beam intensity profiles using the inverse Radon transform. Such a phase-space-tomography scheme was proposed in Raymer {\it et al.} (1994) for quantum or classical waves and implemented for the transverse spatial mode of a ``classical" (coherent-state) light beam by McAlister {\it et al.} (1995).

\subsection{Non-interferometric reconstruction}
The non-interferometric method just described is best performed with an array detector to image the probability distributions at different propagation distances\footnote{Scanning a single detector would be prohibitive.}. In addition, for a reliable reconstruction, it is necessary to ensure that the beam waist (its region of minimum spatial extent) occurs well within the measured zone. If this is not the case, then only partial state reconstruction is possible. In the case of a limited scan, the density matrix in momentum representation can be measured everywhere except for a band around the diagonal, whose width decreases as a larger range of longitudinal distances (time-of-flight) is measured (Raymer, 1997b). This was the case in a demonstration of transverse spatial QST of an ensemble of helium atoms (Kurtsiefer \ieac 1997;
Janicke and Wilkens, 1995).

By using lenses (for light or for atoms), the waist region can be brought into range for imaging, thus ensuring a reliable reconstruction (Raymer \ieac 1994;
McAlister \ieac 1995).
Suppose a beam propagates through a lens at $z = d$, with focal length $f$, to a detection plane at $z = D$. The Maxwell-field propagator (\ref{photonprop}), in paraxial approximation, takes the form
\begin{equation}
G(\vec x' ,\vec x;t) = C\exp \left[ {i h(\vec x') + i{\kern 1pt} k_0 \left( {\frac{{\vec x\cdot \vec x}}{L} - \frac{{\vec|x|^2 }}{{2R_C }}} \right)} \right],
\end{equation}
where $L = (D - d)(1 + d/R_0 )$, $R_C  = R_0  + d$, and $R_0 ^{ - 1}  = (D - d)^{ - 1}  - f^{ - 1}$. Here $C$ is a constant and $h(x)$
 is an unimportant phase function.

\begin{figure}
\includegraphics[keepaspectratio,width=0.45\textwidth]{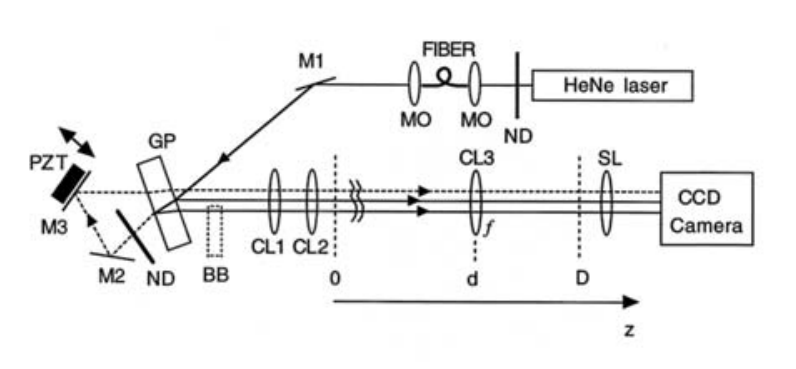}
\vspace*{8pt}
\caption{\label{fig41} Spatial tomography of the classical laser mode. Reflecting glass plate GP and cylindrical lenses CL1, CL2, create a one-dimensional field with two amplitude peaks in the $z = 0$
plane. Cylindrical lens CL3 (oriented $90^\circ$ from CL1 and CL2) is varied in position, and intensity profiles in the $z = D$
plane are imaged and recorded using spherical lens SL and the camera. Profiles are measured for 32 combinations of distances $d$
 and $D$. The piezoelectric transducer (PZT) introduces partial coherence between the two peaks. From McAlister \iea (1995).}
\end{figure}

\begin{figure}
\includegraphics[keepaspectratio,width=0.45\textwidth]{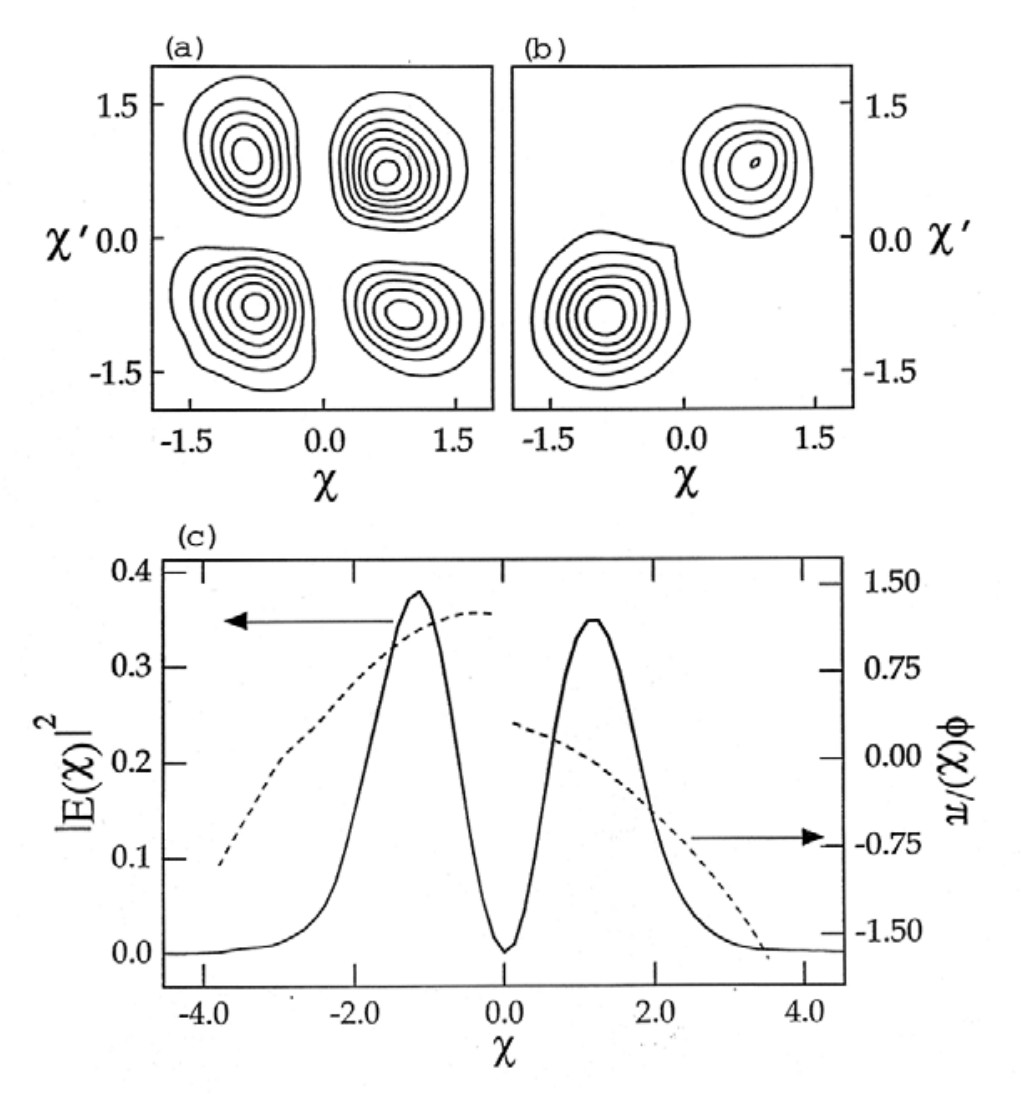}
\vspace*{8pt}
\caption{\label{fig42} Spatial tomography results for the classical laser mode. Equal-separation contours showing the magnitude-squared of the reconstructed field correlation function $\rho(\chi,\chi')$, for (a) the fully coherent field and (b) the partially coherent field.  (c) The intensity profile (solid curve) and phase profile (dashed curve) of the reconstructed complex-wave field obtained in the fully coherent case. Axes are scaled transverse position $\chi  \equiv x/x_0 $, where $x_0$ is a characteristic length. From McAlister \ieac (1995).\vspace{10cm}}
\end{figure}

The first measurements of this type were carried out for macroscopic (``classical") fields from a laser (McAlister \ieac 1995). Figures \ref{fig41} and \ref{fig42} show the setup and the reconstructed data in the object plane, for the case of a two-peaked field distribution created by reflecting the signal beam from a two-sided reflecting glass plate. The field correlation function $\left\langle {E(x_1 )E^*(x_2 )} \right\rangle $
 (which in this case should not interpreted as a quantum density matrix)
 was reconstructed using the method in Raymer {\it et al.} (1994) --- the inverse Radon transform of intensity distributions measured for different lens and detector position combinations.

For the data in Fig.~\ref{fig42}(a), the neutral-density filter ND was replaced by a beam block, so two coherent beams comprised the signal. Four lobes are seen in the reconstructed field correlation function. Figure \ref{fig42}(c) shows the reconstructed field and the phase profile. For the data in Fig.~\ref{fig42}(b), the beam block BB was inserted in the lower beam (as shown) and the second beam component was created by reflection from a mirror M3 mounted on a translator driven by a random voltage, so two mutually incoherent beams comprised the signal. In this case, the off-diagonal lobes are missing in the reconstructed field correlation function, as expected. This experiment verified the method of phase-space tomography for reconstructing spatial field correlations at the macroscopic level. Hansen (2000) applied a similar technique to reconstruct the optical mode emerging from a two-slit interferometer and obtained a Wigner function with negative values, similar to that of Kurtsiefer {\it et al.} (1997). The method has been applied to study light scattering from complex fluids (Anhut \ieac 2003).

Other methods for classical wave-front reconstruction have since been developed (Iaconis and Walmsley, 1996;
Cheng \ieac 2000; Lee \ieac 1999) and applied (Cheng and Raymer, 1999; Lee and Thomas, 2002;
Reil and Thomas, 2005). A method has been elaborated for characterizing single-photon states in terms of a discrete spatial basis (Sasada and Okamoto, 2003; Langford \ieac 2004) and a proposal has been made for generalizing this to arbitrary beams (Dragoman, 2004).

\subsection{Interferometric reconstruction by wave-front inversion}

A technique for continuous-spatial-variable characterization of
single-photon fields was proposed by Mukamel {\it et al.} (2003), and recently implemented by Smith {\it et. al.} (2005). The method uses a parity-inverting Sagnac interferometer to measure the expectation value of the parity operator $\hat\Pi$,\footnote{The parity operator's eigenstates are those with even and odd wavefunction, which correspond, respectively, to the eigenvalues $1$ and $-1$} which, as first shown by Royer (1977), is proportional to the Wigner distribution at the phase space origin:
\begin{equation}
W(0,0) = \frac{1}{\pi }\Tr\left[ {\hat\rho\,\hat\Pi } \right].
\end{equation}

\begin{figure}
\includegraphics[keepaspectratio,width=0.45\textwidth]{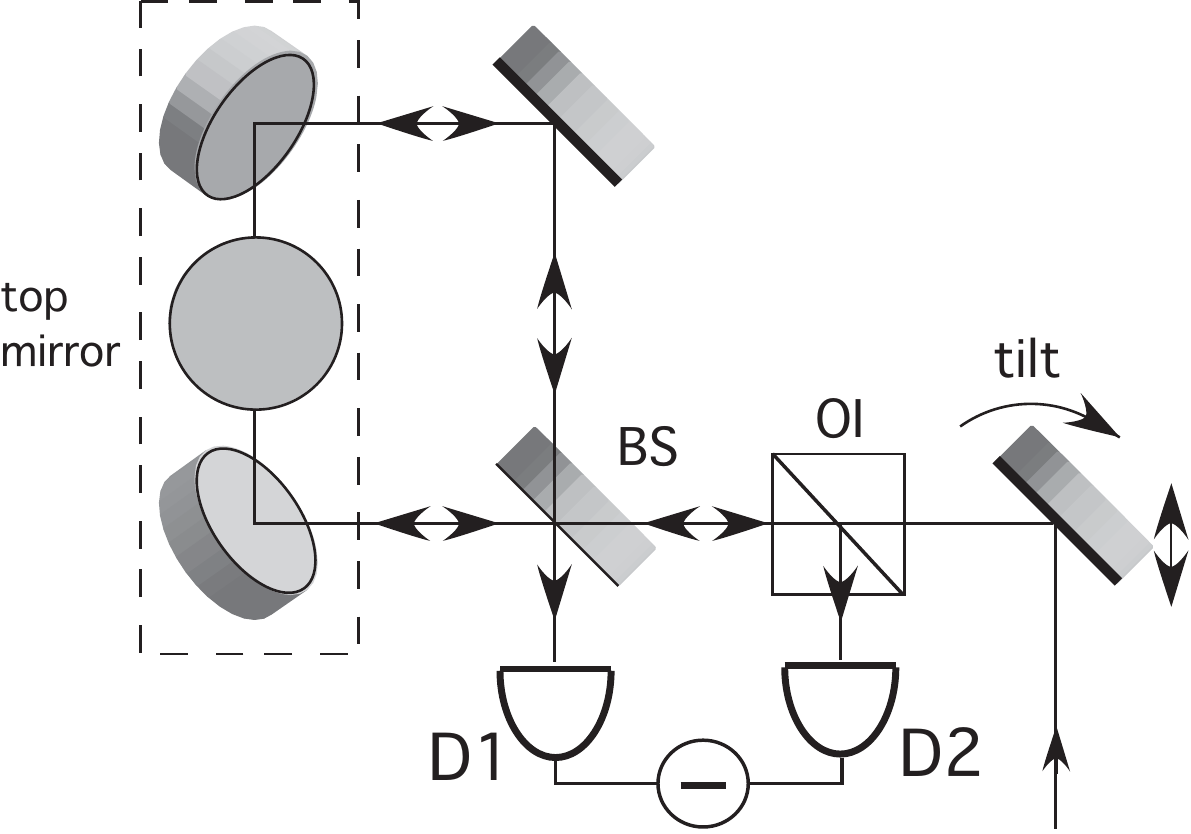}
\vspace*{8pt}
\caption{\label{fig43} Top view of all-reflecting Sagnac interferometer. All mirrors are planar, including beam splitter BS. The beam travels out-of plane to reach the center mirror in the top-mirror configuration (surrounded by dashed lines), which is above the plane of the others. The displacement and tilt of the external steering mirror selects the phase-space point at which the Wigner distribution is measured. Signals from photon-counting detectors D1 and D2
 are subtracted. OI is an optical isolator for directing the reflected signal to D2.}
\end{figure}

The Wigner function at an arbitrary phase-space point can be determined by measuring the parity expectation value of the mode after the latter is displaced in the phase space in a manner similar to that proposed by Banaszek (1999) and discussed in the end of Sec.~\ref{maxlikerror} for Wigner functions in the field quadrature space:
\begin{equation} \label{spacWigdisp}
W(\vec x,\vec k_x ) = \frac{1}{\pi }\Tr\left[ {\hat D^{-1}(\vec x,\vec k_x )\hat\rho\hat D(\vec x,\vec k_x )\,\hat\Pi } \right].
\end{equation}
Experimentally, the displacement $\hat D$ is implemented by physically shifting the mode location by $\vec x$ and tilting its propagation direction by $\vec k_x$ (see Fig.~\ref{fig43}).

The mode parity is measured as follows. One decomposes the signal field into a sum of even and odd terms, $E(\vec x) = E_e (\vec x) + E_o (\vec x)$. Then the Wigner distribution (\ref{spacWigdisp}) evaluates to
\begin{equation}\label{wigoddeven}
W(\vec x,\vec k_x ) = \frac{1}{\pi }\int \left[ {\left\langle {\left| {E_e (\vec x')} \right|^2 } \right\rangle  - \left\langle {\left| {E_o (\vec x')} \right|^2 } \right\rangle } \right]_{\vec{x},\vec{k}_x}\,d^2x',
\end{equation}
the terms in angular brackets being the experimentally measurable mean intensities or photon count rates for a given shift and tilt $(\vec x,\vec k_x)$. This measurement is achieved by means of a dove-prism (Mukamel \ieac 2003) or an
all-reflecting (Smith \ieac 2005) Sagnac interferometer as shown in Fig.~\ref{fig43}. The beam is split at beam splitter BS, after which the two beams travel in different directions around the Sagnac loop. Each beam travels out-of plane to reach the center mirror in the top-mirror configuration, which has the effect of rotating the wave fronts by $\pm 90^\circ$, depending on direction, in the $x$-$y$ plane. The net result is the interference of the original field with its (two-dimensional) parity-inverted image.  Any odd-parity beam [$E( - x, - y) = - E(x,y)$] passes through to detector D1, while any even-parity beam [$E( - x, - y) =  E(x,y)$] reflects back toward the source, and is detected by D2. By subtracting the average count rates integrated over detector faces large enough to capture all signal light, one measures the Wigner distribution at a point in phase space, according to Eq.~(\ref{wigoddeven}). One can also use only one detector, in which case the average counting rate, as a function of $x$ and $k_x$, is proportional to the Wigner distribution plus a constant, which must be subtracted.

In order to apply this technique in the photon-counting regime, one would like to use high quantum-efficiency avalanche photodiodes (APDs) operating in Geiger mode. Unfortunately, these typically have very small detector area (0.1 mm diameter), making them unsuitable for detecting beams with large intrinsic divergence. The experiment by Smith {\it et al.} (2005) used a single large-area photon-counting detector D1. The detector was a photomultiplier tube with 5 mm diameter, 11\% efficient at wavelength 633 nm.

\begin{figure}
\includegraphics[keepaspectratio,width=0.45\textwidth]{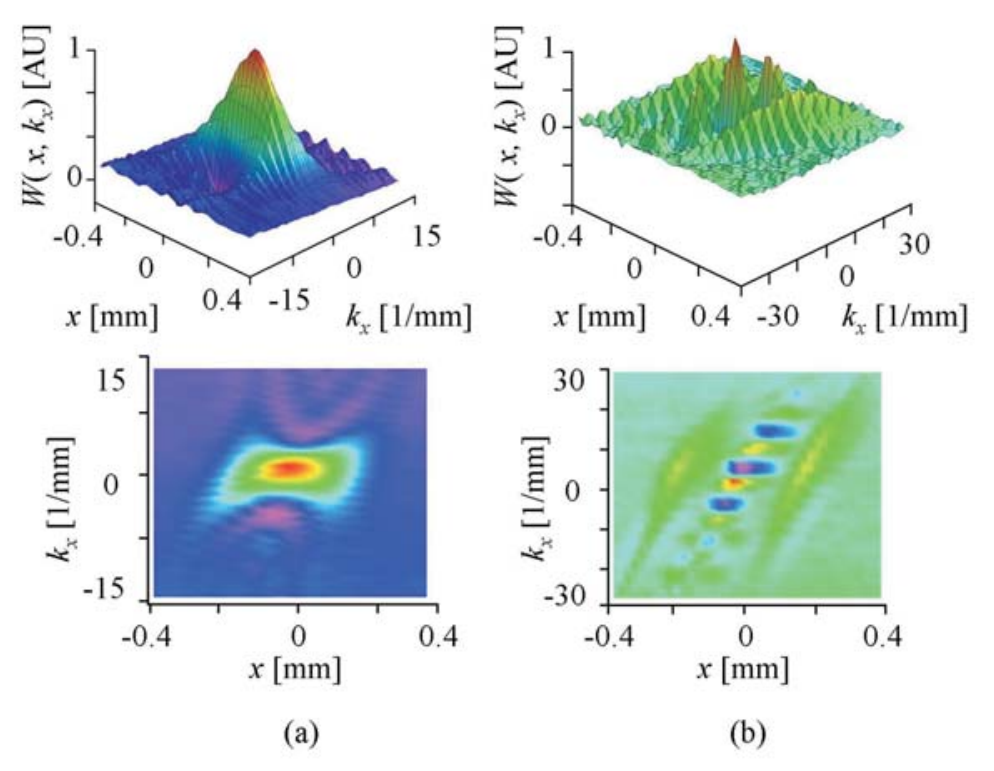}
\vspace*{8pt}
\caption{\label{fig44} Measured Wigner functions for a slightly diverging beam passed through a single slit (left column) or double slit (right column). Each shows a shear associated with beam divergence. In both cases, the interference fringes oscillate positive and negative, as expected for a nonclassical momentum state. In the case of two slits, the fringes can be understood as resulting from a superposition (``Schr\"odinger cat"-like) state of two well-separated components. From Smith \iea (2005).}
\end{figure}

Figure \ref{fig44} shows results obtained for an expanded
laser beam after passing through a single-slit or double-slit aperture
placed in the beam just before the steering mirror. The beam was attenuated so that only a single photon was typically present at any given time. Again, these results are similar to those obtained by Kurtsiefer {\it et al.} (1997) for a beam of helium atoms. Note that although the Wigner functions shown in Fig.~\ref{fig44} are the ``Wigner functions of the single-photon Fock state", they represent quantum objects fundamentally different from that plotted in Fig.~\ref{focktomo}. The latter describes the quantum state of a specific electromagnetic oscillator while the former describes the superposition of electromagnetic oscillators carrying a specific optical state.

The ability to measure quantum states or wave functions for ensembles of single-photon states can be
generalized to two-photon states. As pointed out in Mukamel {\it et al.} (2003) and Smith {\it et al.} (2005), if a photon pair in a position-entangled state
$\hat\rho _{AB}$
 can be separated, then each can be sent into a separate Sagnac interferometer, and subsequently detected. The rate of coincidence counts is proportional to a sum of terms, one of which is the two-photon Wigner distribution,
\begin{equation}
W(\vec x_A ,\vec k_{xA} ,\vec x_B ,\vec k_{xB} ) = \frac{1}{\pi ^2}\Tr[ D^{-1} \hat\rho _{AB} D\,\hat\Pi_A\hat\Pi_B ],
\end{equation}
[where $D\equiv D(\vec x_A,\vec k_{xA},\vec x_B,\vec k_{xB} )$] which can be extracted from the counting data.

If the state is pure, this Wigner function can be transformed into the ``two-photon wave function", defined by (Smith and Raymer, 2007)\footnote{A related quantity, the two-photon coincidence-detection amplitude, has been discussed in Nogueira \iea (2002),
Walborn \iea (2004), Keller and Rubin (1997), and
Scully and Zubairy (1997).}
\begin{equation}
\Psi (x_A ,x_B ) = \left\langle {{\rm{vac}}} \right|\hat E^{( + )}
(x_A )\,\hat E^{( + )} (x_B )\left| {\Psi _{AB} } \right\rangle.
\end{equation}
 Measuring the two-photon wave function would provide a complete characterization of position-entangled states, which are of interest in the context of Einstein-Podolsky-Rosen correlations or Bell's inequalities with photon position and momentum variables (Howell \ieac 2004; Yarnall \ieac 2007).

\section{Summary and outlook}
Prior to the beginning of the present century, quantum physics of light has been developing along two parallel avenues: ``discrete-variable" and ``continuous-variable" quantum optics. The continuous-variable community dealt primarily with the wave aspect of the electromagnetic field, studying quantum field noise, squeezing and quadrature entanglement. Homodyne detection was the primary tool for field characterization. The discrete-variable side of quantum optics concentrated on the particle aspect of light: single photons, dual-rail qubits, and polarization entangled states. These objects were usually measured with single-photon detectors.

These two aspects of quantum optics had little overlap with each other in terms of methodology, but experienced significant mutual influence. Novel results in the discrete-variable domain, such as demonstration of entanglement, quantum tomography, quantum teleportation, etc., were frequently followed by their continuous-variable analogs and vice versa.

Theoretically, the difference between these two domains boils down to the choice of the basis in which states of an optical oscillator are represented: either quadrature (position or momentum) or energy eigenstates. From the experimental point of view, parametric down-conversion, the workhorse of quantum optical state production, can operate in either the weak or strong pumping modes. In the former case, we obtain discrete photon pairs, in the second, squeezing or quadrature entanglement.

The division of quantum optics is thus caused not by fundamental but by pragmatic reasons. It is just that, until recently, our technology allowed us to generate only two classes of quantum states, giving us access to two small islands in the vast ocean of the optical Hilbert space.

Developments of the last decade allowed us to overcome this separation. By applying a traditionally continuous-variable quantum characterization method (homodyne tomography) to discrete-variable quantum states (photons and qubits), researchers have constructed the bridge between the two islands, and then extended it by engineering and characterizing quantum states that belong to neither domain --- such as displaced and photon-added states, squeezed Fock states and Schr\"odinger cats.

In this review, we covered technological developments that led quantum optics to this breakthrough, placing a particular accent on continuous-wave tomography. We discussed new state-reconstruction algorithms, the technology of time-domain homodyne detection,  preparation of high-purity photons and qubits, and methods of quantum state engineering. We also reviewed methods of characterizing the modal structure of a quantum-optical state.

Extrapolating the last years' results into the future, we can isolate certain open problems and future directions along which the field can be expected to develop.

\emph{Reliable state-reconstruction algorithms.} We have focused on maximum likelihood estimation (MaxLik) because it is straightforward to implement and offers improvements over the inverse-linear-transform techniques such as inverse Radon. However, MaxLik is probably not the last word in QST algorithms. It can underperform if only a small amount of data is available. In some cases this technique can yield zero probabilities for certain state components, which are not justified (Blume-Kohout, 2006). We predict that future reconstruction algorithms will combine maximum-likelihood with maximum-entropy and Bayesian methods (Fuchs and Schack, 2004). One attempt at such integration has already been reported (\v{R}eh\'{a}\v{c}ek and Hradil, 2004).

Within MaxLik itself, particularly in application to OHT, there are a number of open questions. To what extent does a bias in the tomography scheme (non-unity sum of the POVM elements, see Sec.~\ref{maxlikerror}) influence the reliability of state reconstruction? What is the optimal point for truncating the Hilbert space that would allow sufficiently complete but noise-free reconstruction? Does there exist a simple and reliable method for evaluating errors in quantum state estimation?

\emph{Faster, low-noise homodyne detectors.} As discussed in Sec.~\ref{tdhd}, there is a compromise between the bandwidth of the homodyne detector and its signal-to-noise ratio. Detectors with higher bandwidths can accommodate higher laser repetition rates, permitting acquisition of larger data sets and eventually analysis of more complex states of light. Future study in this area is well-deserved, also given applications of faster homodyne detectors is continuous-variable quantum cryptography, with a promise of significant secret key transfer rate enhancement.

\emph{Applications of OHT in discrete quantum-information processing.} Most of the optical protocols tested so far employed dual-rail qubits as quantum information carriers (Kok \ieac 2007). Accordingly, photon counting has been the method of choice for state measurement. As discussed in the Introduction, homodyne tomography provides much more complete information about a state of light (and thus performance of a quantum gate), but it is not yet commonly employed due to the relative complexity of its implementation. A goal for future research would be to simplify basic elements of homodyne detection --- mode matching, local oscillator phase variation, data acquisition --- to an extent that would make OHT not much more complicated than photon counting. Homodyne tomography should then be applied for characterizing complex discrete-variable quantum states and protocols. Perhaps one of the first steps would be characterization of an entangled state of two dual-rail qubits --- akin to that in James \iea (2001), but using OHT.

Another important QI-related application is testing protocols on interfacing quantum information between light and stationary media. Examples of such are the experiments of Julsgaard \iea (2004) and Appel \iea (2008), which utilized homodyne tomography to study quantum memory for light. Homodyne measurements on light transmitted through an atomic ensemble permit tomographic reconstruction of its collective spin state (Feenholz \ieac 2008), which is useful, for example, for characterizing quantum tomography from light onto atoms (Sherson \ieac 2006).

\emph{Continuous-variable process tomography.} While measuring superoperators associated with a certain quantum process has been investigated theoretically (Chuang and Nielsen, 1997) and experimentally (Altepeter {\it et al.}, 2003) for discrete variables for quite some time, the progress in the continuous-variable domain has been non-existent. This seems to be an important open problem, whose solution holds a promise to provide much more complete data about quantum processes than current methods.

\emph{Quantum-state engineering,} i.e.~synthesis of arbitrary quantum states of light using nonclassical primitives (squeezed or Fock states), linear optics and conditional measurements. There exist a number of proposals for tackling this objective (reviewed in detail in Dell'Anno \ieac 2006), for example by using coherent displacements and photon subtraction operations (Dakna \ieac 1999a,b; Fiur\'{a}\v{s}ek \ieac 2005), repeated parametric down-conversion (Clausen \ieac 2001) and continuous-variable postselection (Lance \ieac 2006). To date, we have mastered quantum state engineering at the single-photon level: we can create any linear combination of the vacuum and single-photon Fock state. The next step is to bring this to the two-photon level. This can be done, for example, by applying modified photon addition operations (Sec.~\ref{PAsec}) to single-rail qubits.

Quantum optical engineering, as well as any other complex manipulation of light, requires high quality of the ``raw material'', i.e. initial squeezed and Fock states. Here we can see two possibilities for progress. On one hand, parametric down-conversion sources need to be improved to generate spectrally and spatially unentangled signal and idler photons, as well as pulsed squeezing in a single spectral mode (Wasilewski \ieac 2006). On the other hand, it would be great to eliminate down-conversion altogether and employ solid-state, on-demand sources (Grangier \ieac 2004). At present, such sources compromise between efficiency and spatiotemporal purity, and thus cannot be employed in scalable quantum-optical engineering. We hope that the situation will change in the near future. Additionally, there may exist a possibility for improving the efficiency of such sources by means of linear optics and conditional measurements (Berry \ieac 2006, 2007).

\emph{Fundamental tests and new quantum protocols} that are not restricted by either discrete- or continuous domains of quantum optics. Examples are loophole-free nonlocality tests (Garc\'{\i}a-Patr\'{o}n \ieac 2004; Nha and Carmichael, 2004) and purification of continuous-variable entanglement (Opatrn\'{y} \ieac 2000; Browne \ieac 2003). All ``building blocks'' of these protocols have already been experimentally demonstrated, but a task to put them together in operational setups remains on the agenda.

In summary, more work is needed before we gain full control over the optical Hilbert space. It is however worth the effort: if we have seen so many wonders within the boundaries of the two small islands colonized so far, who can predict what surprises await us in the vast expanses of the whole ocean?

\section*{Acknowledgements}
We acknowledge numerous essential contributions of our collaborators listed in the references. The work of A. L. is supported by NSERC, CFI, CIAR, QuantumWorks and AIF. The work of M. R. was supported by NSF. We thank Gina Howard for assistance in preparing the manuscript.


\begin{thebibliography}{}


\bibitem[\mbox{}]{MMpaper} Aichele, T., A. I. Lvovsky and S. Schiller, 2002, Eur. Phys. J. D {\bf 18}, 237.
\bibitem[\mbox{}]{agarwal91} Agarwal, G. S., and K. Tara, 1991, Phys. Rev. A {\bf 43}, 492.
\bibitem[\mbox{}]{alt03} Altepeter, J. B., D. Branning, E. Jeffrey, T. C. Wei, P. G. Kwiat, R. T. Thew, J. L. O'Brien, M. A. Nielsen, and A. G. White, 2003, Phys. Rev. Lett. {\bf 90}, 193601.
\bibitem[\mbox{}]{alt04} Altepeter J. B., D. F. V. James, and P. G. Kwiat, 2004, in {\it Quantum State Estimation, Lect. Notes Phys.}, edited by M. Paris and J. \v{R}eh\'{a}\v{c}ek (Springer, Berlin Heidelberg), Vol. 649, p. 113.
\bibitem[\mbox{}]{anh03} Anhut T., B. Karamata, T. Lasser, M. G. Raymer and L. Wenke, 2003, in {\it Coherence Domain Optical Methods and Optical Coherence Tomography in Biomedicine VII}, edited by V. V. Tuchin, J. A. Izatt, and J. G. Fujimoto, Proceedings of SPIE, Vol. 4956, p. 120.
\bibitem[\mbox{}]{app07} Appel, J., D. Hoffman, E. Figueroa, A. I. Lvovsky, 2007, Phys. Rev. A {\bf 75}, 035802.
\bibitem[\mbox{}]{app08} Appel, J., E. Figueroa, D. Korystov, A. I. Lvovsky, 2008, Phys. Rev. Lett. (to be published).
\bibitem[\mbox{}]{art05} Artiles, L., Gill, R. D. and Gu\c t\u a, M. I., 2005, J. R. Statist. Soc B {\bf 67}, 109.
\bibitem[\mbox{}]{QSpaper}  Babichev, S. A., J. Ries and A. I. Lvovsky, 2003, Europhys. Lett. {\bf 64}, 1.
\bibitem[\mbox{}]{RSP}  Babichev, S. A., B. Brezger, and A. I. Lvovsky, 2004, Phys. Rev. Lett. {\bf 92}, 047903.
\bibitem[\mbox{}]{qubittomo}  Babichev S. A., J. Appel, and A. I. Lvovsky, 2004, Phys. Rev. Lett. {\bf 92}, 193601.
\bibitem[\mbox{}]{bachorsbook} Bachor, H.-A., T. C. Ralph, 2004, {\it A Guide to Experiments in Quantum Optics} (Wiley-VCH, Weinheim).
\bibitem[\mbox{}]{B1} Banaszek, K., 1998, Phys. Rev. A {\bf 57}, 5013.
\bibitem[\mbox{}]{B2} Banaszek, K., 1998, Acta Phys. Slov. {\bf 48}, 185.
\bibitem[\mbox{}]{B3} Banaszek, K., 1999, Phys. Rev. A {\bf 59} 4797.
\bibitem[\mbox{}]{NLPhoton2}  Banaszek, K., and K. Wodkiewicz, 1999, Phys. Rev. Lett. {\bf 82}, 2009.
\bibitem[\mbox{}]{B5} Banaszek, K., G. M. D'Ariano, M. G. A. Paris, and M. F. Sacchi, 1999, Phys. Rev. A {\bf 61}, 010304.
\bibitem[\mbox{}]{ban70} Band, W., and J. L. Park, 1970, Found. Phys., {\bf 1}, 133.
\bibitem[\mbox{}]{ban71} Band, W., and J. L. Park, 1971, Found. Phys., {\bf 1}, 339.
\bibitem[\mbox{}]{ban79} Band, W., and J. L. Park, 1979, Am. J. Phys., {\bf 47}, 188.
\bibitem[\mbox{}]{bec93} Beck, M., D. T. Smithey, and M. G. Raymer, 1993, Phys. Rev. A {\bf 48}, 890.
\bibitem[\mbox{}]{Bennett}  Bennett, C. H., G. Brassard, C. Cr\'{e}peau, R. Jozsa, A. Peres, and W. K. Wootters, 1993, Phys. Rev. Lett. {\bf 70}, 1895.
\bibitem[\mbox{}]{ben84}  Bennett, C. H., and G. Brassard, 1984, in {\it Proceedings of IEEE International Conference on Computers, Systems and Signal Processing, Bangalore}, (IEEE, New York), p. 175.
\bibitem[\mbox{}]{ber06} Berry, D. W., A. I. Lvovsky, B. C. Sanders, 2007, J. Opt. Soc. Am. B {\bf 24}, 189.
\bibitem[\mbox{}]{ber07} Berry, D. W., A. I. Lvovsky, B. C. Sanders, 2007, Opt. Lett. {\bf 31}, 107 (2006)
\bibitem[\mbox{}]{ber87}  Bertrand, J., and P. Bertrand, 1987, Found. Phys. {\bf 17}, 397.
\bibitem[\mbox{}]{bir96}   Bialynicki-Birula, I., 1996, in \emph{Progress in Optics XXXVI}, edited by E. Wolf, (Elsevier, Amsterdam).
\bibitem[\mbox{}]{bla01} Blansett, E. L., M. G. Raymer, G. Khitrova, H. M. Gibbs, D. K. Serkland, A. A. Allerman, and K. M. Geib, 2001, Opt. Express {\bf 9}, 312.
\bibitem[\mbox{}]{bla05} Blansett, E. L., M. G. Raymer, G. Cui, G. Khitrova, H. M. Gibbs, D. K. Serkland, A. A. Allerman, and K. M. Geib, 2005, IEEE J. Quant. Electron. {\bf 41}, 287.
\bibitem[\mbox{}]{blu06}Blume-Kohout, R., 2006, quant-ph/0611080
\bibitem[\mbox{}]{boh58} Bohr, N., 1958, from ``Quantum Physics and Philosophy", reprinted in \emph{Niels Bohr Collected Works, Foundations of Quantum Physics II}, Vol. 7, edited by J. Kalckar (Elsevier, Amsterdam, 1996).
\bibitem[\mbox{}]{EPRexp3}  Bowen, W. P., R. Schnabel, P. K. Lam, and T. C. Ralph, 2003, Phys. Rev. Lett. {\bf 90}, 043601.
\bibitem[\mbox{}]{HomoTomoSingle} Breitenbach, G., S. Schiller, and J. Mlynek, 1997, Nature (London) {\bf 387}, 471.
\bibitem[\mbox{}]{EPRdistill}  Browne, D.E., J. Eisert, S. Scheel and M.B. Plenio, 2003, Phys. Rev. A {\bf 67}, 062320.
\bibitem[\mbox{}]{but05} Butucea, C., Gu\c t\u a, M. I. and Artiles, L., 2005, math/0504058
\bibitem[\mbox{}]{maxent} Bu\v{z}ek, V., and G. Drobny, 2000, J. Mod. Opt. {\bf 47}, 2823.
\bibitem[\mbox{}]{buz95} Bu\v{z}ek, V., and P. Knight, 1995, in {\it Progress in Optics XXXIV}, edited by E. Wolf (North-Holland, Amsterdam)
\bibitem[\mbox{}]{che00} Cheng, C.-C., M. G. Raymer, and H. Heier, 2000, J. Mod. Opt., {\bf 47}, 1237.
\bibitem[\mbox{}]{che99}  Cheng, C.-C., and M. G. Raymer, 1999, Phys. Rev. Lett. {\bf 82}, 4807.
\bibitem[\mbox{}]{cho05} Chou, C. W., H. de Riedmatten, D. Felinto, S. V. Polyakov, S. J. van Enk and H. J. Kimble, 2005, Nature (London) {\bf 438}, 828
\bibitem[\mbox{}]{chu97} Chuang, I. L. and Nielsen, M. A., 1997, J. Modern Optics, {\bf 44}, 732.
\bibitem[\mbox{}]{welscheng} Clausen, J.,  H. Hansen, L. Knoll, J. Mlynek, and D.-G. Welsch, 2001, Appl. Phys. B {\bf 72}, 43.
\bibitem[\mbox{}]{coc99} Cochrane, P. T., G. J. Milburn, and W. J. Munro, 1999, Phys. Rev. A {\bf 59}, 2631
\bibitem[\mbox{}]{Cramer} Cram\'{e}r, H., 1946 {\it Mathematical methods of statistics} (Princeton University Press).
\bibitem[\mbox{}]{photonsub} Dakna, M., T. Anhut, T. Opatrn\'{y}, L. Kn\"{o}ll, and D.-G. Welsch, 1997, Phys. Rev. A {\bf 55}, 3184.
\bibitem[\mbox{}]{dak99a} Dakna, M., J. Clausen, L. Knöll, D.-G.Welsch, 1999a, Phys. Rev. A {\bf 59} 1658.
\bibitem[\mbox{}]{dak99b} Dakna, M., J. Clausen, L. Knöll, D.-G.Welsch, 1999b, Phys. Rev. A {\bf 60} 726.
\bibitem[\mbox{}]{dar97}  D'Ariano, G. M., 1997, in {\it Quantum Optics and the Spectroscopy of Solids}, edited by T. Hakioglu and A. S. Shumovsky (Kluwer, Dordrecht), p. 139.
\bibitem[\mbox{}]{dar94} D'Ariano, G. M.,C. Macchiavello, and M. G. A. Paris, 1994, Phys. Rev. A {\bf 50}, 4298.
\bibitem[\mbox{}]{dar96}  D'Ariano, G. M.,  and H. P. Yuen, 1996, Phys. Rev. Lett. {\bf 76}, 2832.
\bibitem[\mbox{}]{dar00}  D'Ariano, G. M., M. F. Sacchi, and P. Kumar, 2000, Phys. Rev. A {\bf 60}, 013806.
\bibitem[\mbox{}]{dar04}  D'Ariano, G. M., M. G. A. Paris, and M. F. Sacchi, 2004, in {\it Quantum State Estimation, Lect. Notes Phys.}, edited by M. Paris and J. \v{R}eh\'{a}\v{c}ek (Springer, Berlin Heidelberg), Vol. 649, p. 297.
\bibitem[\mbox{}]{del06} Dell'Anno, F., S. De Siena, F. Illuminati, 2006, Phys. Rep. {\bf 428}, 53.
\bibitem[\mbox{}]{dem77} Dempster, A. P., Laird, N. M., and Rubin, 1977, D. B., J. R. Statist. Soc. B {\bf 39}, 1.
\bibitem[\mbox{}]{vogelnc2} Di\'osi, L., 2000, Phys. Rev. Lett. {\bf 85}, 2841.
\bibitem[\mbox{}]{dor03} Dorrer, C., D. C. Kilper, H. R. Stuart, G. Raybon, and M. G. Raymer, 2003, IEEE Photon. Tech. Lett. {\bf 15}, 1746.
\bibitem[\mbox{}]{dra04} Dragoman, D., 2004, Appl. Opt., {\bf 43}, 4208.
\bibitem[\mbox{}]{dun95} Dunn, T. J., I. A. Walmsley, and S. Mukamel, 1995, Phys. Rev. Lett. {\bf 74}, 884.
\bibitem[\mbox{}]{fan57} Fano U., 1957, Rev. Mod. Phys. {\bf 29}, 74.
\bibitem[\mbox{}]{fer08} Fernholz, T., H. Krauter, K. Jensen, J. F. Sherson, A. S. Sorensen, and E. S. Polzik, ``Spin Squeezing of Atomic Ensembles via Nuclear-Electronic Spin Entanglement'', preprint arXiv:0802.2876.
\bibitem[\mbox{}]{fiu05} Fiur\'{a}\v{s}ek, J., R. Garc\'{\i}a-Patr\'{\i}on, and N. J. Cerf, 2005, Phys. Rev. A {\bf 72}, 033822.
\bibitem[\mbox{}]{fuc02} Fuchs C. A., 2002, ``Quantum Mechanics as Quantum Information (and only a little more)", preprint quant-ph/0205039
\bibitem[\mbox{}]{fuc04} Fuchs, C. A. and R. Schack, 2004, in {\it Quantum State Estimation, Lect. Notes Phys.}, edited by M. Paris and J. \v{R}eh\'{a}\v{c}ek (Springer, Berlin Heidelberg), Vol. 649, p. 113.
\bibitem[\mbox{}]{FunkThesis} Funk, A., 2004, Ph.D. thesis (University of Oregon).
\bibitem[\mbox{}]{gar04} Garc\'ia-Patr\'on, R., J. Fiur\'{a}\v{s}ek, N. J. Cerf, J. Wenger, R. Tualle-Brouri, and P. Grangier, 2004, Phys. Rev. Lett. {\bf 93}, 130409
\bibitem[\mbox{}]{ger72} Gerchberg, R. W. and W. O. Saxon, 1972, Optik {\bf 35}, 237.
\bibitem[\mbox{}]{Gran86} Grangier, P., G. Roger and A. Aspect, 1986, Europhys. Lett. {\bf 1}, 173.
\bibitem[\mbox{}]{gra04} Grangier, P., B. C. Sanders, J. Vucovic (Eds.), 2004, New J. Phys. {\bf 6}, Focus Issue on Single Photons on Demand
\bibitem[\mbox{}]{gre95} Greenberger, D. M., M. A. Horne, and A. Zeilinger, 1995, Phys. Rev. Lett. {\bf 75}, 2064.
\bibitem[\mbox{}]{GriceWalmsley} Grice, W. P., and I. A. Walmsley, 1996, J. Mod. Opt. {\bf 43}, 795.
\bibitem[\mbox{}]{gricedecorr} Grice W. P., A. B. Uen, and I. A. Walmsley, 2001, Phys. Rev. A {\bf 64}, 063815.
\bibitem[\mbox{}]{gg} Grosshans, F., and P. Grangier, 2001, Eur. Phys. J. D {\bf 14}, 119.
\bibitem[\mbox{}]{cvcrypt2} Grosshans, F., and Ph. Grangier, 2002,  Phys. Rev. Lett. {\bf 88}, 057902.
\bibitem[\mbox{}]{gut06} Gu\c t\u a, M. and Artiles, L. (2006), math/0611117.
\bibitem[\mbox{}]{HansenThesis} Hansen, H., 2000, Ph.D. thesis (Universit\"at Konstanz).
\bibitem[\mbox{}]{FockHD} Hansen, H., T. Aichele, C. Hettich, P. Lodahl, A. I. Lvovsky, J. Mlynek, and S. Schiller, 2001, Opt. Lett. {\bf 26}, 1714.
\bibitem[\mbox{}]{her80} Herman, G. T., 1980, {\it Image Reconstruction from Projections: The Fundamentals of Computerized Tomography} (Academic Press, New York).
\bibitem[\mbox{}]{NLPhoExp} Hessmo, B., P. Usachev, H. Heydari, and G. Bj\"ork, 2004, Phys. Rev. Lett. {\bf 92}, 180401.
\bibitem[\mbox{}]{Mandel86} Hong, C. K., and L. Mandel, 1986, Phys. Rev. Lett. {\bf 56}, 58.
\bibitem[\mbox{}]{kimbleqed} Hood, C. J., T. W. Lynn, A. C. Doherty, A. S. Parkins, and H. J. Kimble, 2000, Science {\bf 287}, 1447.
\bibitem[\mbox{}]{boyd} Howell, J. C., R. S. Bennink, S. J. Bentley, and R. W. Boyd, 2004, Phys. Rev. Lett. {\bf 92} , 210403.
\bibitem[\mbox{}]{Hradil97} Hradil, Z., 1997, Phys. Rev. A, {\bf 55}, 1561.
\bibitem[\mbox{}]{Hradil99} Hradil, Z., J. Summhammer, and H. Rauch, 1999, Phys. Lett. A, {\bf 261} 20.
\bibitem[\mbox{}]{hra04} Hradil, Z., \v Reh\'{a}\v cek, J., \v Fiur\`{a}\v sek, J. and Je\v zek, M., 2004, in {\it Quantum State Estimation, Lect. Notes Phys.}, edited by M. Paris and J. \v{R}eh\'{a}\v{c}ek (Springer, Berlin Heidelberg), Vol. 649, p. 235.
\bibitem[\mbox{}]{hra06} Hradil, Z., D. Mogilevtsev, and J. \v{R}eh\'{a}\v{c}ek, 2006, Phys. Rev. Lett. {\bf 96}, 230401.
\bibitem[\mbox{}]{iac96} Iaconis, C., and I. A. Walmsley, 1996, Opt. Lett. {\bf 21}, 1783.
\bibitem[\mbox{}]{jan95} Janicke, U., and M. Wilkens, 1995, J. Mod. Opt. {\bf 42}, 2183.
\bibitem[\mbox{}]{JacobsKnight} Jacobs, K., and P. L. Knight, 1996, Phys. Rev. A {\bf 54}, 3738.
\bibitem[\mbox{}]{White01} James, D. F. V., P. G. Kwiat, W. J. Munro and A. G. White, 2001, Phys. Rev. A, {\bf 64}, 052312.
\bibitem[\mbox{}]{jul04} Julsgaard, B., J. Sherson, J. I. Cirac, J. Fiur\'{a}\v{s}ek and E. S. Polzik, 2004, Nature {\bf 432}, 482
\bibitem[\mbox{}]{rub97} Keller, T. E., and M. H. Rubin, 1997, Phys. Rev. A {\bf 56}, 1534.
\bibitem[\mbox{}]{InvBernoulli} Kiss, T., U. Herzog, and U. Leonhardt, 1995, Phys. Rev. A {\bf 52}, 2433.
\bibitem[\mbox{}]{Klyshko1} Klyshko, D. N., 1988, Phys. Lett. A {\bf 128}, 133.
\bibitem[\mbox{}]{Klyshko2} Klyshko, D. N., 1988, Phys. Lett. A {\bf 132}, 299.
\bibitem[\mbox{}]{Klyshko3} Klyshko, D. N., 1988, Sov. Phys. Usp., {\bf 31}, 74.
\bibitem[\mbox{}]{klm} Knill, E. ,R. Laflamme, and G. J. Milburn, 2001, Nature (London) {\bf 409}, 46;
\bibitem[\mbox{}]{Koashi2001} Koashi, M., T. Yamamoto, and N. Imoto, 2001, Phys. Rev. A {\bf 63}, 030301.
\bibitem[\mbox{}]{kok07} Kok, P., W. J. Munro, K. Nemoto, T.C. Ralph, J. P. Dowling, and G.J. Milburn, 2007, Rev. Mod. Phys. {\bf 79}, 135.
\bibitem[\mbox{}]{kuh94} Kuhn, H., D.-G. Welsch and W. Vogel, 1994, J. Mod. Opt. {\bf 41}, 1607.
\bibitem[\mbox{}]{kur97} Kurtsiefer, C., T. Pfau, and J. Mlynek, 1997, Nature (London) {\bf 386}, 150.
\bibitem[\mbox{}]{kok07} Lance, A. M., H. Jeong, N. B. Grosse, T. Symul, T. C. Ralph, and P. K. Lam, 2006, Phys. Rev. A {\bf 73}, 041801
\bibitem[\mbox{}]{White} Langford, N. K., R. B. Dalton, M. D. Harvey, J. L. O'Brien, G. J. Pryde, A. Gilchrist, S. D. Bartlett, and A. G. White, 2004, Phys. Rev. Lett. {\bf 93}, 053601.
\bibitem[\mbox{}]{lee95} Lee, C. T., 1995, Phys. Rev. A {\bf 52}, 3374.
\bibitem[\mbox{}]{lee02}  Lee, K. F., and J. E. Thomas, 2002, Phys. Rev. Lett. {\bf 88}, 097902.
\bibitem[\mbox{}]{lee99} Lee, K. F., F. Reil, S. Bali, A. Wax, and J. E. Thomas,  1999, Opt. Lett. {\bf 24}, 1370.
\bibitem[\mbox{}]{leo97}  U. Leonhardt, 1997, {\it Measuring the Quantum State of Light} (Cambridge University Press, Cambridge).
\bibitem[\mbox{}]{leo93} Leonhardt, U., and H. Paul, 1993, Phys. Rev. A {\bf 48}, 4598.
\bibitem[\mbox{}]{leo96} Leonhardt, U., and M. G. Raymer, 1996, Phys. Rev. Lett. {\bf 76}, 1985.
\bibitem[\mbox{}]{leo96b} Leonhardt, U., M. Munroe, T. Kiss, Th. Richter, and M. G. Raymer, 1996, Opt. Commun. {\bf 127}, 144.
\bibitem[\mbox{}]{lo00} Lo, H. K., 2000, Phys. Rev. {\bf A 62}, 012313.
\bibitem[\mbox{}]{lod07} Lodewyck, J. \ieac 2007, Phys. Rev. A {\bf 76}, 042305.
\bibitem[\mbox{}]{lu97} Lu, C.-Y., X.-Q. Zhou, O. G\"uhne, W.-B. Gao, J. Zhang, Z.-S. Yuan, A. Goebel, T. Yang, J.-W. Pan, 2007, Nature Phys. 3, 91.
\bibitem[\mbox{}]{lundcat} Lund, A. P., H. Jeong, T. C. Ralph, and M. S. Kim, 2004, Phys. Rev. A {\bf 70}, 020101.
\bibitem[\mbox{}]{Fock} Lvovsky, A. I., H. Hansen, T. Aichele, O. Benson, J. Mlynek, and S. Schiller, 2001, Phys. Rev. Lett. {\bf 87}, 050402.
\bibitem[\mbox{}]{DFS} Lvovsky, A. I., and S. A. Babichev, 2002, Phys. Rev. A {\bf 66}, 011801.
\bibitem[\mbox{}]{catalysis} Lvovsky, A. I., and J. Mlynek, 2002, Phys. Rev. Lett. {\bf 88} 250401.
\bibitem[\mbox{}]{Shapiro} Lvovsky, A. I., and J. H. Shapiro, 2002, Phys. Rev. {\bf A 65}, 033830.
\bibitem[\mbox{}]{htmaxlik} Lvovsky, A. I., 2004, J. Opt. B: Q. Semiclass. Opt. {\bf 6} S556.
\bibitem[\mbox{}]{mca95} McAlister, D. F., M. Beck, L. Clarke, A. Mayer, and M. G. Raymer, 1995, Opt. Lett. {\bf 20}, 1181.
\bibitem[\mbox{}]{mca97a}  McAlister, D. F., and M. G. Raymer, 1997, J. Mod. Opt. {\bf 44}, 2359.
\bibitem[\mbox{}]{mca97b}  McAlister, D. F., and M. G. Raymer, 1997, Phys. Rev. A {\bf 55}, 1609.
\bibitem[\mbox{}]{mog07} Mogilevtsev, D., J. \v{R}eh\'{a}\v{c}ek, and Z. Hradil, 2007, Phys. Rev. A {\bf 75}, 012112.
\bibitem[\mbox{}]{mol06} M{\o}lmer, K., 2006, Phys. Rev. A {\bf 73}, 063804.
\bibitem[\mbox{}]{mos07} Mosley, P. J., J. S. Lundeen, B. J. Smith, P. Wasylczyk, A. B. U'Ren, C. Silberhorn, and I. A. Walmsley, 2007, arXiv:0711.1054.
\bibitem[\mbox{}]{mun95} Munroe, M., D. Boggavarapu, M. E. Anderson, and M. G. Raymer, 1995, Phys. Rev. A {\bf 52}, R924.
\bibitem[\mbox{}]{muk03} Mukamel, E., K. Banaszek, I. A. Walmsley, and C. Dorrer, 2003, Opt. Lett. {\bf 28}, 1317.
\bibitem[\mbox{}]{nee07} Neergaard-Nielsen, J. S.,  B. Melholt Nielsen, H. Takahashi, A. I. Vistnes, and E. S. Polzik, 2007, Opt. Express {\bf 15}, 7940.
\bibitem[\mbox{}]{nee06} Neergaard-Nielsen, J. S.,  B. Melholt Nielsen, C. Hettich, K. M{\o}lmer, and E. S. Polzik, 2006, Phys. Rev. Lett. {\bf 97}, 083604.
\bibitem[\mbox{}]{new68} Newton, R. G., and  B. L. Young, 1968, Ann. Phys. (New York), {\bf 49}, 393.
\bibitem[\mbox{}]{nha04} Nha, H. and H. J. Carmichael, 2004, Phys. Rev. Lett. {\bf 93}, 020401.
\bibitem[\mbox{}]{nic74} Nicholson, P. W., 1974, {\it Nuclear Electronics} (Wiley, New York).
\bibitem[\mbox{}]{nie07a} Nielsen, A. E. B. and K. M{\o}lmer, 2007a, Phys. Rev. A {\bf 75}, 023806 (2007).
\bibitem[\mbox{}]{nie07b} Nielsen, A. E. B. and K. M{\o}lmer, 2007b, Phys. Rev. A {\bf 75}, 043801 (2007).
\bibitem[\mbox{}]{nog02} Nogueira, W. A. T., S. P. Walborn, S. Pádua, and C. H. Monken, 2002, Phys. Rev. A {\bf 66}, 053810.
\bibitem[\mbox{}]{oliverstroud} Oliver, B. J., and C. R. Stroud, 1989, Phys. Lett., A {\bf 135},  407.
\bibitem[\mbox{}]{opa96} Opatrn\'{y}, T., D.-G. Welsch, and W. Vogel, 1996, Acta Phys. Slovaca {\bf 46}, 469.
\bibitem[\mbox{}]{opa00} Opatrn\'{y}, T., G. Kurizki, D.-G. Welsh, Phys. Rev. A. 61, 032302 (2000)
\bibitem[\mbox{}]{Ou97} Ou, Z. Y., 1997, J. Opt. B: Qu. Semiclass. Opt. {\bf 9}, 599.
\bibitem[\mbox{}]{EPRexp1} Ou, Z. Y., S. F. Pereira, H. J. Kimble, and K. C. Peng, 1992, Phys. Rev. Lett. {\bf 68}, 3663.
\bibitem[\mbox{}]{our06a} Ourjoumtsev, A., R. Tualle-Brouri, J. Laurat, P. Grangier, 2006a, Science {\bf 312}, 83.
\bibitem[\mbox{}]{our06b} Ourjoumtsev, A., R. Tualle-Brouri, P. Grangier, 2006b, Phys. Rev. Lett. {\bf 96}, 213601.
\bibitem[\mbox{}]{our07} Ourjoumtsev, A., H. Jeong, R. Tualle-Brouri, P. Grangier, 2007, Nature (London) {\bf 448}, 784.
\bibitem[\mbox{}]{ScissorsImoto} \"{O}zdemir, S. K., A. Miranowicz, M. Koashi, and N. Imoto, 2002, Phys. Rev. A {\bf 66}, 053809.
\bibitem[\mbox{}]{zav07b} Parigi, V., A. Zavatta, and M. Bellini, 2007, Science {\bf 317}, 1890.
\bibitem[\mbox{}]{par04} Paris, M., and J. \v{R}eh\'{a}\v{c}ek (Eds.) 2004, {\it Quantum State Estimation, Lect. Notes Phys.} 649 (Springer, Berlin Heidelberg).
\bibitem[\mbox{}]{pati01} Pati, A. K., 2001, Phys. Rev. {\bf A 63}, 014302.
\bibitem[\mbox{}]{pau95} Paul, H., U. Leonhardt, and G. M. D'Ariano, 1995, Acta Phys. Slovaca {\bf 45}, 261.
\bibitem[\mbox{}]{detloop} Pearle, P., 1970, Phys Rev. D {\bf 2}, 1418.
\bibitem[\mbox{}]{Scissors} Pegg, D. T., L. S. Phillips, and S. M. Barnett, 1998, Phys. Rev. Lett. {\bf 81}, 1604.
\bibitem[\mbox{}]{rad88} Radeka, V., 1988, Ann. Rev. Nucl. Part. Sci. {\bf 38}, 217.
\bibitem[\mbox{}]{ral03} Ralph, T. C., 2003, Phys. Rev. A {\bf 68}, 042319 (2003).
\bibitem[\mbox{}]{Rao} Rao, C. R., 1945, {\it Bull. Calcutta Math. Soc.} {\bf 37}, 81.
\bibitem[\mbox{}]{rar95} Rarity, J. G., 1995, Ann. NY Acad. Sci. {\bf 755}, 624.
\bibitem[\mbox{}]{ray97a} Raymer, M. G., 1997, Contemp. Phys. {\bf 38}, 343.
\bibitem[\mbox{}]{ray97b} Raymer, M. G., 1997, J. Mod. Opt. {\bf 44}, 2565.
\bibitem[\mbox{}]{ray94}  Raymer, M. G., M. Beck and D. F. McAlister, 1994, Phys. Rev. Lett. {\bf 72}, 1137.
\bibitem[\mbox{}]{ray95} Raymer, M. G., J. Cooper, H. J. Carmichael, M. Beck, and D. T. Smithey, 1995, J. Opt. Soc. Am. B {\bf 12}, 1801.
\bibitem[\mbox{}]{ray96} Raymer M. G., D. F. McAlister, and U. Leonhardt, 1995, Phys. Rev. A {\bf 54}, 2397.
\bibitem[\mbox{}]{microcavity} Raymer, M. G., J. Noh, K. Banaszek, and I.A. Walmsley, 2005, Phys. Rev. A {\bf 72}, 023825
\bibitem[\mbox{}]{ray04} Raymer, M. G., and M. Beck, 2004, in {\it Quantum State Estimation, Lect. Notes Phys.}, edited by M. Paris and J. \v{R}eh\'{a}\v{c}ek (Springer, Berlin Heidelberg), Vol. 649, p. 235.
\bibitem[\mbox{}]{Rehacek01} \v{R}eh\'{a}\v{c}ek, J., Z. Hradil, and M. Je\v{z}ek, 2001, Phys. Rev. A {\bf 63}, 040303.
\bibitem[\mbox{}]{reh04} \v{R}eh\'{a}\v{c}ek, J. and Z. Hradil, 2004, physics/0404121.
\bibitem[\mbox{}]{reh07} \v{R}eh\'{a}\v{c}ek, J., Z. Hradil, Knill, E. and Lvovsky, A.I., 2007, Phys. Rev. A {\bf 75}, 042108.
\bibitem[\mbox{}]{lee05}  Reil, F.  and J. E. Thomas, 2005, Phys. Rev. Lett. {\bf 95}, 143903.
\bibitem[\mbox{}]{vogelnc3} Richter, Th., and W. Vogel, 2002, Phys. Rev. Lett. {\bf 89}, 283601.
\bibitem[\mbox{}]{roh07} Rohde, P. P., A. P. Lund, 2007, quant-ph/0702064
\bibitem[\mbox{}]{roy77} Royer, A., 1977, Phys. Rev. A {\bf 15}, 449.
\bibitem[\mbox{}]{yamamotohom} Santori, C., D. Fattal, J. V. Caronkovi, 2002, Nature (London) {\bf 419}, 594.
\bibitem[\mbox{}]{Japanese} Sasada, H., and M. Okamoto, 2003, Phys. Rev. A {\bf 68}, 012323.
\bibitem[\mbox{}]{sas06} Sasaki, M. and S. Suzuki, 2006, Phys. Rev. A {\bf 73}, 043806
\bibitem[\mbox{}]{scu97}  Scully, M. O., and M. S. Zubairy, 1997,{\it Quantum Optics} (University Press, Cambridge)
\bibitem[\mbox{}]{sch96}  Schiller, S., G. Breitenbach, S. F. Pereira, T. Müller, and J. Mlynek, 1996, Phys. Rev. Lett. {\bf 77}, 2933.
\bibitem[\mbox{}]{sch35} Schr\"odinger, E., 1935, Naturwissenschaften {\bf 48}, 807.
\bibitem[\mbox{}]{she06} Sherson, J. F., H. Krauter, R. K. Olsson, B. Julsgaard, K. Hammerer, J. I. Cirac, and E. S. Polzik, 2006, Nature {\bf 443}, 557.
\bibitem[\mbox{}]{cvcrypt1} Silberhorn, Ch., T. C. Ralph, N. L\"{u}tkenhaus, and G. Leuchs, 2002, Phys. Rev. Lett. {\bf 89}, 167901.
\bibitem[\mbox{}]{sip95} Sipe, J. E., 1995, Phys. Rev. A {\bf 52}, 1875.
\bibitem[\mbox{}]{slu85}  Slusher, R. E., L. W. Hollberg, B. Yurke, J. C. Mertz, and J. F. Valley, 1985, Phys. Rev. Lett. {\bf 55}, 2409.
\bibitem[\mbox{}]{smi05} Smith, B. J., B. Killett, M. G. Raymer, I. A. Walmsley, and K. Banaszek, 2005, Opt. Lett. {\bf 20}, 3365.
\bibitem[\mbox{}]{smi07} Smith, B. J., M. G. Raymer, 2007, New J. Phys. {\bf 9}, 414
\bibitem[\mbox{}]{smi92} Smithey, D. T., M. Beck, M. Belsley, and M. G. Raymer, 1992, Phys. Rev. Lett. {\bf 69}, 2650.
\bibitem[\mbox{}]{smi93a} Smithey, D. T., M. Beck, M. G. Raymer, and A. Faridani, 1993, Phys. Rev. Lett. {\bf 70}, 1244.
\bibitem[\mbox{}]{smi93b} Smithey, D. T., M. Beck, J. Cooper, M. G. Raymer and A. Faridani, 1993, Physica Scripta {\bf T48}, 35.
\bibitem[\mbox{}]{smi93c} Smithey, D. T., M. Beck, J. Cooper, and M. G. Raymer, 1993, Phys. Rev. A {\bf 48}, 3159.
\bibitem[\mbox{}]{NLPhoton1} Tan, S. M., D. F. Walls, and M. J. Collett, 1991, Phys. Rev. Lett. {\bf 66}, 252.
\bibitem[\mbox{}]{the02} Thew, R. T., K. Nemoto, A. G. White, and W. J. Munro, 2002, Phys. Rev. A {\bf 66}, 012303.
\bibitem[\mbox{}]{tit66} Titulaer, U. M., and R. J. Glauber, 1966, Phys. Rev. {\bf 145}, 1041.
\bibitem[\mbox{}]{tor05} Torres, J. P., F. Maci\`{a}, S. Carrasco, and L. Torner, 2005, Opt. Lett. {\bf 30}, 314.
\bibitem[\mbox{}]{ure05} U'Ren, A. B., C. Silberhorn, K. Banaszek, I. A. Walmsley, R. Erdmann, W. P. Grice, and M. G. Raymer, 2005, Laser Phys., {\bf 15}, 146.
\bibitem[\mbox{}]{ure03} U'Ren, K. Banaszek, and I. A. Walmsley, 2003, Quantum Inf. Comput. {\bf 3}, 480
\bibitem[\mbox{}]{ure06} U'Ren, A. B., R. K. Erdmann, M. de la Cruz-Gutierrez, and  I. A. Walmsley, 2006, Phys. Rev. Lett. {\bf 97}, 223602.
\bibitem[\mbox{}]{ure07} U'Ren, A. B., Y. Jeronimo-Moreno, and H. Garcia-Gracia, 2007, Phys. Rev. A {\bf 75}, 023810.
\bibitem[\mbox{}]{Usami} Usami, K., Y. Nambu, Y. Tsuda, K. Matsumoto and K. Nakamura, 2003, Phys. Rev. A {\bf 68} 022314.
\bibitem[\mbox{}]{van01} van Enk, S. J. and O. Hirota, 2001, Phys. Rev. A {\bf 64}, 022313.
\bibitem[\mbox{}]{vanenk05} Van Enk, S. J., 2005, Phys. Rev. A {\bf 72}, 022308
\bibitem[\mbox{}]{vai05} Vaidman, L., 1995, Phys. Rev. Lett. {\bf 75}, 2063.
\bibitem[\mbox{}]{van07} S. J. van Enk, N. Lutkenhaus, and H. J. Kimble, 2007, Phys. Rev. A {\bf 75}, 052318.
\bibitem[\mbox{}]{YardiLee} Vardi, Y. and D. Lee, 1993, J. R. Statist. Soc B {\bf 55} 569.
\bibitem[\mbox{}]{vas00}  Vasilyev, M., S. K. Choi, P. Kumar, and G. M. D'Ariano, 2000, Phys. Rev. Lett. {\bf 84}, 2354.
\bibitem[\mbox{}]{vogelnc1} Vogel, W., 2000, Phys. Rev. Lett. {\bf 84}, 1849.
\bibitem[\mbox{}]{vog89} Vogel, K.,  and H. Risken, 1989, Phys. Rev. A {\bf 40}, 2847.
\bibitem[\mbox{}]{vos02} Voss, P., T.-G. Noh, S. Dugan, M. Vasilyev, P. Kumar, and G. M. D'Ariano, 2002, J. Mod. Opt. {\bf 49}, 2289.
\bibitem[\mbox{}]{wak07} Wakui, K., H. Takahashi, A. Furusawa, and M. Sasaki, 2007, Opt. Express {\bf 15}, 3568.
\bibitem[\mbox{}]{wal04}  Walborn, S. P., A. N. de Oliveira, R. S. Thebaldi, and C. H. Monken, 2004, Phys. Rev. A {\bf 69}, 023811.
\bibitem[\mbox{}]{walt03} Walton, Z. D., M. C. Booth, A. V. Sergienko, B. E. A. Saleh, and M. C. Teich, 2003, Phys. Rev. A {\bf 67}, 053810.
\bibitem[\mbox{}]{walt04} Walton, Z. D., A. V. Sergienko, B. E. A. Saleh, and M. C. Teich, 2004, Phys. Rev. A {\bf 70}, 052317.
\bibitem[\mbox{}]{was06} Wasilewski, W., A. I. Lvovsky, K. Banaszek, C. Radzewicz, 2007, Phys. Rev. A {\bf 73}, 063819
\bibitem[\mbox{}]{wel99} Welsch, D.-G., W. Vogel, and T. Opatrny, 1999, in {\it Progress in Optics}, ed. by E. Wolf (North Holland, Amsterdam), Vol. XXXIX, p. 63.
\bibitem[\mbox{}]{gra04a} Wenger, R., R. Tualle-Brouri, and P. Grangier, 2004, Opt. Lett. {\bf 29}, 1267.
\bibitem[\mbox{}]{gra04b} Wenger, R., R. Tualle-Brouri, and P. Grangier, 2004, Phys. Rev. Lett. {\bf 92}, 153601.
\bibitem[\mbox{}]{wig32}  Wigner. E. P., 1932, Phys. Rev. {\bf 40}, 749.
\bibitem[\mbox{}]{wu86} Wu, L. A., H. J. Kimble, J. L. Hall, and H. Wu, 1986, Phys. Rev. Lett. {\bf 57}, 2520.
\bibitem[\mbox{}]{yar07} T. Yarnall, A. F. Abouraddy, B. E. A. Saleh, and M. C. Teich, 2007, Phys. Rev. Lett. {\bf 99}, 170408 (2007)
\bibitem[\mbox{}]{bellinihd} Zavatta, A., M. Bellini, P. L. Ramazza, F. Marin, and F. T. Arecchi, 2002, J. Opt. Soc. Am. B {\bf 19}, 1189.
\bibitem[\mbox{}]{SPACS} Zavatta, A., S. Viciani, and M. Bellini, 2004, Science {\bf 306}, 660.
\bibitem[\mbox{}]{bellinipra} Zavatta, A., S. Viciani, and M. Bellini, 2004, Phys. Rev. A {\bf 70}, 053821.
\bibitem[\mbox{}]{zav05a} Zavatta, A., S. Viciani, and M. Bellini, 2005a, Phys. Rev. A {\bf 72}, 023820.
\bibitem[\mbox{}]{zav05} Zavatta, A., S. Viciani, and M. Bellini, 2005b, Laser Phys. Lett. {\bf 3}, 3.
\bibitem[\mbox{}]{zav07a} Zavatta, A., V. Parigi, and M. Bellini, 2007, Phys. Rev. A {\bf 75}, 052106.
\bibitem[\mbox{}]{zha04} Zhang, Z. M., 2004, Modern Phys. Lett. {\bf 18}, 393.
\bibitem[\mbox{}]{ZZW95} Zukowski, M., A. Zeilinger, and H. Weinfurter, 1995, Ann. NY Acad. Sci. {\bf 755}, 91.



\end{thebibliography}
\end{document}